\documentclass[11pt,twoside]{book} 
\usepackage{asp2008n}
\usepackage{times}
\usepackage{lscape}
\usepackage{epsf}

\usepackage{graphicx}
\usepackage{amssymb}
\usepackage{longtable}
\usepackage[figuresright]{rotating}
\usepackage{multirow}

\newcommand{\simless}{\mathbin{\lower 3pt\hbox {$\rlap{\raise 5pt\hbox{$\char'074$}}\mathchar"7218$}}}

\mainmatter

\newlength{\deftabcolsep}
\begin{document}

\title{Star Formation in the Rosette Complex}   %%% Fill in title

\author{Carlos G. Rom\'an-Z\'u\~niga}   %%% Fill in author names
\affil{Centro Astron\'omico Hispano Alem\'an, \\
Jes\'us Durb\'an Rem\'on 2-2, Almer\'ia 04004, Spain}

\author{Elizabeth A. Lada}
\affil{Astronomy Department, University of Florida, \\
211 Bryant Space Sciences Center, Gainesville, FL 32611, USA}    %%% Fill in author affiliations

\begin{abstract} The Rosette Complex in the constellation of Monoceros is a magnificent
laboratory for the study of star formation. The region presents an
interesting scenario, in which an expanding HII region generated by the large
OB association NGC~2244 is interacting with a giant molecular cloud. Inside
the cloud a number of stellar clusters have formed recently. In this
chapter we present a review of past and present research on the region,
and discuss investigations relevant to the physics of the nebula and
the molecular cloud. We also review
recent work on the younger embedded clusters and individual nebulous objects
located across this important star forming region.

\end{abstract}

%%% MAIN BODY OF TEXT GOES HERE. CONSULT "INSTRUCTIONS FOR AUTHORS USING
%%% LATEX2E MARKUP", SECTIONS 2.3-2.6 FOR HELP WITH EQUATIONS, FIGURES,
%%% AND TABLES.

%\section{}   %%% Top level section head (remove "%" symbol)
%\subsection{}   %%% Second level section head (remove "%" symbol)
%\subsubsection{}   %%% Lowest level section head (remove "%" symbol)
%\section*{}	%%% Unnumbered top level section head (remove "%" symbol)
%\subsection*{}   %%% Unnumbered second level section head (remove "%" symbol)

\section{Historical Perspective}

\begin{figure}[!ht]
\includegraphics[width=\textwidth]{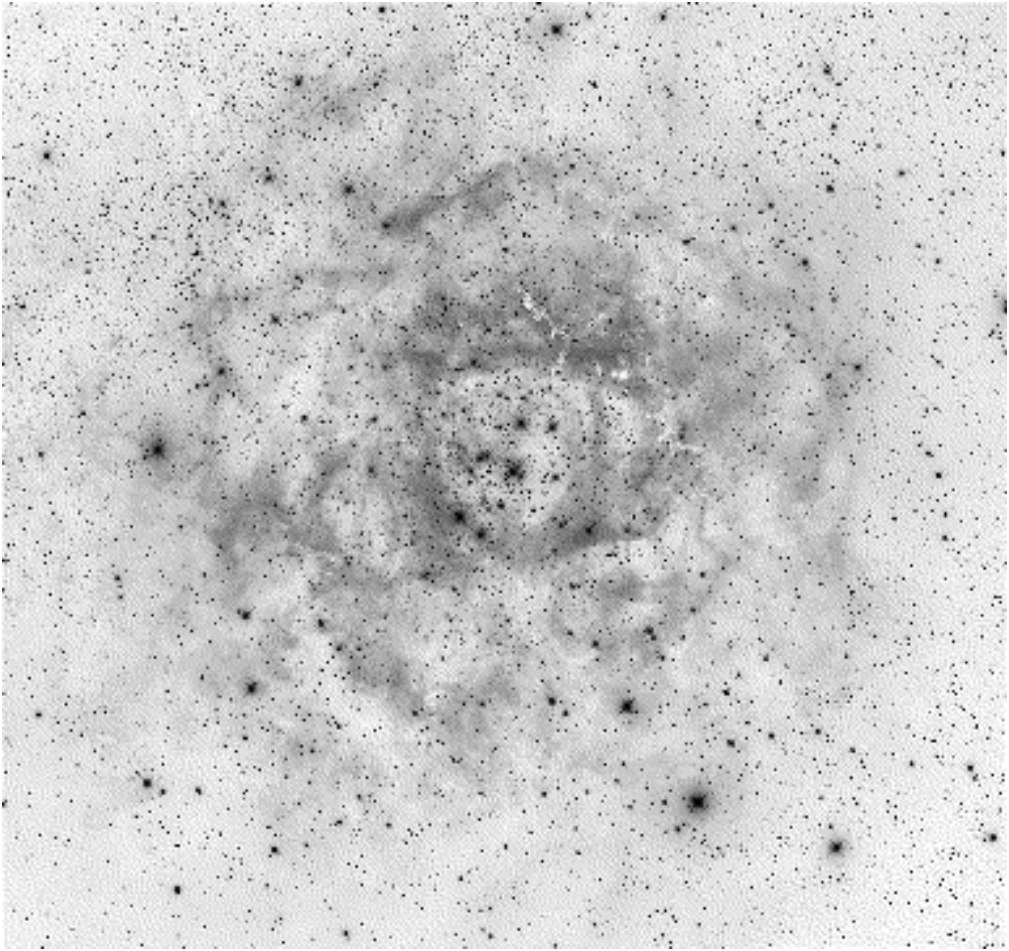}
 \caption{The majestic {Rosette Nebula} in Monoceros. Image courtesy of Robert Gendler.}
\label{f:photo}
\end{figure}

 {The Rosette Complex} ($l=207.0$, $b=-2.1$) is located near the anti-center of the Galactic Disk
in the constellation of Monoceros. The region is very popular, partly because of the staggering beauty of
its main feature: a very extended emission nebula which hosts
a large central HII region, evacuated by the winds of a central
OB association (see Figure \ref{f:photo}).

\begin{figure}[!ht]
\includegraphics[draft=False,width=1.0\textwidth]{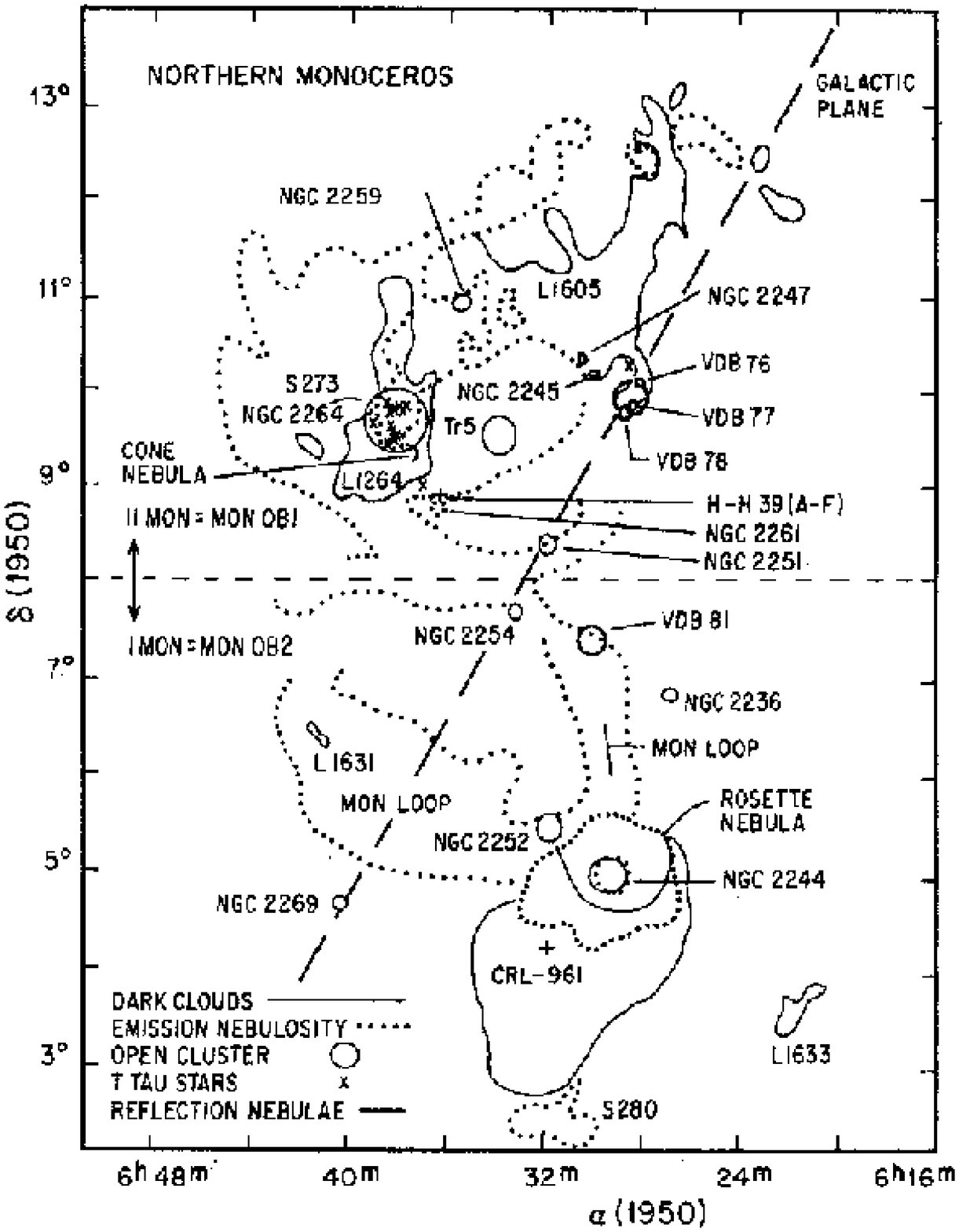}
 \caption{The location of the {Rosette} in the context of
the Monoceros Complex, from P\'erez (1991)}
\label{f:perez3}
\end{figure}

The complex is part of a much larger structure known as the
 Northern Monoceros Region. This region comprises the {Mon OB1 Cloud}
 (host of {NGC~2264} and the {Cone Nebula),} the {Monoceros Loop} {(NGC 2252),} and the {Mon OB2 Cloud}
 in which the {Rosette} is one of the most prominent features (see Figure \ref{f:perez3}).

 The catalog name for the {Rosette} can be somewhat confusing because it is not unique:
 The nebula itself is usually cataloged as {NGC~2237} or {NGC~2246}
 (especially by amateur observers), although {NGC~2237}
 originally referred to the brightest patch at its western side and {NGC~2246}
originally pointed to a bright zone at the eastern side.
 In addition, while the central cluster is usually known as {NGC~2244,} it has also
 been cataloged as {NGC~2239.} However, this latter designation historically referred to
 the brightest star in the region, {12 Monocerotis.}

The cluster was first noticed by Flamsteed in the late 17th century
and later reported by William Herschel --- who did not notice the nebulosity --- and
John Herschel, who discovered several conspicuous nebulosity features
and reported them in his general catalog
\citep[][NGC 2239 = GC 1420]{hershel1864}.
A contemporary report on nebulous emission was done by Albert Marth
\citep[][NGC 2238 = GC 5361 = Marth 99]{lassell1867},
 who described a ``small, faint star in nebulosity."

 Other parts of the nebula {(NGC 2237} and {NGC 2246)} were reported by
\citet{swift2} who cataloged the object as being ``pretty bright [pB],
very, very large [vvL] and diffuse [diff]."  Afterwards, the region
 was formally known as the ``Swift Nebula," until the name {``Rosette"}
 became more popular. The total extent of the {Rosette} was not
determined until the first photographic plates were obtained by
\citet{barnard1894}.

 Two of the first applications of {Rosette Nebula} photographic data were
made by \citet{hubble22} in his study of diffuse nebulae associated
with massive stars, and \citet{minkowski49}, who published a
photographic study along with a first discussion on the expansion of
the HII region by the winds of O stars in the central
cluster. Minkowski also pointed out the existence of elephant trunks
and dark globules. He estimated the mass of the nebula to be $10^4
M_\odot$ and suggested that it could be ``surrounded and pro\-ba\-bly
embedded in obscuring material,'' thus proposing the existence of the
companion molecular cloud.

 The first formal radio wavelength observation of the {Rosette Nebula}
was reported by \citet{kokrauss55}. They surveyed the area with the
Ohio State University antenna at 187.5 MHz with a HPBW resolution of
1.3$\arcmin$, and calculated a blackbody temperature of 200 K for the
nebula. Later, \citet{menon62} used the 85 ft. Tatel telescope to make
a very complete study of the nebula at 3 GHz (10 cm). His analysis
 showed convincingly that the {Rosette Nebula} is an ionization bounded
Str\"omgren sphere and he discussed how it could have formed from a
fraction of a large dark cloud with a total mass of approximately
1.4$\times 10^4$~M$_\odot$.

\section{The Rosette Nebula and the Young Cluster NGC~2244}

\subsection{The Rosette Nebula}

\citet{oguraishida81} tabulated the different methods used to
determine the age of the nebula. These varied from studies of the
properties of the central cavity \citep{kahnmenon61,lasker66} to
evolutionary models of the HII region based on the luminosity of the
stars \citep{hjellming68}. Other methods involve time scales of
radiation pressure \citep{mathews66,mathews67}, estimates of the
formation time for dark globules in the nebula \citep{herbig74}, and
the separation of [OIII] emission lines \citep{smith73}. The median
value of all these age estimates is approximately $3\pm 1\times 10^6$
yr.

A series of studies by \citet{celnik1,celnik2,celnik3} discussed the
 global physical characteristics of the {Rosette Complex.} The first two
of these are dedicated to the nebula, while the third one is a model
of the interaction with the molecular cloud, thus we defer its
discussion to Section 3. In the first of the \citeauthor{celnik1}
articles, he presented a map of the H$\alpha$ emission in the nebula
region, and calculated a total integrated flux density of $5\times
10^{-11}$~W$\cdot$m$^{-2}$ within 60$\arcmin$ from the center of the
HII cavity. He suggested that the emission is contained in a more or
less symmetric ring with a peak at 16$\arcmin$ from the center.

\begin{table}[!htb]
%\caption
\begin{center}
{\small
{Table 1.~~~ Distance Estimates to the Rosette (NGC~2244)}
\smallskip
\begin{tabular}{llc}
\tableline
\noalign{\smallskip}

{Author} & {Value (pc)} & {Method} \\

\noalign{\smallskip}
\tableline
\noalign{\smallskip}

Johnson (1962)         & 1660      & Photoelectric Photometry \\
Ogura \& Ishida (1981) & 1420      & Visual Photometry \\
P\'erez et al. (1987)   & 1670      & Visual Photometry \\
Park \& Sung (2002)    & 1660      & Visual Photometry \\
Hensberge et al. (2002) & 1390      & Spectroscopy \\

\noalign{\smallskip}
\tableline
\end{tabular}
}
\end{center}
\label{t:distances}
\end{table}
\stepcounter{table}

In the second paper, \citeauthor{celnik2} reported radio continuum
observations (1410 and 4750 MHz) from which he was able to determine
that the nebula is bound by ionization, forming a spherical shell with
radius of $\sim$20-30~pc \citep[in agreement with the 3.7, 7.7 and
14.3 GHz observations of][]{kaidanovskii80} and a total ionized matter
mass of 2.3$\times$10$^4$ M$_\odot$ \citep[almost twice the estimate
of 1.1$\times$10$^4$ M$_\odot$ by][]{menon62}. Using the H112$\alpha$
and He112$\alpha$ (4619 and 4621 Mhz) radio recombination lines (RRL)
Celnik calculated a He$^+$ abundance of 0.12$\pm$0.03 and a non-LTE
average electron temperature for the nebula of $T_e=5800\pm700$~K ---
almost $1100$~K above the LTE --- with no evidence for a significant
gradient with respect to the radial distance from the center.

However, other studies revealed that the LTE temperature of the
 {Rosette} was definitely not uniform. For example \citet{pedlar73} found
$T_e$ values between 6400$\pm$1300 and 8160$\pm$1100~K (average of
7900 K) across the nebula using H166$\alpha$ RRL (1424 MHz). In a
similar study, \citet{viner79} calculated non-LTE values between
3900$\pm$700 and 6000$\pm$500 K from H100$\alpha$ RRL (6480 MHz)
observations. The 25 MHz observations (decameter length) of
\citet{krymkin78} suggested a median $T_e$ of 3600 K. Later,
\citet{deshpande84} using continuum absorption observations at 34.5
MHz combined with 2700 MHz observations by \citet{graham82} argued
that the electron temperature might vary from approximately 5000 K in
the southeastern quadrant of the nebula --- which contains more dust
from the interface with the molecular cloud --- to roughly 8000 K at
the northwestern regions. Finally, \citet{tsivilev02} presented
observations of the H92$\alpha$ RRL at three points coincident with
crucial positions of the 4750 MHz map of Celnik which revealed a
higher average temperature of 7980$\pm$580~K. They also discussed all
 other RRL measurements of the {Rosette Nebula} available in the
literature, and found $T_e$ measurements to vary from 3000 to 10050~K.

The investigation by \citet{shipman1} revealed a good fit to a
$T\propto r^{-\alpha}$, $\alpha=0.4$ model for the temperature
gradient in the nebula cavity which, interestingly, could not be
adjusted to the observed IRAS emission. Instead, they found that this
temperature gradient was better adjusted to $\alpha=0.05$ for
$r<47\arcmin$ and $\alpha=0.2$ for $47\arcmin<r<65\arcmin$ (see also
Section 3.2).

The pioneering observations by \citet{gosakher82} of the HI emission
 at 21 cm made with the RATAN-600 telescope revealed that the {Rosette
 Complex} and the {Monoceros Loop} were enclosed in a thin HI envelope 130
pc in diameter, possibly expanding at 20 km$\cdot$s$^{-1}$, suggesting
that the supernova event that gave origin to the Monoceros remnant
occurred about 1.8$\times 10^4$ yr ago. The study of \citet{guseva84}
discussed in detail how the supernova event in Monoceros could in fact
be responsible for the collision of clouds that triggered star
 formation in {NGC 2264} (1 Mon) and the {Rosette Complex} (2 Mon). Later,
 \citet{kuba93} made a new, local 21 cm map of the {Rosette Complex}
using the Arecibo telescope. They found that neutral gas in the
 {Rosette Complex} is distributed in three main regions which form a
rough, extended shell of about 45~pc in radius around the nebula and
extend beyond the molecular cloud. This shell would have a center of
expansion at $(\alpha,\delta)=(6^h31^m48^s,4^d59^\prime 12^{\prime
\prime}$, J2000), and has a mass close to 2$\times$10$^4$ M$_\odot$,
which implies a budget of kinetic energy for the shell expansion of
approx. 4$\times$10$^{48}$ ergs, or roughly 2\% of the total energy
 available from the stars in {NGC~2244.}

\subsection{NGC~2244}

 The prominent OB association {NGC 2244} contains more than 70 high mass
sources with spectral types O and B (7 and 24 sources respectively,
see Table \ref{t:obtable}). The stars in the cluster are presumed
responsible for the evacuation of the central part of the
nebula. \citet{flynn65} used a Fabry-Perot interferometer to study the
velocity distribution of the low density gas in the cavity and its
relation to the cluster. A subsequent analysis by \citet{smith68}
suggested an initial streaming velocity of 13 km$\cdot$s$^{-1}$ at the
ionization front. Later, in the study of \citet{fountain79}, 700
positions in the nebula were analyzed with multi-slit echelle
spectroscopy in H$\alpha$, and these data revealed large systematic
velocities of up to 20 km$\cdot$s$^{-1}$ at the cavity, a sign of
rapid evacuation of material due to powerful stellar winds. A similar
result was found by \citet{smith73}.

The cluster itself has been the subject of many detailed studies over
the years. The distance to this young cluster (and therefore to the
entire complex) has been estimated many times with slightly different
results. Table 1 is a compilation of these values,
from which the most commonly used is 1600 to 1700~pc.

\subsubsection{Optical Studies}

 Some of the first visual photometric studies of {NGC~2244} were made by
\citet{kirillova58}, who constructed the first HR diagram of the
cluster and assigned to it a distance modulus of 12.1 mag, and by
\citet{johnson62}, who estimated the mean color excess in the cluster
to be $E(B-V)=0.46$ for $R=A_V/E(B-V)=3.0$.  This was confirmed by
\citet{turner76} and later \citet{oguraishida81}, who suggested a
value of $R=3.2\pm0.15$.  \citet{oguraishida81} also proposed an age
of 4$\pm$1 Myr and a star formation efficiency of 22\% for the
cluster. Later, \citet{marschalletal82} completed a proper motion
 study of 287 stars in the {NGC 2244} area. They confirmed membership for
113 objects, 52 of them from the list of \citet{oguraishida81}.

A study that combined photometry as well as spectroscopy was completed
 by \citet{perez1}. They found that some members of {NGC~2244} presented
anomalous values of R, possibly suggesting the coexistence of main
sequence stars with very young objects--likely T Tauri stars. This was
confirmed in the {\it uvby$\beta$} photometry study \citep{perez2}, in
which 4 members presented evidence of being true pre-main sequence
 (PMS) objects. \citet{perez2} also confirmed the age of {NGC~2244} to be
below 4 Myr but spread towards younger values, thus confirming a model
of continuous formation.

\begin{figure}[!ht]
\includegraphics[draft=False,width=1.0\textwidth]{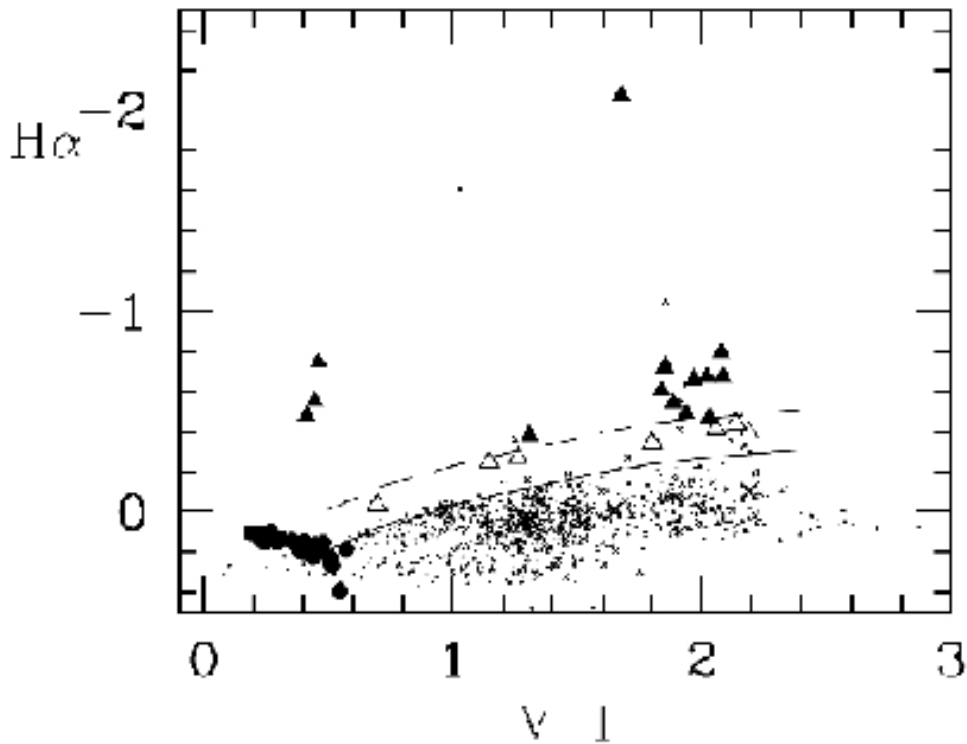}
 \caption{An H$\alpha$ vs. V-I diagram for {NGC~2244.} The solid line
represents a ZAMS relation while the dashed line is a selection
limit. Filled triangles are PMS stars while open triangles are PMS
candidates. Bright members are marked with dark filled circles. X
symbols are X-ray sources and dots are non-members. From
\citet{parkandsung02}. \label{f:halphavi}}
\end{figure}

A study of great importance was performed by
\citet{parkandsung02}. They obtained UBVI and H$\alpha$ photometry for
the cluster. They were able to determine membership for a total of 30
cluster sources and to extend the list of known PMS candidates to
21. They subsequently identified members coincident with ROSAT point
source catalogs and compared their properties to the rest of the
sample. Six of the PMS candidates were confirmed as X-ray sources. In
Figure \ref{f:halphavi}, we show the \citeauthor{parkandsung02}
 diagram of H$\alpha$ emission vs. V-I color for {NGC~2244.} In this
figure, PMS stars are clearly located above the main sequence.

Using evolutionary models, \citeauthor{parkandsung02} showed that most
of the PMS stars and PMS candidates in their sample appear to have
masses close to 1~M$_\odot$ and an approximate mean age of 0.4 to 0.9
Myr. Later, they estimated the main sequence turn-off age of the
cluster to be 1.9 Myr, thus suggesting that the cluster has not
stopped forming stars yet.

Another important determination by Park \& Sung is the Initial Mass
 Function (IMF) of {NGC~2244.} They found it has a flat ($\Gamma=-0.7$)
IMF slope in the range $0.5\leq \log{m}\leq2.0$, which compares well
with the $\Gamma=-0.8$ slope obtained by \citet{massey95}. By
comparing directly to the IMF model of \citet{scalo86} and to the
 observed mass function of {NGC~2264,} \citeauthor{parkandsung02}
 suggested that the shape of the stellar mass spectrum of {NGC 2244} was
dominated by high mass stars. However, they cautioned about the lack
of sensitivity at the intermediate and low mass ranges in their
 sample. More recently, \citet{junfeng08}, showed that {NGC~2244}
actually has a standard IMF shape regardless of the large number of
massive stars.

\subsubsection{Spectroscopic Studies}

 The most complete spectroscopic study of {NGC~2244} was done by
\citet{verstesis}, and it has been widely used in the literature. In
particular, \citeauthor{parkandsung02} used \citet{verstesis} data to
identify the spectral types of candidate T Tauri stars in
 {NGC~2244.} Most of available spectral classifications for stars in
 {NGC~2244} are compiled by \citet{oguraishida81}, but additional data
can be found in \citet{perez1} and \citet{massey95} (see also
\citeauthor{chen04}, \citeyear{chen04} and references therein).

\begin{table}[!htb]
\caption{OB Members of the Young Cluster NGC~2244}
\smallskip
\begin{center}
{\footnotesize
\begin{tabular}{l@{\hskip6pt}c@{\hskip6pt}c@{\hskip6pt}r@{\hskip6pt}r@{\hskip6pt}r@{\hskip6pt}r}
\tableline
\noalign{\smallskip}

{Star ID}\tablenotemark{a} & {RA}\tablenotemark{b} & {DEC}\tablenotemark{b}
& {Sp. Type}\tablenotemark{c} & {V\tablenotemark{d}} & {K}
& {Other}\tablenotemark{e}\\
\cline{2-3} \cline{5-6}
{} & \multicolumn{2}{c}{J2000} & {} &  \multicolumn{2}{c}{[mag]}
& {} \label{t:obtable}\\
\noalign{\smallskip}
\tableline
\noalign{\smallskip}

            HD~46223  &   06 32 09.32  & +04 49 24.6  &      O4V((f))  &   7.32 &   6.68 &	 OI81-203 \\
            HD~46150  &   06 31 55.52  & +04 56 34.3  &      O5V((f))  &   6.75 &   6.44 &	 OI81-122 \\
            HD~46485  &   06 33 50.95  & +04 31 31.6  &       O7V      &   8.20 &   7.45 &	 OI81-387 \\
            HD~46056  &   06 31 20.87  & +04 50 03.9  &      O8V((f))  &   8.16 &   7.82 &	  OI81-84 \\
            HD~46149  &   06 31 52.54  & +05 01 59.1  &    O8.5V((f))  &   7.59 &   7.25 &	   PS-160 \\
           HD~258691  &   06 30 33.31  & +04 41 27.6  &      O9V((f))  &   9.70 &   7.93 &	 OI81-376 \\
            HD~46202  &   06 32 10.48  & +04 57 59.7  &      O9V((f))  &   8.20 &   7.72 &	   PS-305 \\
           HD~259238  &   06 32 18.22  & +05 03 21.7  &       B0V      &  11.10 &  10.28 &	       PS-348 \\
            HD~46106  &   06 31 38.40  & +05 01 36.3  &     B0.2V      &   7.95 &   7.62 &    PS-75; OI81-115 \\
               MJD95  &   06 31 37.08  & +04 45 53.7  &     B0.5V      &  15.15 &  12.20 &		  --- \\
           HD~259135  &   06 32 00.61  & +04 52 41.0  &     B0.5V      &   8.54 &   8.12 &   PS-226; OI81-200 \\
     GSC~00154-00234  &   06 33 37.49  & +04 48 47.0  &     B0.5V      &  11.89 &   8.64 &		  --- \\
           HD~259012  &   06 31 33.46  & +04 50 39.7  &       B1V      &   9.35 &   8.79 &     PS-43; OI81-80 \\
           HD~259105  &   06 31 52.00  & +04 55 57.3  &       B1V      &   9.42 &   8.95 &   PS-155; OI81-128 \\
     GSC~00154-02337  &   06 32 06.13  & +04 52 15.3  &     B1III      &   9.73 &   9.39 &   PS-269; OI81-201 \\
            HD~46484  &   06 33 54.41  & +04 39 44.6  &       B1V      &   7.65 &   6.87 &	     OI81-389 \\
         BD+05~1281B  &   06 31 58.93  & +04 55 39.9  &     B1.5V      &  10.38 &   9.74 &   PS-214; OI81-193 \\
           HD~259172  &   06 32 02.59  & +05 05 08.6  &       B2V      &  10.71 &  10.09 &   PS-240; OI81-167 \\
     GSC~00154-02247  &   06 33 06.56  & +05 06 03.4  &        B2      &  12.85 &  11.20 &	     OI81-345 \\
     GSC~00154-02504  &   06 31 31.47  & +04 50 59.6  &     B2.5V      &  10.64 &  10.24 &     PS-34; OI81-79 \\
     GSC~00154-02141  &   06 31 47.89  & +04 54 18.1  &     B2.5V      &  11.66 &  10.87 &   PS-123; OI81-130 \\
     GSC~00154-02187  &   06 31 58.91  & +04 56 16.2  &    B2.5Vn      &  11.26 &  10.54 &   PS-213; OI81-190 \\
     GSC~00154-01007  &   06 32 09.84  & +05 02 13.4  &     B2.5V      &  11.22 &  10.43 &   PS-300; OI81-172 \\
     GSC~00154-01016  &   06 32 24.24  & +04 47 03.7  &     B2.5V      &  11.41 &  10.58 &   PS-398; OI81-274 \\
     GSC~00154-01247  &   06 33 50.56  & +05 01 37.6  &     B2.5V      &  11.12 &   9.99 &	     OI81-392 \\
     GSC~00154-01753  &   06 32 15.49  & +04 55 20.4  &        B3      &  12.01 &  10.96 &   PS-327; OI81-194 \\
               MJD95  &   06 32 22.49  & +04 55 34.2  &       B3V      &  15.39 &  13.48 &	       PS-383 \\
           HD~259268  &   06 32 23.04  & +05 02 45.7  &        B3      &  11.09 &  10.33 &	       PS-390 \\
           HD~259300  &   06 32 29.39  & +04 56 56.1  &      B3Vp      &  10.79 &   9.34 &	       PS-441 \\
               MJD95  &   06 33 10.16  & +04 59 49.9  &       B3V      &  14.98 &  12.39 &		  --- \\
     GSC~00154-02164  &   06 32 51.79  & +04 47 16.2  &       B5V      &  12.88 &  11.53 &	     OI81-334 \\

\noalign{\smallskip}
\tableline
\noalign{\smallskip}
\multicolumn{7}{l}{\parbox{0.9\textwidth}{\footnotesize $^a$~Star identification. HD catalog is used when available, followed in priority by GSC and
BD catalogs. Stars listed as MJD95 are identified by position in the list of \citet{massey95}.} }\\[1ex]
\multicolumn{7}{l}{\parbox{0.9\textwidth}{\footnotesize $^b$~Coordinates from the 2MASS Point Source Catalog}}\\[1ex]
\multicolumn{7}{l}{\parbox{0.9\textwidth}{\footnotesize $^c$~Spectral types as listed by \citet{junfeng08}, except GSC~00154-02164}}\\[1ex]
\multicolumn{7}{l}{\parbox{0.9\textwidth}{\footnotesize $^d$~Visual (V) photometry from \citet{oguraishida81}, Massey et al (1995) and \citet{parkandsung02}.
Near-infrared (K) photometry from the 2MASS Point Source Catalog.}}\\[1ex]
\multicolumn{7}{l}{\parbox{0.9\textwidth}{\footnotesize $^e$~Correspondence with the lists of \citet{oguraishida81} [OI] and \citet{parkandsung02} [PS].}}\\[1ex]

\end{tabular}
}
\end{center}
\end{table}

A low resolution, single slit investigation by \citet{hensberge1} of 2
 members and 3 field stars in the region of {NGC~2244} yielded evidence
that these were chemically peculiar, possibly magnetic
stars. \citet{hensberge2} performed spectroscopic analysis of the
 binary member {V578 Mon,} which resulted in an estimated distance
slightly lower than other photometric estimates (see Table 1).~They
also calculated the age of the system to be
2.3$\pm$0.2 Myr. Another study by \citet{bagnulo04} revealed that the
source NGC2244-334 (a He-weak B5~V type star) has a very large
longitudinal magnetic field ($-9$ kG), which is the second largest
found in a non-degenerate star. They suggest that given the age of
 {NGC~2244,} this source might also be among the youngest Ap-Bp stars
ever observed, suggesting that these kind of sources can show magnetic
fields from their birth. \citet{lispec02} presented low resolution
spectra for a sample of X-ray counterparts from the ROSAT PSPC survey
\citep[see also][]{gregorio}.~They were able to confirm that five
sources had strong H$\alpha$~emission.~Two of the stars were found to
be Herbig Ae/Be and two others appear to be WTTS.~These data indicate
that X-rays are an efficient tracer of young populations.~

\subsubsection{Near Infrared}

\begin{figure}[!tb]
\includegraphics[draft=False,width=1.00\textwidth]{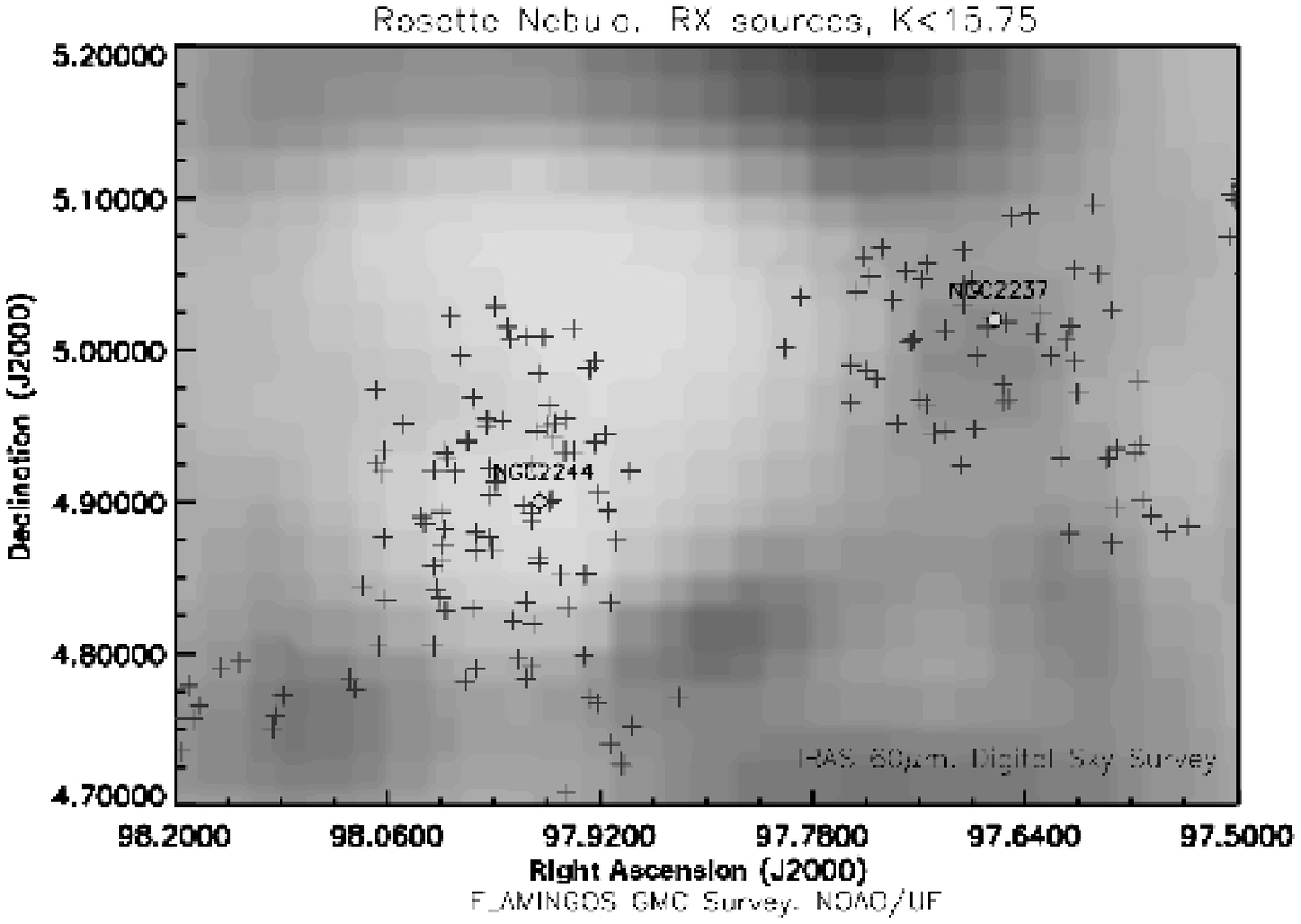}
\caption{Infrared Excess sources with significant local surface
 densities in the central part of the {Rosette Nebula,} from the
 FLAMINGOS survey. The clusters {NGC~2244} and a possible companion
 cluster at the west region of the nebula, {NGC~2237,} are traced by this
population. The background levels indicate 60~$\mu$m emission from
IRAS, in steps of 5.0 Jy. Adapted from \citet{romanzetal1}.}
\label{f:ngc2244andtwin}
\end{figure}

Recent surveys in the near-infrared have permitted investigation of
 the extension and structure of {NGC~2244.} \citet{li05} analyzed 2MASS
 data and suggested that {NGC~2244} had a second component located
approximately 6.6 pc west of the core center. Data from the FLAMINGOS
Survey \citep{romanzetal1} confirmed the existence of this second
association, which is coincident with the area originally labeled as
 {NGC~2237.} In Figure \ref{f:ngc2244andtwin} we show the locations of
infrared excess (IRX) sources with local surface densities above the
field levels. The extinction in the region is rather low ($\langle
A_V\rangle\approx3.0$) compared to the molecular cloud areas but the
distribution of excess sources is well defined and trace the location
 and extension of {NGC~2244} and the satellite cluster.

\begin{figure}[!htb]
\includegraphics[draft=False,width=1.00\textwidth]{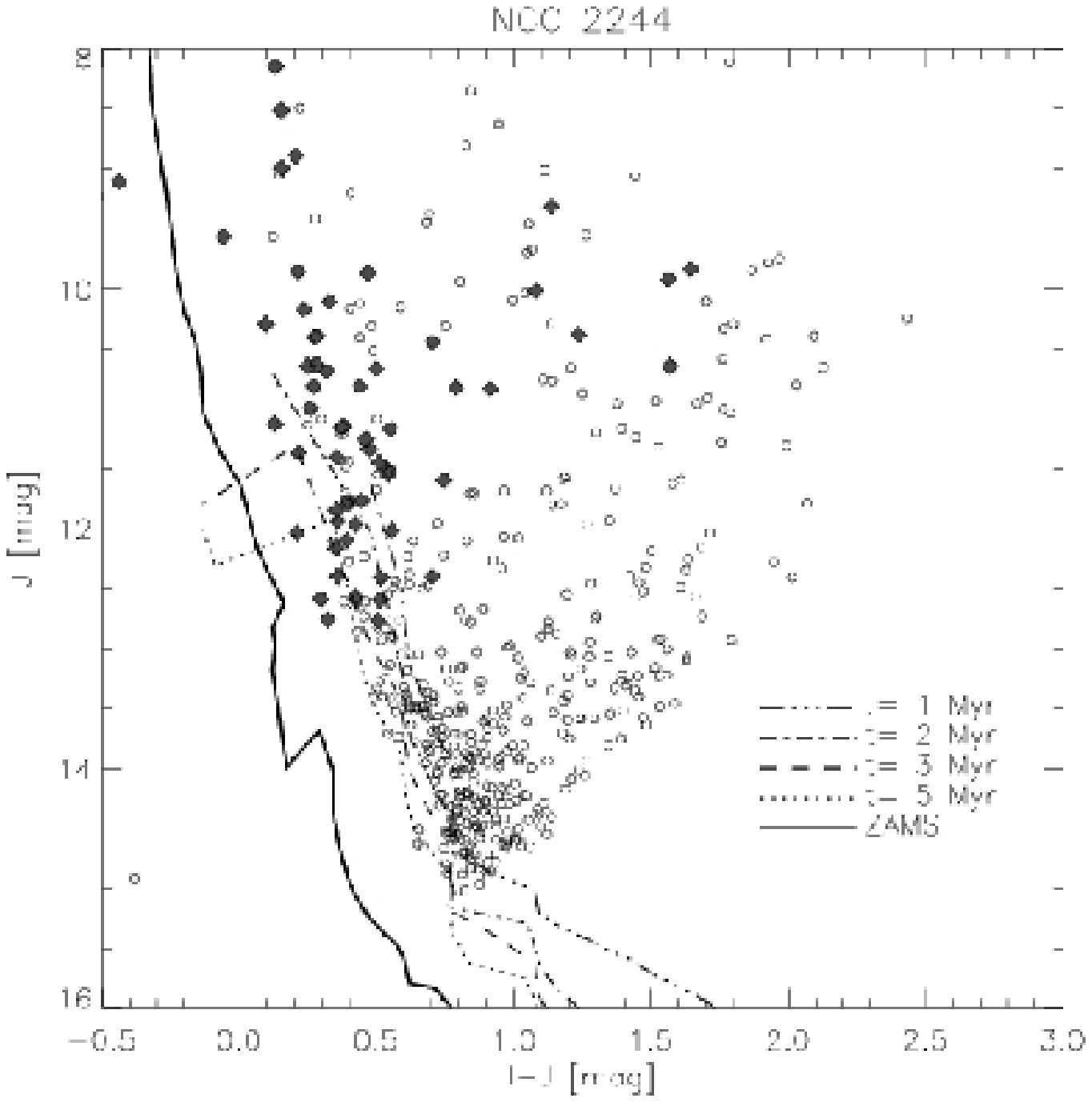}
 \caption{An optical-infrared color-magnitude diagram for {NGC~2244.} The
solid diamond symbols have membership probabilities higher than
60\%. The broken lines are model PMS isochrones at 1, 2, 3 and 5 Myr
from D'Antona and Mazzitelli (1997). The thick solid line is the
ZAMS. \label{f:opt_nir_cmd}}
\end{figure}

 Using a cross-referenced catalog of {NGC~2244} (Rom\'an-Z\'uniga \&
Lada, in preparation) which includes data from the 2MASS and FLAMINGOS
near-infrared surveys and the survey of \citet{parkandsung02}, we were
able to construct two optical-infrared photometric diagrams: Figure
\ref{f:opt_nir_cmd} is an optical-infrared color-magnitude diagram in
which probable members (P=0.6 or more) from the study of
\citet{marschalletal82} are shown with dark symbols. The 1,2,3 and 5
Myr isochrones from the pre-main sequence evolutionary models of
\citet{dm97} are also shown. The 1-3 Myr isochrones are in good
agreement with our data (at least for those stars which are less
affected by reddening).  Notice that the \citeauthor{dm97} models are
calculated only for stars with 3 M$_\odot$ or less.

\begin{figure}[!htb]
\includegraphics[draft=False,width=1.00\textwidth]{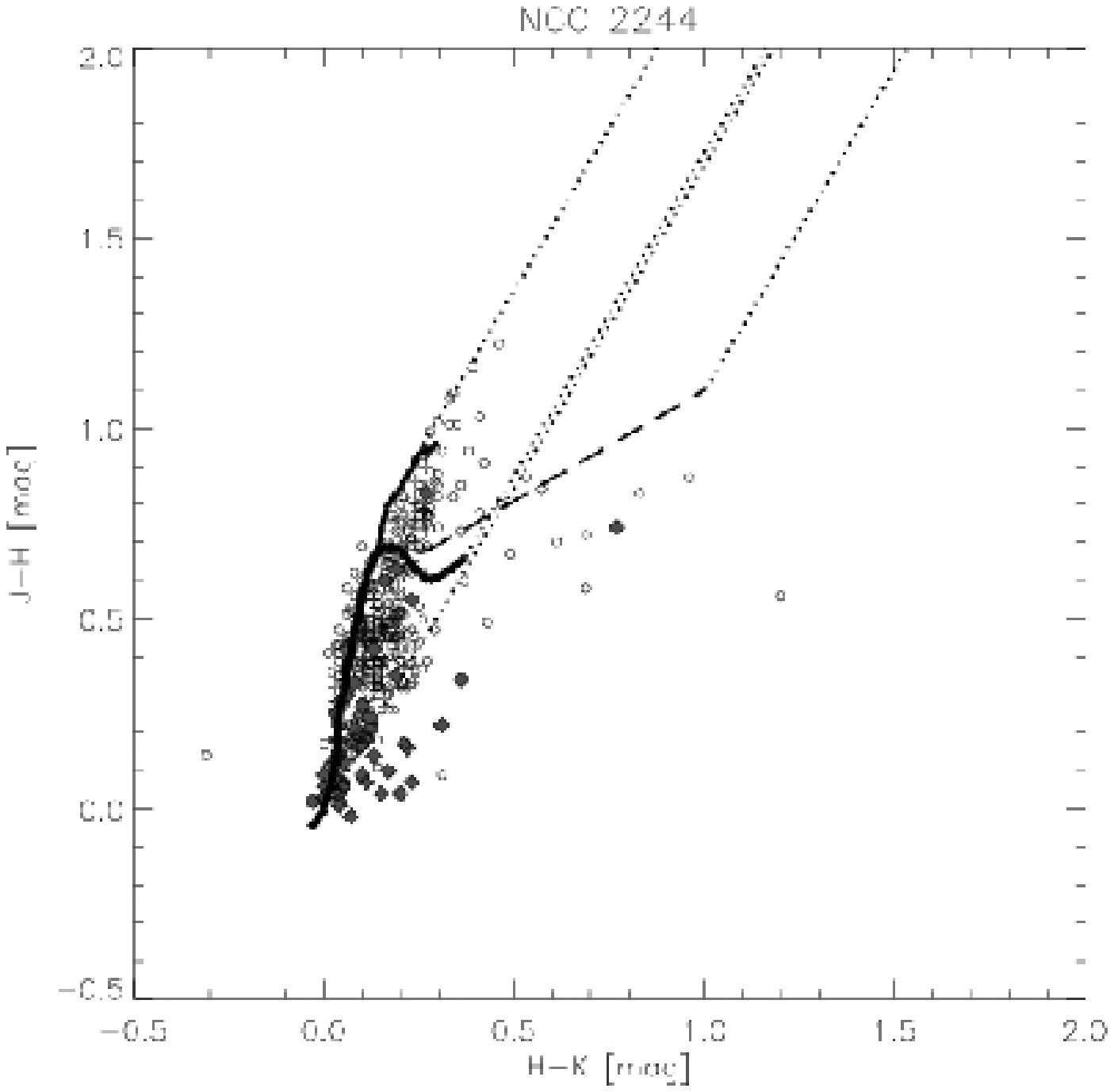}
 \caption{A near-infrared color-color diagram of {NGC 2244.} The solid diamond symbols have membership probabilities higher than 60\%. The solid black lines indicate the ZAMS for dwarfs and giants, while the dotted lines indicate the reddening band from the law of Cohen (1981). \label{f:nir_nir_ccd}}
\end{figure}

Figure \ref{f:nir_nir_ccd} is an infrared color-color diagram of
 {NGC~2244.} The scarcity of sources along the reddening band shows that
 the extinction towards {NGC~2244} is rather low, as the parental
material has been almost completely evacuated. A few of the compiled
stars have large H-K colors and are located to the right of the
reddening band with similar slope and direction as the classical T
Tauri locus \citep{mecahi97}, but with slightly bluer J-H colors
likely caused by nebular gas extinction
\citep{heraudeau96,Ishii02}. Another interesting feature is that some
of the cluster members (solid symbols) form a locus that protrudes to
the right of the reddening band, coincident with the observed locus
for Be stars \citep{howells01,subra05},which suggests the presence of
circumstellar material around some of the massive members of the
association.

In Table ~\ref{t:obtable} we present a list of the O and B members of
 {NGC~2244,} as listed in the X-ray studies by Townsley et al. (2003) and
Wang et al. (2007). The list includes identifications, positions,
spectral types, B and K photometry and a cross identification with the
lists of \citet{oguraishida81} and \citet{parkandsung02}.

\subsubsection{Mid-Infrared Observations}

 {NGC 2244} was observed with the Infrared Array Camera (IRAC) and the
Multiband Imaging Photometer (MIPS) on board the Spitzer Space
Telescope in 2004 and 2005, respectively. \citet{balog07} obtained
from these observations a catalog with more than a thousand high
quality sources detected simultaneously at the four IRAC bands (3.6,
4.5, 5.8 and 8.0 $\mu$m), plus a total of 279 MIPS counterparts at
 24~$\mu$m, succeeding in completing a census of {NGC 2244} in the
mid-infrared down to a mass limit of $\sim$0.8 M$_\odot$. The main
purpose of the study of \citeauthor{balog07} was the investigation of
circumstellar disk survival in the OB environment.

\begin{figure}[!htb]
\includegraphics[draft=False,width=1.0\textwidth]{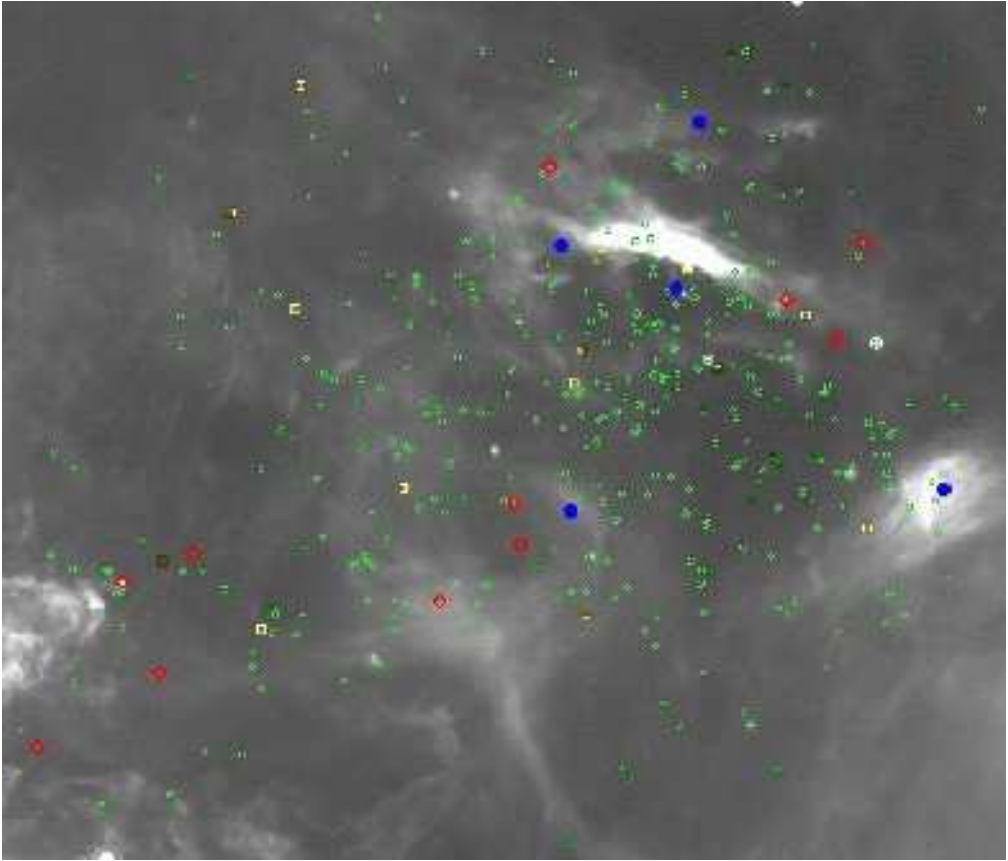}
\caption{Distribution of candidate Class I sources (diamonds), Class
II sources (circles), Class I-II sources (squares) and O stars (solid
 circles) in {NGC 2244,} overlaid on a 24 $\mu$m image obtained with MIPS
onboard the Spitzer Space Telescope. Adapted from
\citet{balog07}. \label{f:irac}}
\end{figure}

A total of 337 and 25 sources were successfully identified as Class II
and Class I sources, respectively (with 213 and 20 sources identified
likewise in the IRAC-MIPS combined catalog). This implies that the
 ratio of Class I to Class II sources in {NGC 2244} is less than 7\%,
much lower than the ratios observed in star forming regions with
smaller proportions of massive stars. \citeauthor{balog07} found that
Class I sources are preferentially located at the outer regions of the
cluster, with none located near the central parsec, opposite from the
Class II sources which appear to concentrate around the OB
center. These two aspects suggest that within the cluster star
formation might have stopped, but it continues outside the HII region
(see Figure \ref{f:irac}).

Later, Balog et al. divided the detections into stars with and without
infrared excess to give an estimate of the disk ratio in the
cluster. They found that the disk ratio within 0.5 pc from the center
of the cluster is 27\%, then it increases to about 45\% within 0.5 and
2.5 pc. The average disk fraction is thus above 40\%, in close
agreement with \citet{halala01} which suggest that the disk fraction
in a cluster decreases from 80 to 50\% within the first 2-3 Myr of
life of a cluster\footnote{For comparison, the disk fraction observed
in the near-infrared is 12\% (see Section 4.1), but such discrepancy
is due to the fact that IRAC is more successful at detecting
circumstellar material because the cold and hot dust that comprises
circumstellar disks emits preferentially within 3 and 10 $\mu$m}. The
 disk fraction in {NGC~2244} calculated by Balog et al. is not very
different ($\Delta\approx 10$\%) from those found in clusters of
similar age without O stars, plus they found a steep decrease of the
photoevaporation rate as a function of distance from the cluster
center, allowing them to conclude that the effect of high mass stars
on disks is only significant at very close distances to the hot stars.

\subsubsection{X-Ray Studies}

 {NGC~2244} is an important target for X-ray studies due to the interest
in investigating the nature of massive stars as sources of high energy
photons. Observations made with the Einstein space telescope by
 \citet{leahy85} revealed that the OB stars in {NGC 2244} were embedded
in a bubble of hot, low-density gas with diffuse X-ray emission at a
temperature of 1 keV, possibly originated by hot shocks in the stellar
winds of O stars. The ROSAT Consortium observations (\citeyear{rosat})
 yielded 34 X-ray sources in {NGC~2244,} with typical X-ray luminosities
of $10^{30}-10^{32}$ ergs~s$^{-1}$. Six of these X-ray sources are PMS
candidates as reported by \citet{parkandsung02}. \citet{berghofer02}
 also studied {NGC~2244 ROSAT} sources and found that objects with the
faintest X-ray emission have very high X-ray to optical luminosity
ratios. They noted that the number of X-ray emitters associated with
 H$\alpha$ emission in {NGC~2244} is remarkable.

\begin{figure}[!ht]
\includegraphics[draft=False,width=1.00\textwidth]{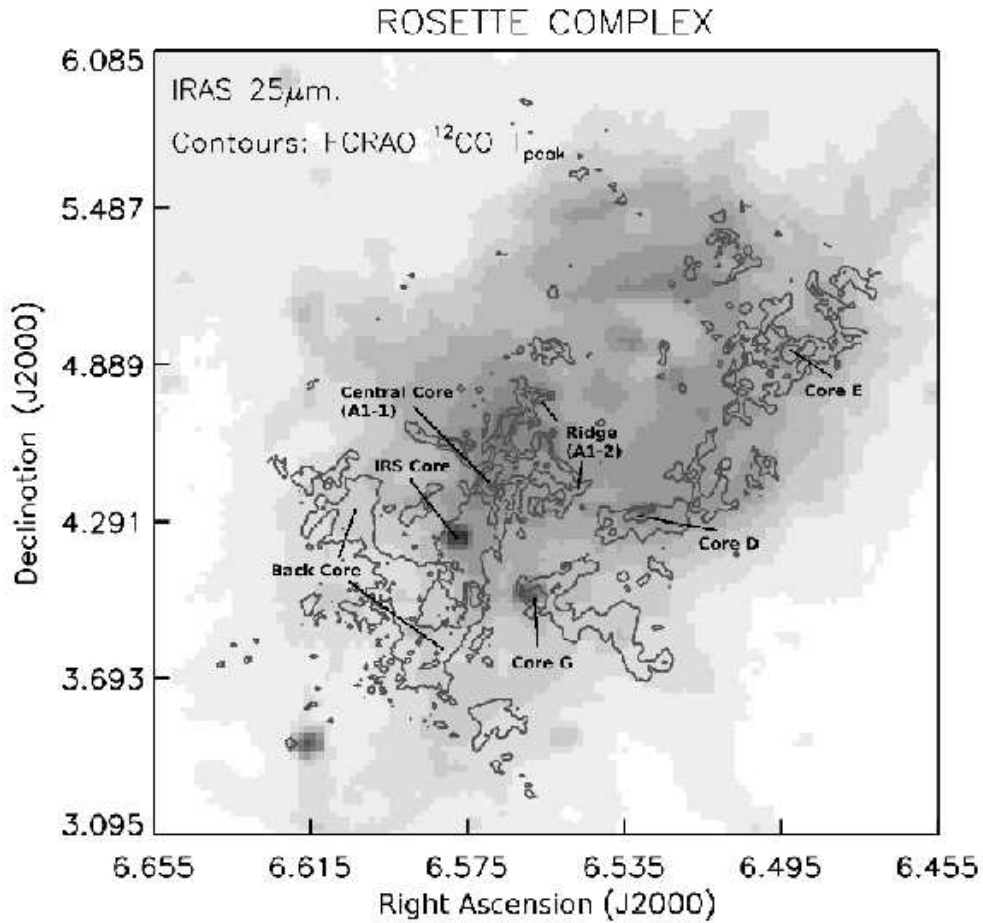}
 \caption{A map of the {Rosette Complex.} The background image is a plate
of the IRAS survey at 25~$\mu$m. The contours represent $^{12}$CO
levels (as $T_{peak}$) from the survey of \citet{HWB06}. Labels
indicate the main regions of the molecular complex, originally
identified by Blitz \& Thaddeus (1980). \label{blitzcloud}}
\end{figure}

A more recent study by \citet{chen04} was dedicated also to analyze
 the coincidence of ROSAT with stars of {NGC~2244.} A careful re-sampling
of the ROSAT data combined with new optical imaging and spectroscopic
data yield 47 X-ray peaks with stellar counterparts in the Nebula
region. Although it was estimated that about half of the counterparts
were late type foreground stars, they could confirmed X-ray emission
for the rest. Specifically, X-ray emission was confirmed for the 2
 earliest type stars in the cluster {(HD~46223} and {HD~46150} with types
05 and O4 respectively), for a very young binary
 {(J063148.30+045820.5),} for the source {HD~259210} which coincides with
the location of high speed knots identified by \citet{meaburnwalsh86},
 and for the source {J063309.61+044624.5} located near a massive nebular
pillar identified by \citet{schneps80} (see also Section 5.2). Taken
together, all these investigations give strength to the hypothesis
 that X-ray emitters in {NGC~2244} are either very massive or very young
in the case of late types.

 Probably the most sensitive X-ray observations of {NGC 2244} to date
were obtained with the Advanced CCD Imaging Spectrometer (ACIS) on
board the Chandra space Observatory. The point source analysis by
\citet{junfeng08} is complementary to the study of \citet{townsley03}
which was dedicated to investigate the diffuse emission in the HII
region (see Section 3.2). Within the ACIS mosaic,
\citeauthor{junfeng08} were able to extract, with high reliability,
over 800 X-ray point sources within 20 arcmin from
$(\alpha,\delta)=(6^h31^m59^s,+4^d55^\prime 36^{\prime \prime}$,
J2000). The X-ray sources have luminosities between 10$^{29.4}$ and
10$^{32.0}$ ergs$\cdot$s$^{-1}$ in the hard band (2-8 keV). All
sources with spectral types earlier than B1 were detected.

A total of 712 Chandra sources in the \citeauthor{junfeng08} survey
were found to be counterparts of near-infrared sources detected in the
FLAMINGOS (see Section 4) and 2MASS surveys, down to a mass limit
close to 0.1~M$_\odot$. Among this X-ray selected sample, the fraction
of stars with mass $>0.5$~M$_\odot$ with significant K-band excess
(likely indicative of a circumstellar disk) is about 6\%, and three of
the sources were positively identified as potential Class I
 sources. The unobscured population in {NGC 2244} is about 1.2 times
 larger than that of the {ONC} \citep{coup} and both the XLF and the KLF
(constructed from the uniformly sampled 2MASS catalog) appears to
suggest a normal Salpeter IMF for stars with mass $>0.5$~M$_\odot$. By
 comparing to the {ONC} distribution, they estimate that the total
 population of {NGC 2244} could be around 2000 members. The radial source
density profile of X-ray sources appears to have a relaxed structure
around the center of the cluster and suggests a low probability of
significant mass segregation.

\section{The Rosette Molecular Cloud and Embedded Populations}

\subsection{Structure of the Rosette Molecular Cloud: CO studies}

 First attempts to detect CO emission associated with the {Rosette}
nebula were unsuccessful as they pointed at the nebula region, which
is mostly composed of neutral and ionized gas. The observations
reported by~\citet{bt80}, which targeted the southeast adjacent region
of the nebula, were the first successful detections of molecular gas
in the Rosette.~Their NRAO survey mapped over 80\% of the $^{12}$CO
emission in the area of the cloud with a 1$\arcmin$ beam size, and
yielded information about its large scale distribution.~They estimated
the angular extent of the cloud to be 3.5$\deg$ (98 pc at a distance
of 1600 pc) and labeled the most prominent sub-structures (see Figure
\ref{blitzcloud}).

Particularly important regions are the {\it Monoceros Ridge} (region
A1-2) which is literally a region of gas compression at the
cloud-nebula interface; the {\it central core} (A1-1), which hosts the
most massive clumps in the cloud and is the strongest region of star
formation; the cores D and G, which are separated from the main body
of the cloud but have ongoing star cluster formation; the {\it IRS
 core}, which hosts the massive proto-binary {AFGL-961} (see Sections 4
and 5.1); the {\it back core} B, which is more loose in structure than
the regions near the nebula; and the {\it arm} or E core, which
despite its brightness contains no significant star formation (no IRAS
sources, or near-infrared clusters have been found in this core so
far).

\begin{figure}[!ht]
\includegraphics[draft=False,width=1.00\textwidth]{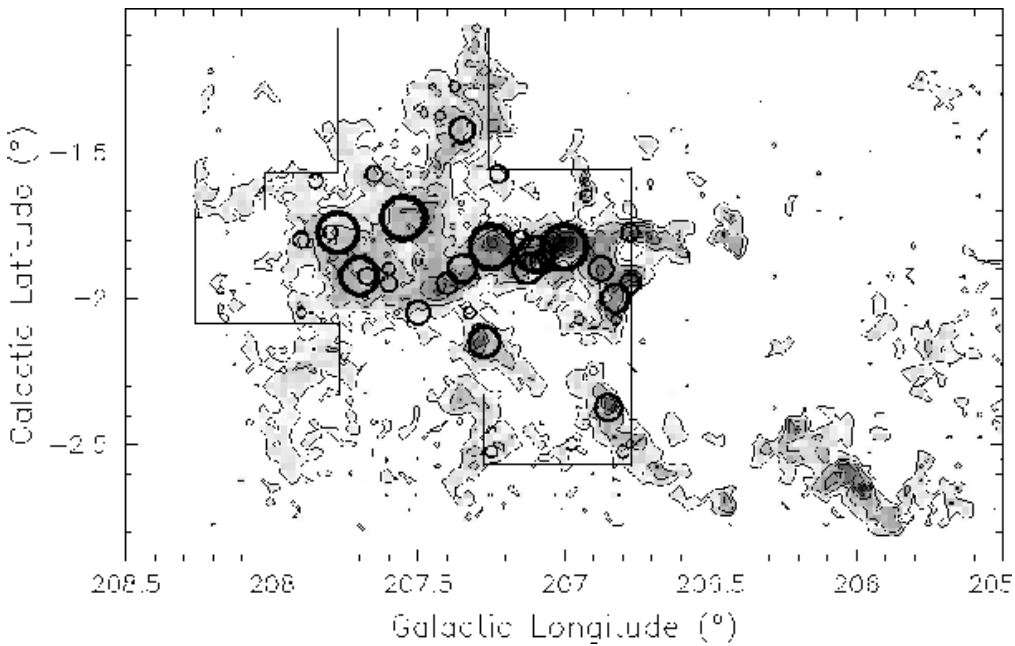}
 \caption{Locations and relative sizes of clumps in the {Rosette
 Molecular Cloud} from \citet{wbs95}. The size of the symbol is
proportional to the mass of the clump. The contours represent CO
emission from the survey of \citet{bs86}, and the solid line delimits
their coverage in $^{13}$CO. \label{wbs95_f7}}
\end{figure}

In a subsequent study, \citet{bs86} mapped the $^{12}$CO and
$^{13}$CO emission with improved sensitivity at the AT\&T Bell Labs,
uncovering the high degree of clumpiness of the cloud. \citet{wdgb94}
made use of the data from \citeauthor{bs86} and listed a total of 95
clumps. The clumps with evidence of star formation had larger peak
temperatures, larger densities and also were more gravitationally
 bound compared to clumps from the {Maddalena Complex}
\citep{maddathad85}, a cloud with very low star formation. Later,
 \citet{wbs95} showed that about half of the clumps in the {Rosette
 Molecular Cloud} were gravitationally bound and the rest were supported
by pressure from the inter-clump medium, which was shown to be mostly
atomic and about 40 times less dense. In Figure \ref{wbs95_f7} we show
the locations and relative sizes of the clumps from \citet{wbs95}.

From the clump central velocities \citet{wbs95} found that the cloud
has a well defined velocity gradient of about 0.08
km$\cdot$s$^{-1}\cdot$pc$^{-1}$. Also, the negative correlation
between clump mass and clump to clump velocity dispersion suggested
that the system is still far from equipartition even though it is
dynamically evolved. \citet{wbs95} also found the star forming
activity to be more intense in the ridge and core areas, near the
interface cloud-nebula: clumps located near the nebula presented
larger excitation temperatures, average densities and star forming
efficiencies, all clues of cloud evolution. Other properties of the
clumps (mass, sizes and line widths) did not show any significant
variations along the cloud.

Another CO study was done by \citet{schneiderco}. The observations
focused on the central part of the cloud, detailing the structure of
the midplane star forming cores. They paid special attention to the
 IRS core, where the source {AFGL-961} is located (see also Section 5.1)
and pointed out the ample blue wing emission due to the powerful
outflow from this object. In a complementary study,
\citet{schneidercs} examined the CII emission (158~$\mu$m) at the
ridge, the central core and the IRS core. They found weak but
significant C$^+$ emission deep into the molecular cloud cores and
suggested that the distribution agrees well with a clumpy molecular
cloud exposed to a low level UV radiation field. The penetration of UV
photons in the cloud is apparently facilitated by a high density
contrast clump-interclump medium.

 The clump mass spectrum in the {Rosette Molecular Cloud} has the form
$dN/dM \propto M^{-x}$ where $x\approx1.6$, depending on the range,
bin size and beam resolution used.  The exponent in this power law is
similar to other clouds \citep{blitz93}, but what is more important,
it is much shallower than the one corresponding to the stellar IMF
($x=2.35$).  This shows that although there are more small clumps by
number, most of the mass is contained in only a few big clumps, while
for stars both numbers and total mass are dominated by the lowest mass
bins. Interestingly enough, the power law index is in fact similar to
that corresponding to the mass distribution function for embedded
clusters, which is suggestive of a uniform star formation efficiency
for most star forming cores \citep{ll03}.

The more recent survey of \citet{HWB06}, obtained with the wide field
array SEQUOIA at the FCRAO 14m telescope achieved resolutions of
45$\arcsec$ at 115 GHz and 47$\arcsec$ at 110 GHz. The maps revealed
``textural variations'' in the $^{12}$CO emission across the complex,
with a brighter emission component within the nebula projected radius
(approx. 30 pc from the center as defined by \citeauthor{celnik2},
\citeyear{celnik2}) and weaker, more extended emission outside this
ionization edge. \citet{HWB06} suggest that the weaker emission is
probably due to subthermally excited material with lower
densities. They also calculated the total molecular mass of the cloud
to be $1.6\times 10^5$~M$_\odot$ from $^{12}$CO, and found a LTE mass
of $1.16\times 10^5$~M$_\odot$ from $^{13}$CO. Moreover, they were
able to apply a Principal Component Analysis \citep{heyershloerb97} to
 determine the turbulent flows and the turbulence scale in the {Rosette
 Molecular Cloud.} The analysis revealed more significant variations in
the velocity structure of the cloud at the regions located within the
ionization than in the more diffuse, external component. This fact
confirms the interaction of the cloud and the HII region. They
suggested, however, that the interaction is still very localized, and
have not affected the global dynamics of the cloud yet.

\subsection{Interaction with the Rosette Nebula}

\citet{celnik3} focused on comparing his H$\alpha$ map and radio
continuum observations of the nebula (see Section 2.1) with the CO map
of the molecular cloud from \citet{bt80}. Celnik constructed a complex
model of the distribution of the main CO cores (see Fig
\ref{wbs95_f7}) in the context of the HII region and estimated the
rotation center of the cloud at
$(\alpha,\delta)=(6^h32^m39^s,+4^d19^\prime 43^{\prime \prime}$,
J2000). Finally, he re-calculated the mass of the entire complex by
adding the total mass of ionized atoms, stars, dust and molecular gas,
resulting in $3.3\times 10^5$~M$_\odot$.

Another important study was done by \citet{coxetal90}, who made use of
the available infrared emission data from the IRAS satellite (12, 25,
60 and 100~$\mu$m). \citeauthor{coxetal90} were able to study in great
detail the distribution of dust and compared this to the distributions
of ionized and molecular gas.~Additionally, they were able to estimate
a total infrared luminosity of roughly 1.1$\times$10$^6$ L$_\odot$, or
 about 50\% of the available luminosity from the cluster {NGC~2244.~Warm}
dust (usually present near an OB association) typically emits strongly
at the four IRAS bands. However, \citeauthor{coxetal90} showed that in
 the {Rosette,} while the 60 and 100~$\mu$m emission are quite strong at
regions of ionized and neutral gas (nebula), the 12~$\mu$m emission is
preferentially located beyond the limits of the ionization front
(molecular cloud), suggesting a heavy rate of destruction of dust
grains from UV radiation from the cluster. Surprisingly, the 25~$\mu$m
emission was found to be significant in some parts of the ionized
nebula, possibly due to the existence of a second type of dust
particle that is more resistant to UV photons. This was also suggested
by \citet{shipman1}, who found that the maximum temperature in the
shallow temperature gradient found in the nebula (see Section 2.1)
seems too low to sublimate ice mantles in grains and too low for
grains to emit significantly at 12 or 25~$\mu$m --- a problem possibly
solved with a second type of grain in the region. Later,
\citet{shipman2} suggested that line emission could be contributing
strongly to the IR emission of the nebula, and suggested once more
that the presence of a ``hot dust" component is necessary to model
this emission, especially at 25~$\mu$m.

\begin{figure}[!htb]
\centering
\includegraphics[draft=False,width=1.00\textwidth]{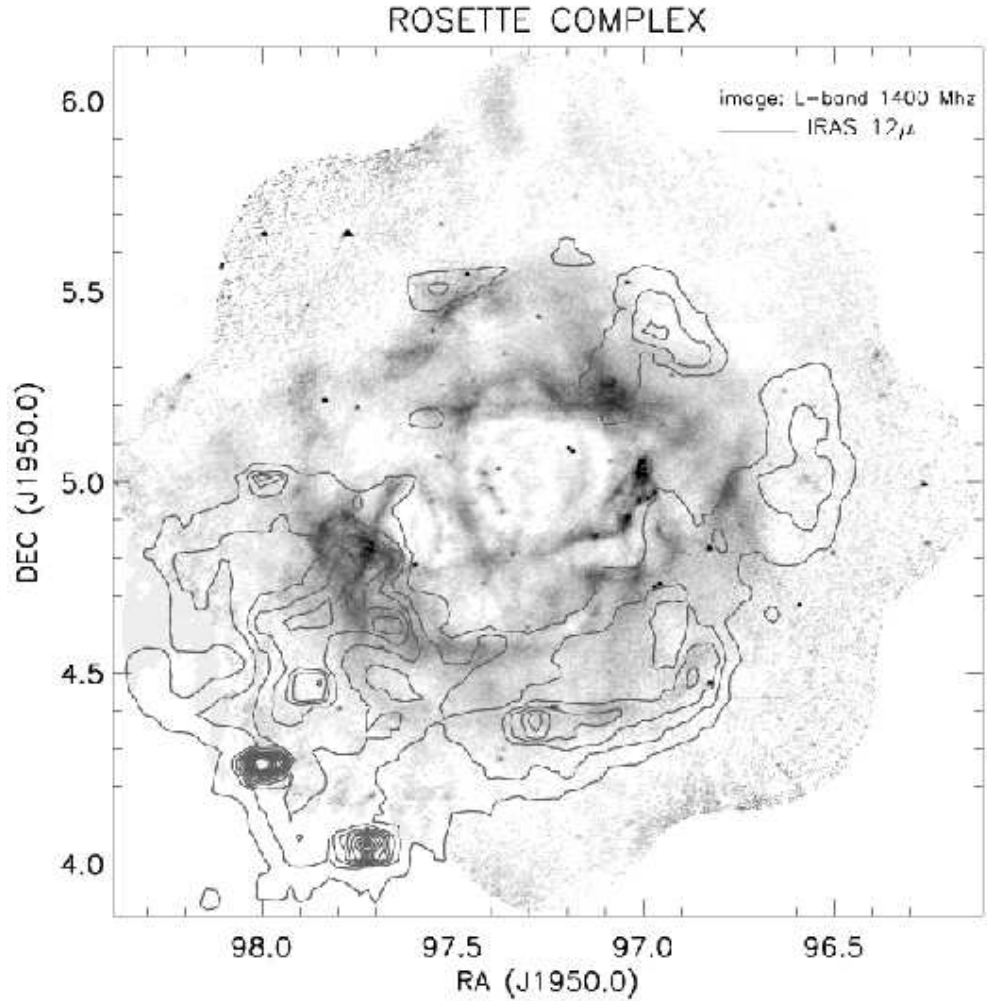}
\caption{IRAS 12~$\mu$m emission map superimposed on a 1400 MHz radio
continuum emission map. Credit: \citeauthor{holdaway99} (1999) and
NRAO. \label{f:LO12mu}}
\end{figure}

Figure \ref{f:LO12mu} shows the superposition of the 12~$\mu$m
emission from the IRAS survey and a 1400 MHz radio continuum emission
map from \citet{holdaway99}.~The infrared contours indicate that the
warm dust emission defines a shell that encloses the ionization front,
showing the effect of heavy dust destruction by the nebula. The
superposition of these maps defines very clearly the region where the
HII bubble impacts the molecular cloud.

\citet{patel93} observed the SE quadrant of the nebula, which
coincides with the regions located south and east of the Monoceros
Ridge, where molecular globules were identified by \citet{block90},
\citet{bdm92} and \citet{sugitani91}. The CO and $^{13}$CO
($J=1\rightarrow 0$) data of \citeauthor{patel93} confirmed that this
part of the cloud contained nine well defined cometary globules, which
appear to be blue-shifted by approximately 6 km$\cdot$s$^{-1}$ with
 respect to the {Rosette Molecular Cloud.} The globules have sizes of
1-3~pc and masses of 50 to 300~M$_\odot$ and four of them are
coincident with IRAS sources and local maxima in the $^{13}$CO
emission. \citeauthor{patel93} suggest that these globules cannot be
formed as Rayleigh-Taylor instabilities as in the case of globules
 (elephant trunks) in the NW quadrant of the {Rosette Nebula} (see
Section 5), but rather by a mechanism of radiation driven implosion,
caused by penetration of UV radiation from the nebula cluster which
favors D-type ionization fronts inside massive clumps, causing the
formation of the globules. A follow up study by \citet{whiteetal97}
also included CO, $^{13}$CO, and C$^{18}$O observations of a similar
region, revealing that the Globule 1 from \citeauthor{patel93} is
indeed the largest cometary globule in the area; this object has a
prominent head-tail structure with a projected length of almost 1.3~pc
which points away from the center of the Nebula. They also confirmed
the agreement with a RDI model and were able to calculate an age of
0.4 Myr for the globule, suggesting that it has almost reached the
stage of maximum compression and is close to a quasi-static cometary
phase.

\subsubsection{High Energy Studies}

 When the {Rosette Molecular Cloud} was confirmed as a region of star
formation, \citet{gregorio} used ROSAT data again, this time to map
 the {MonR2} cluster and the {Rosette Molecular Cloud} areas in order to
confirm a correlation between star forming cores and clusters of
 X-rays sources. They found strong X-ray emission in {NGC~2244,} at the
ridge of the cloud (A1-2 in fig~\ref{blitzcloud}), and at the cloud
core area (A1-1), but the resolution was poor and individual sources
could not be resolved. They suggested that molecular cores known to
have active star formation but failing to show significant X-ray
emission, could be predominantly forming low-mass stars. They also
suggested that detectable X-ray counterparts are in most cases
 Herbig~Ae/Be or T Tauri stars, as found in {NGC~2244}
\citep{lispec02,junfeng08}.

\begin{figure}[!htb]
\includegraphics[draft=False,width=1.00\textwidth]{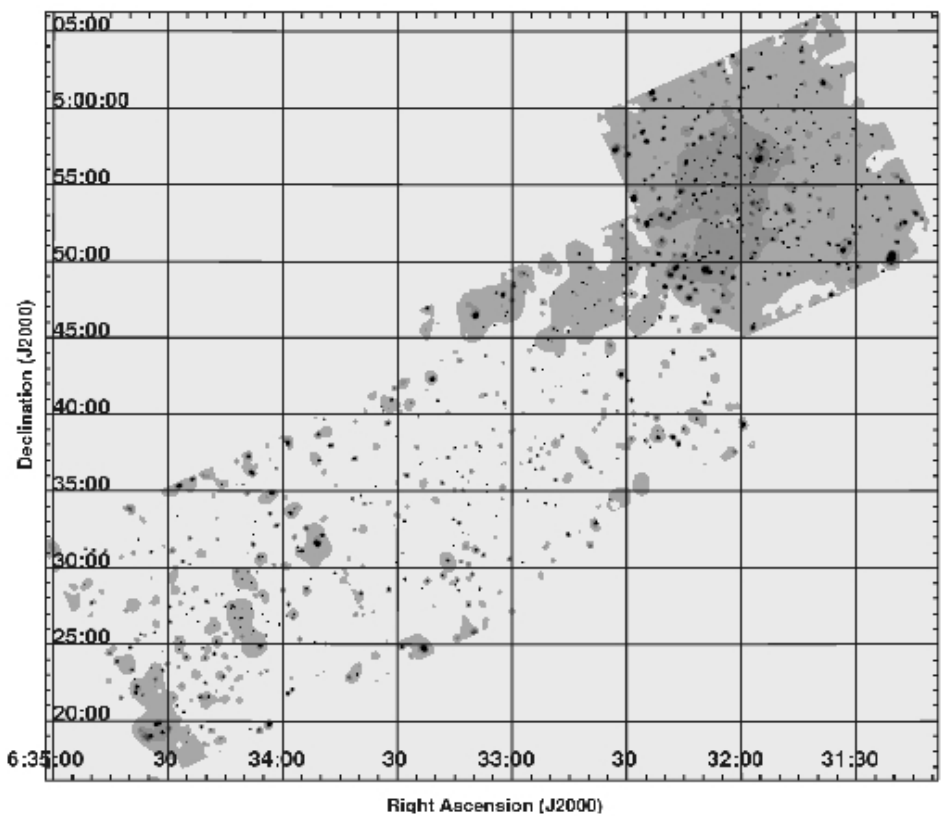}
 \caption{A 0.5-2 keV Chandra image of the {Rosette Complex.} The
emission has been smoothed to highlight the soft diffuse emission that
originates in the nebula and propagates into the molecular
cloud. Credit: \citeauthor{townsley03} (2003) and NASA/Chandra X-Ray
Observatory.}
\label{f:chandra}
\end{figure}

 The more recent observations of the {Rosette Complex} done with Chandra
\citep{townsley03} have resolutions of only a few arcseconds, thus
allowing for the detection of X-ray counterparts for 75\% of the OB
 members of {NGC~2244.} One of the most interesting results of this study
was the confirmation of a second, soft diffuse emission which probably
originates from the O star winds and is later brought to
thermalization by wind-wind interactions or by the shock with the
surroundings, in this case the molecular cloud (see Figure
\ref{f:chandra}). This X-ray plasma surrounds the OB association and
fills the nebula cavity completely.

 {The Rosette Nebula} may not only be interacting with its companion
 molecular cloud. The location of the {Rosette Nebula} near the edge of
 the {Monoceros Loop} \citep{davies63} --- also known as the {Monoceros
 Supernova Remnant ---} at the edge of the NE quadrant, has motivated a
number of studies aimed at investigating the high energy photon
emission in that region. Deep H$\alpha$+[NII] photographic plates by
\citet{davies78} suggested a correlation between a filamentary
 structure observable in H$\alpha$ emission and a {Rosette} nebula
feature observable in 240 MHz radio waves. This feature was proposed
as evidence of loop-nebular interaction and confirmed by decameter
\citep{odegard86} and diffuse X-ray emission \citep{leahy86}
observations.  Later, high energy (100 MeV) $\gamma$-ray images from
EGRET \citep{jaffe97}, revealed a feature partly coincident with the
filaments and apparently significant (7$\sigma$) over expected diffuse
emission. If real, this $\gamma$-ray feature would possibly be due to
the interaction of charged particles with the dense ambient medium at
the shock region. The HEGRA system of atmospheric Cerenkov telescopes
at La Palma Observatory was used to calculate the cosmic ray emission
from the loop-nebula interaction region, and found one possible TeV
emission source (3EG J0634+0521) with low statistical significance
\citep{aharonian04}. More recently, the HESS telescope array in
Namibia was used again to map the SNR-Nebula interaction region and
 found one gamma ray source, {HESS J0632+057,} located close to the rim
 of the {Monoceros Loop} \citep{aharonian07}.

\nopagebreak

\subsection{The Rosette Molecular Cloud: Embedded Populations}

\subsubsection{Near-infrared Observations}

The coincidence of massive clumps and luminous IRAS sources pointed
out by the study of \citet{wbs95} strongly suggested that star
formation had already taken place across the molecular cloud. However,
the poor spatial resolution of the IRAS point source survey did not
allow the resolution of individual members of an embedded
population. Early near-infrared studies \citep[e.g.][]{perez1} did not
cover the molecular cloud areas, and optical photometric studies were
incapable of detecting obscured populations.

An exploratory near-infrared survey (JHK) by \citet{pl97} that made
use of the imager SQIID finally confirmed the existence of embedded
clusters in some of the most massive clumps from the list of
\citeauthor{wbs95} that were associated with an IRAS
source. \citeauthor{pl97} were able to identify by visual inspection
of their images seven deeply embedded clusters with bright
nebulosities. Other massive clumps did not contained a cluster, which
suggests that high mass is a necessary but not a sufficient condition
for cluster formation.

 Complete spatial coverage of the {Rosette Complex} in the near-infrared
was first accomplished with the release of the All-Sky 2MASS survey
catalogs. The 2MASS survey was a major gain in data uniformity but
 unfortunately not in sensitivity. Due to the distance to the {Rosette}
(1.6$\pm$0.2 kpc), the 2MASS completeness limit (K=14.3~mag) is not
deep enough to study the low mass end of the IMF. \citet{lismith3}
studied the distribution of 2MASS sources with large H-K colors and
they were able to suggest three general areas of cluster
formation. The extensions of these star formation regions were defined
visually, and could not be compared with the clusters of \citet{pl97}
because at that point the extents and membership of individual
clusters were not determined.

The survey of \citet{romanzetal1} aimed for deeper observations using
FLAMINGOS on the KPNO-2.1m. The project was done as part of a
FLAMINGOS/NOAO survey of Giant Molecular Clouds \footnote{see
http://flamingos.astro.ufl.edu/sfsurvey/sfsurvey.html}. Observations
 of the {Rosette} covered the regions surveyed by \citeauthor{pl97} and
extended to the Back Core and the nebula. Identification of young
embedded populations was done by selecting objects with near-infrared
excess characteristic of circumstellar emission. Also, a Nearest
Neighbors analysis was applied to this color selected sample to
facilitate the detection of embedded clusters as regions with
significant local surface densities.

\begin{figure}[!ht]
\includegraphics[draft=False,width=1.0\textwidth]{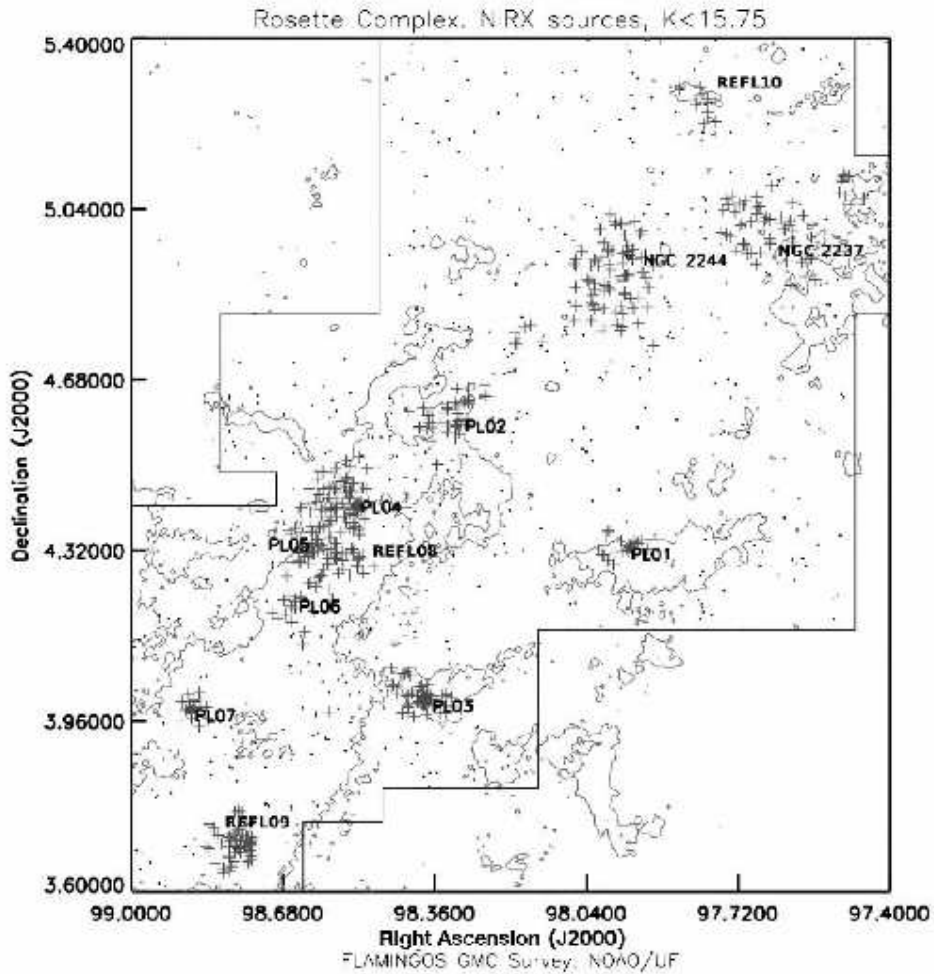}
 \caption{Identification of young clusters in the {Rosette Complex.} Plus
symbols indicate the location of NIRX objects brighter than K$<$15.75
and local surface densities higher than the background average. Black
dots are sources with near-infrared excess colors and local surface
densities below the background average. The contour levels indicate a
baseline level (0.8 K$\cdot$km$\cdot$s$^{-1}$) of $^{13}$CO emission
from the survey of \citet{HWB06} used to define the projected surface
of the cloud. The thin solid line marks the coverage of the FLAMINGOS
survey.}
\label{f:identification}
\end{figure}

The technique allowed for a clear confirmation of the location and
extensions of the seven bright clusters reported in \citeauthor{pl97},
 labeled as {PL01} to {PL07,} but also indicated the existence of another
two clusters associated with massive clumps in the molecular cloud,
 labeled as {REFL08} and {REFL09.} The sources with near-infrared excess
 also traced well the extension of {NGC~2244} and a possible new cluster
at the western edge of the HII bubble, located at the region
 historically identified as {NGC~2237,} (see also Section 2.2). Finally,
a small cluster was detected at the northeast end of the nebula and
 was labeled {REFL10.} The location of the clusters and their
distribution of near-infrared excess sources is shown in Figure
\ref{f:identification}.

 The sizes (equivalent radii) of clusters in the {Rosette} are
relatively large, appear to be anti-correlated with their mean average
extinctions, and with the near-infrared excess fractions. If larger
near-infrared excess fractions indicate slightly younger ages, then
the anti-correlation might be indicative of rapid cluster evolution:
clusters could form as compact units then expanded in a timescale
shorter than the T Tauri phase.

From the direct counting of near-infrared excess stars, it was
possible to determine the fraction of the young stellar population
that is forming in clusters at the present day. The 9 largest clusters
 embedded in the {Rosette Molecular Cloud} occupy approximately 242
$(\arcmin )^2$ or approximately 9\% of the main molecular cloud areas
covered by the FLAMINGOS survey. After correcting for background stars
the embedded clusters account for 86$\pm$5\% of the embedded
population in the cloud. These fractions are comparable to what was
found in other nearby molecular clouds like Orion or Perseus
\citep{la91,ca00}, where clusters comprise 50 to 96\% of the young
stars that can be observed.

\begin{table}[!ht]
\caption{Young Clusters in the Rosette Complex}
\smallskip
\begin{center}
{\footnotesize
\begin{tabular}{lcc@{\hskip8pt}c@{\hskip8pt}c@{\hskip8pt}c@{\hskip8pt}c@{\hskip8pt}c@{\hskip8pt}c}
\tableline
\noalign{\smallskip}
{ID}\tablenotemark{a} &
{RA}\tablenotemark{b} &
{DEC}\tablenotemark{b} &
{Clump}\tablenotemark{c} &
{$R_{clump}$}\tablenotemark{d} &
{No. IRX}\tablenotemark{e} &
{No. Class I}\tablenotemark{f} &
{$R_{eq}$}\tablenotemark{g}\\
 \cline{2-3}
{} &
\multicolumn{2}{c}{(J2000)} &
{} &
{[pc]} &
{K$<$15.75} &
{M$<$0.4 M$_\odot$} &
{[pc]} \label{t:clusters}\\
\noalign{\smallskip}
\tableline
\noalign{\smallskip}

PL01    	& 06 31 49.32 & +04 19 34.7 & 11  &  2.21 &  29$\pm$5 & 9  & 1.16 \\
PL02    	& 06 33 16.78 & +04 35 32.1 & 18  &  1.33 &  32$\pm$6 & 12 & 1.46 \\
PL03    	& 06 33 33.07 & +04 00 11.2 & 7   &  2.79 &  80$\pm$9 & $>$6  & 1.69 \\
PL04    	& 06 34 13.92 & +04 25 05.4 & 1   &  2.36 &  89$\pm$9 & $<$107\tablenotemark{h} & 1.85 \\
PL05    	& 06 34 30.70 & +04 20 01.8 & 19  &  1.50 &  57$\pm$8 & $<$107 & 1.31 \\
PL06    	& 06 34 38.76 & +04 12 55.4 & 2   &  3.06 &  13$\pm$4 & $>$10 & 0.75 \\
PL07    	& 06 35 29.78 & +03 59 10.9 & 3   &  2.79 &  22$\pm$5 & 27 & 0.88 \\
REFL08   	& 06 34 18.98 & +04 20 03.6 & 17  &  1.71 &  49$\pm$7 & $<$107 & 1.30 \\
REFL09   	& 06 35 07.73 & +03 41 34.7 & 5   &  3.60 &  65$\pm$8 & 12 & 1.49 \\
PouC		& 06:33:12.68 & +04:31:00.5 & --- &  ---  &  ---      & 12 & 0.90 \\
PouD		& 06:33:40.98 & +04:03:56.1 & --- &  ---  &  ---      & 9  & 0.83 \\
REFL10		& 06 31 06.79 & +05 14 50.0 & --- &  ---  &  15$\pm$4 & --- & 1.15 \\
NGC~2237	& 06 31 58.51 & +04 54 35.7 & --- &  ---  &  36$\pm$6 & --- & 1.91 \\
NGC~2244	& 06 30 36.10 & +04 58 50.6 & --- &  ---  &  62$\pm$8 & 176 & 2.30 \\

\noalign{\smallskip}
\tableline
\noalign{\smallskip}

\multicolumn{8}{l}{\parbox{0.9\textwidth}{\footnotesize $a$~ID key: PL---Phelps \& Lada (1997); REFL---Rom\'an-Z\'u\~niga et al. (2006a); Pou---Poulton et al. (2008)}}\\[1ex]
\multicolumn{8}{l}{\parbox{0.9\textwidth}{\footnotesize $^b${Location of peak surface density of near-infrared excess sources}}}\\[1ex]
\multicolumn{8}{l}{\parbox{0.9\textwidth}{\footnotesize $^c${Molecular clump host identification from Williams, Blitz \& Stark (1995)}}}\\[1ex]
\multicolumn{8}{l}{\parbox{0.9\textwidth}{\footnotesize $^d${Clump host equivalent radius}}}\\[1ex]
\multicolumn{8}{l}{\parbox{0.9\textwidth}{\footnotesize $^e${Limited to K$<$15.75~mag.}}}\\[1ex]
\multicolumn{8}{l}{\parbox{0.9\textwidth}{\footnotesize $^f${Adapted from Table 2 and figure 9 of Poulton et al. (2008). Mass detection limit is 0.4 M$_\odot$}}}\\[1ex]
\multicolumn{8}{l}{\parbox{0.9\textwidth}{\footnotesize $^g${Cluster radii defined from IRX local surface density contours, except PouC and PouD, listed as equivalent radius.}}}\\[1ex]
\multicolumn{8}{l}{\parbox{0.9\textwidth}{\footnotesize $^h${Clusters PL04, PL05 and REFL08 are reported as a single cluster in Poulton et al.}}}\\[1ex]
\end{tabular}
}
\end{center}
\end{table}

\subsubsection{Mid-infrared Observations}

Recently, \citet{poulton08} reported the first results of the Spitzer
 telescope survey of the {Rosette Molecular Cloud.} The area surveyed
with IRAC and MIPS covers 1$\times$1.5$^\circ$, similar to the one
studied with FLAMINGOS. The Spitzer observations allowed to identify,
via the fit of SED profiles, a total of 751 young stellar objects with
infrared excess down to a mass limit of 0.4 M$_\odot$. Also, they used
a method of nearest neighbors applied to all sources with infrared
excess to identify embedded clusters, confirming the seven clusters of
\citet{pl97} as the major clusters in the cloud and confirming the
 existence of clusters {REFL08} and {REFL09} (labeled as clusters E and F,
respectively).  Their analysis also revealed two new small clusters or
associations, labeled clusters C and D (we used the labels PouC and
 PouD in this work). These two groups are located near clusters {PL02}
 and {PL03,} respectively.

 The Spitzer survey of the {Rosette} confirmed that {Clusters PL01} to {PL05}
have very dense envelopes of hot dust, evidenced by strong extended
emission in all IRAC and MIPS bands. The Spitzer survey also revealed
 that sources within 15 pc from the center of {NGC 2244} are definitely
older than those deeply embedded in the cloud, with 2\% vs 7\% of
identified Class I sources, respectively.

\begin{figure}[!htbp]
\centering
\includegraphics[draft=False,width=0.42\textwidth]{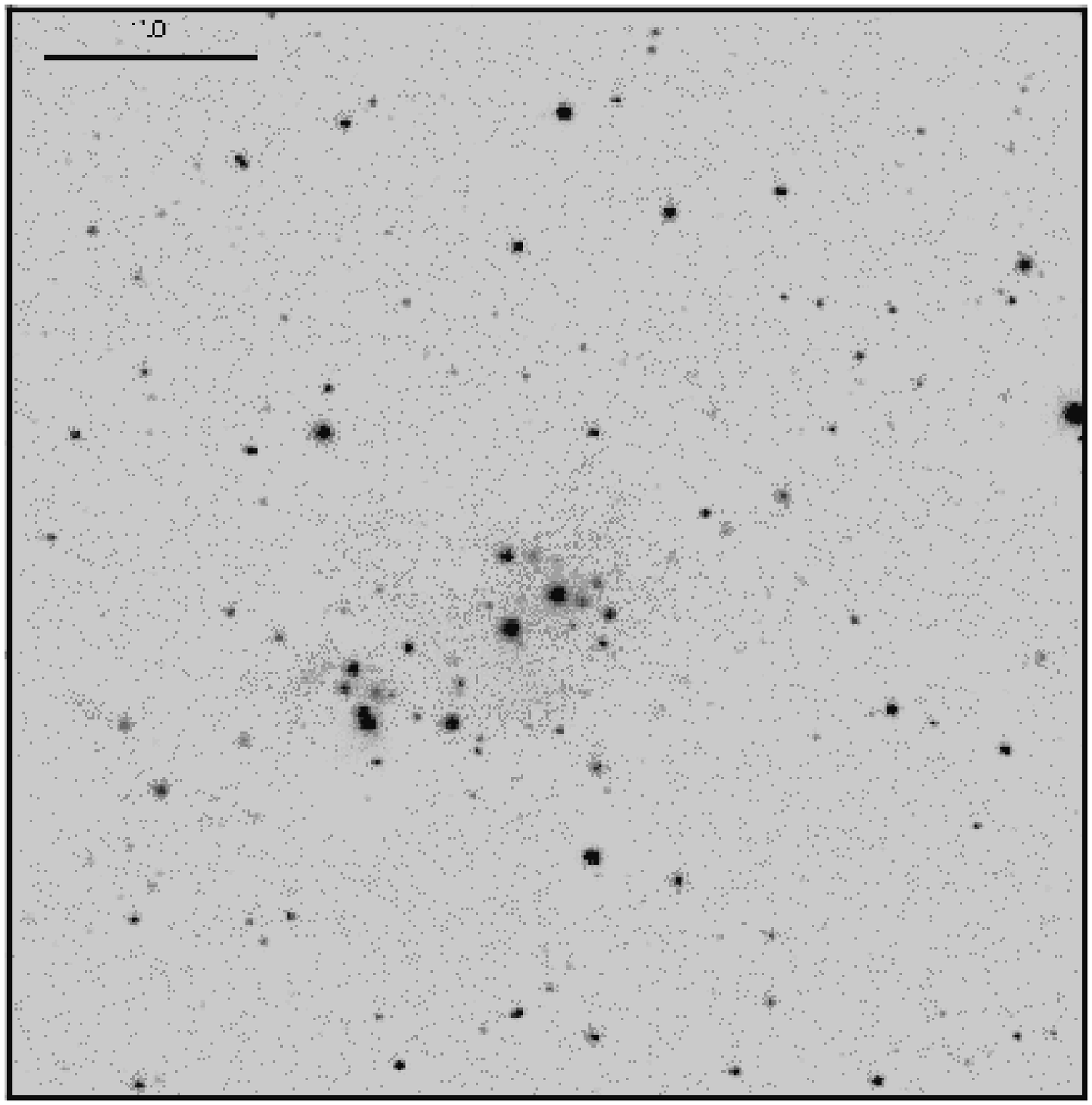}
\includegraphics[draft=False,width=0.425\textwidth]{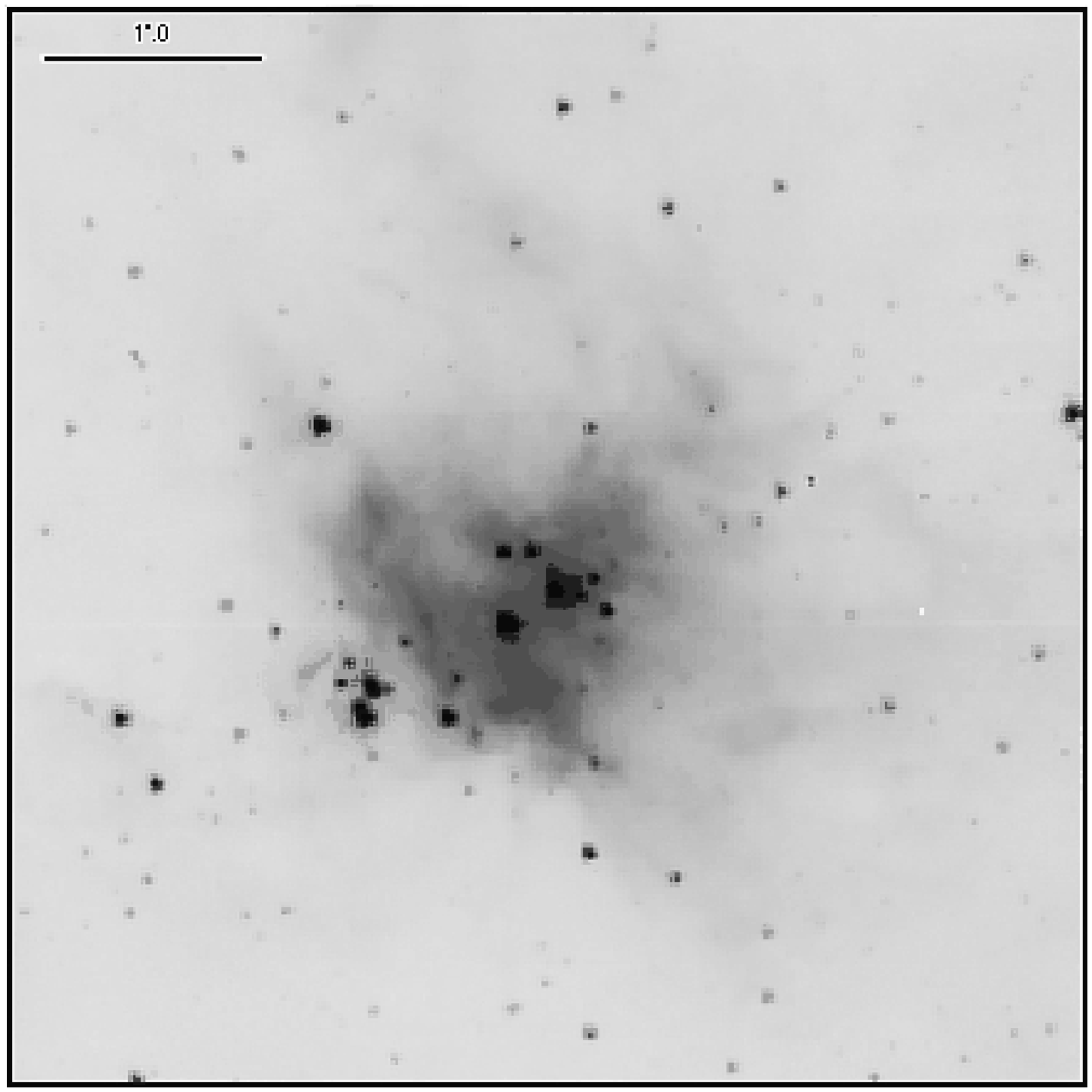}
\includegraphics[draft=False,width=0.42\textwidth]{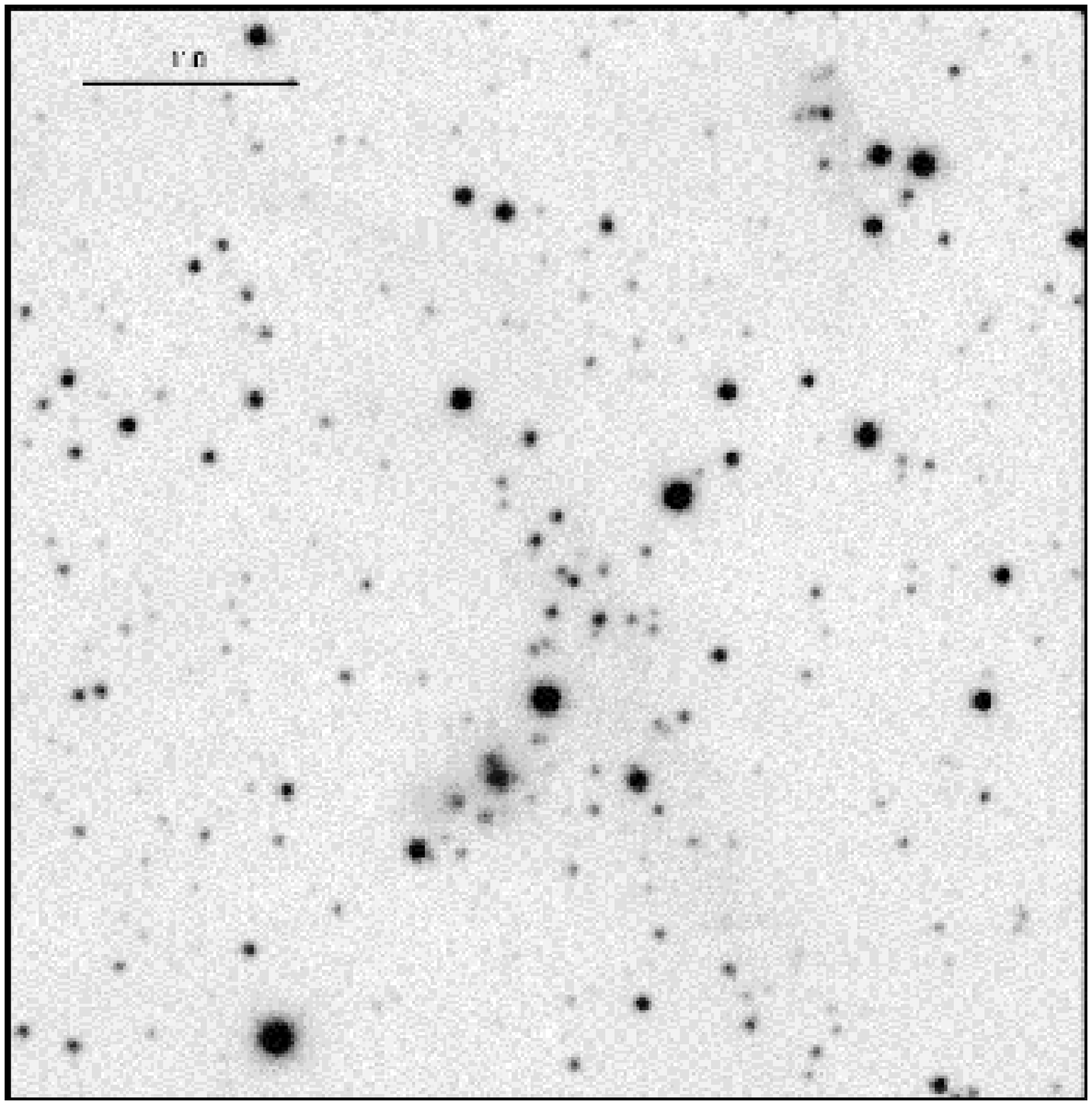}
\includegraphics[draft=False,width=0.425\textwidth]{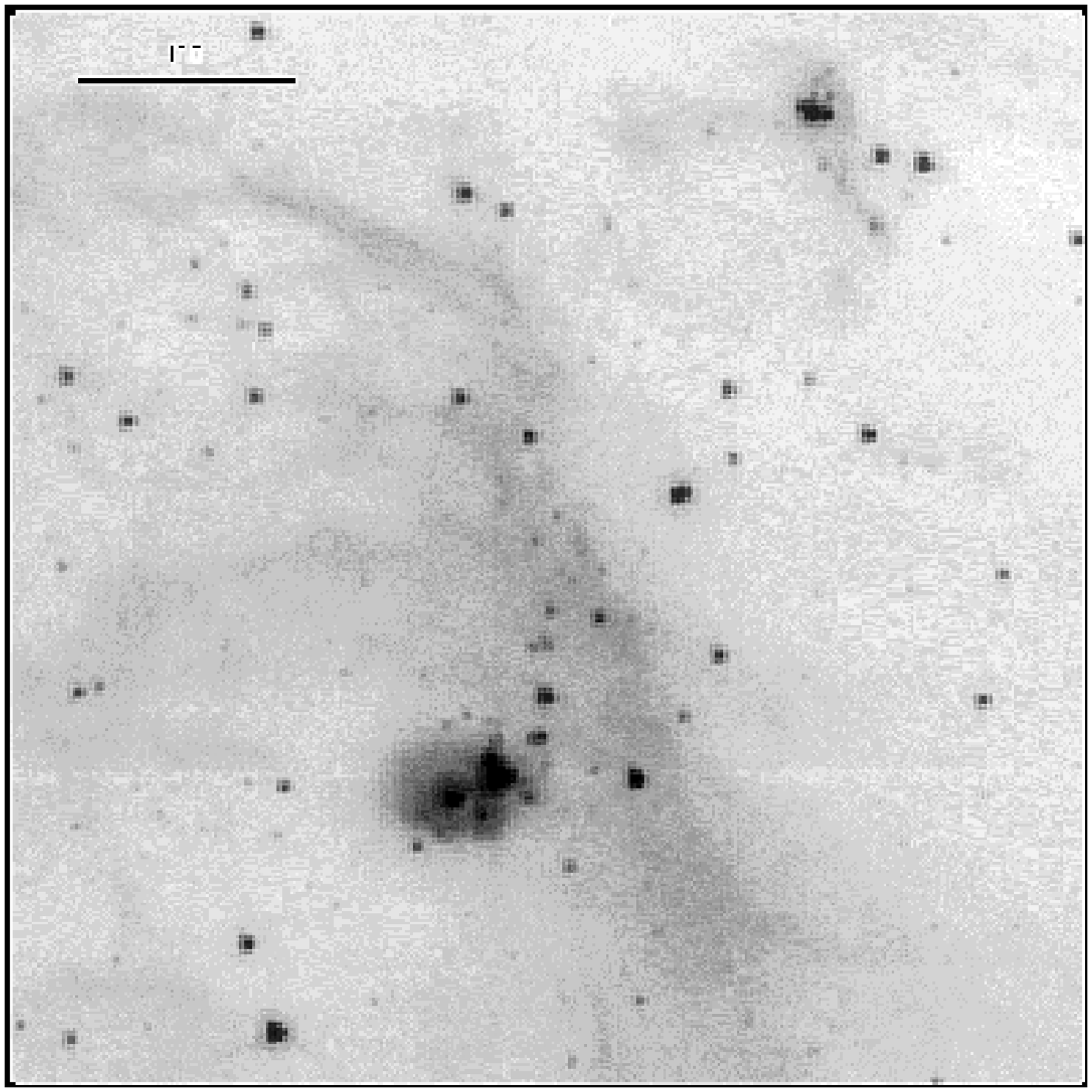}
\includegraphics[draft=False,width=0.42\textwidth]{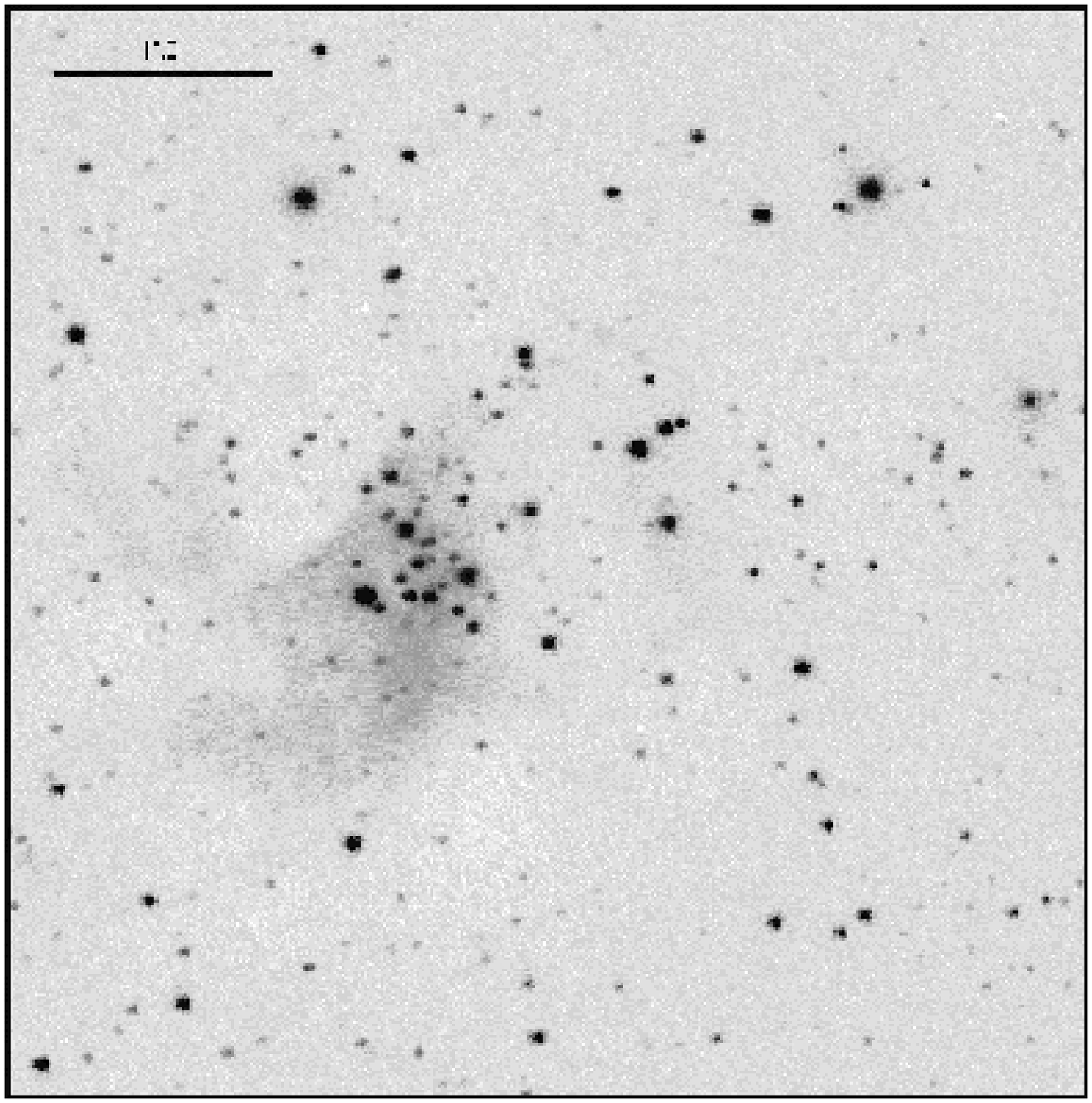}
\includegraphics[draft=False,width=0.425\textwidth]{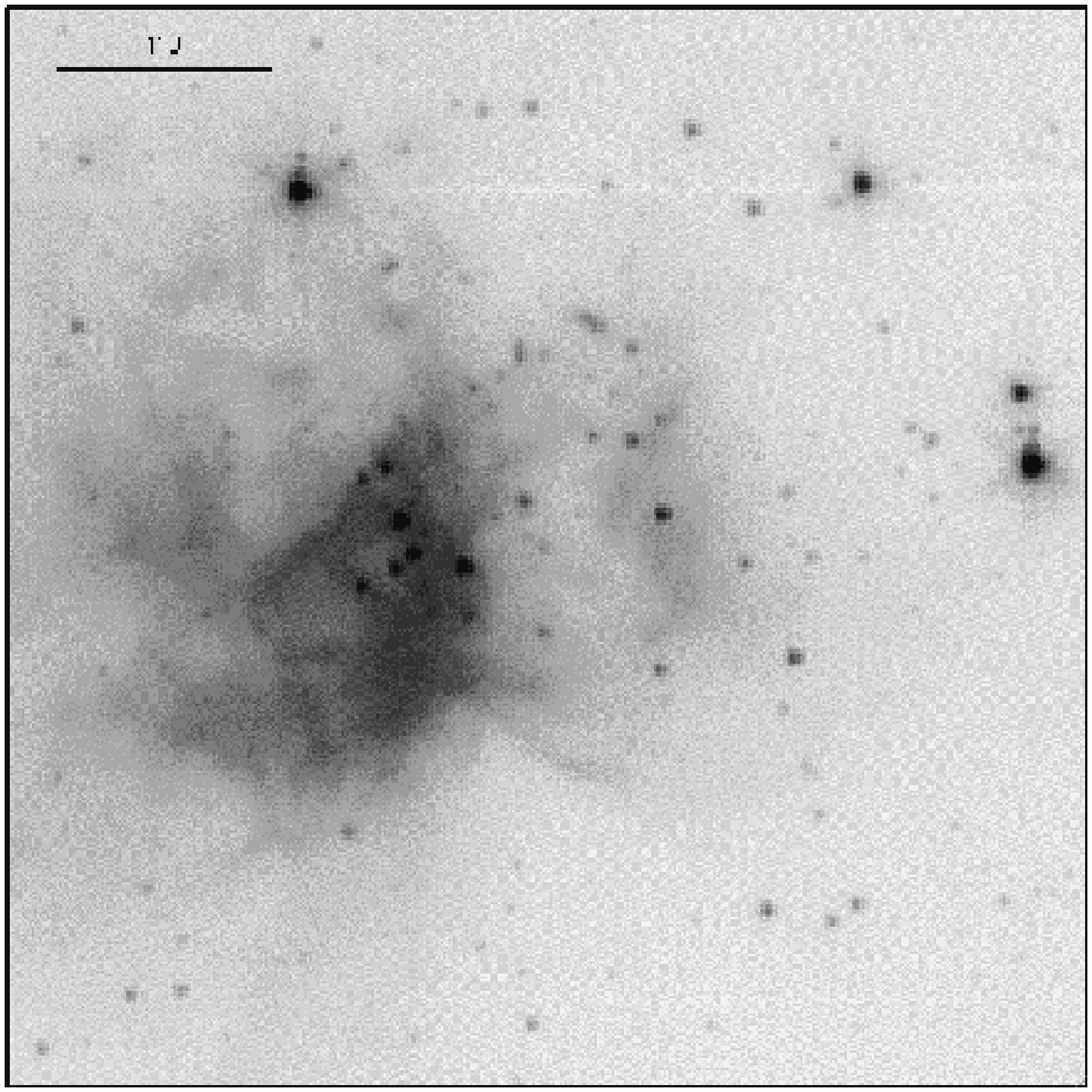}
 \caption{Embedded clusters {PL01,} {PL02} and {PL03} (top, middle and
 bottom, respectively) in the {Rosette Molecular Cloud.} {\em Left
Panels}: False 3 color images (J,H,K) constructed with images from the
FLAMINGOS survey. {\em Right Panels}: False 3 color images
([3.6],[4.5] and [8.0]$\mu$m) constructed with images from the IRAC
survey. All images cover a 5$^\prime \times$5$^\prime$ area.  Scale
bars of 1.0 arcmin are also shown on top. \label{f:pics1} }
\end{figure}

\begin{figure}[!htbp]
\centering
\includegraphics[draft=False,width=0.42\textwidth]{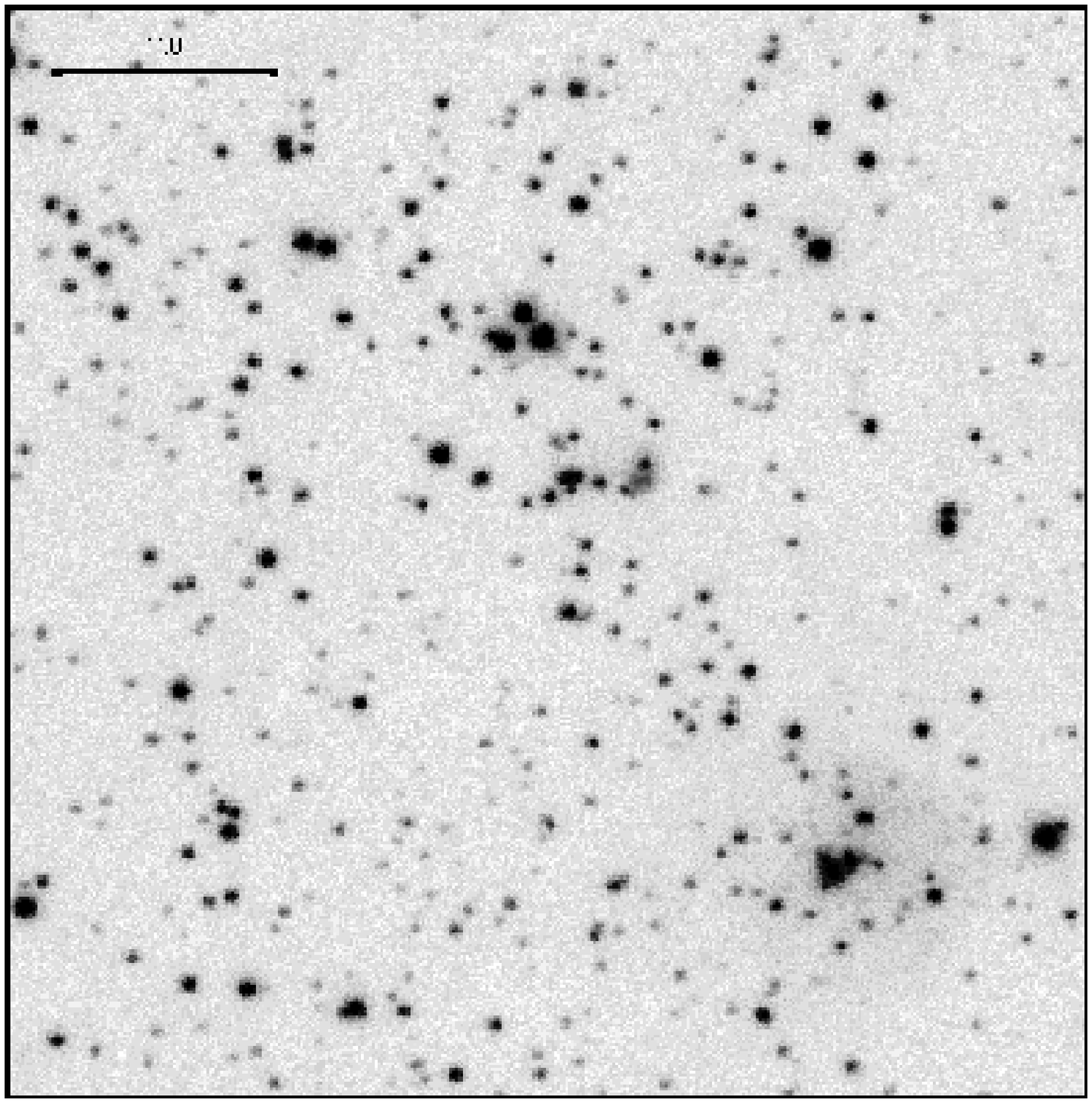}
\includegraphics[draft=False,width=0.425\textwidth]{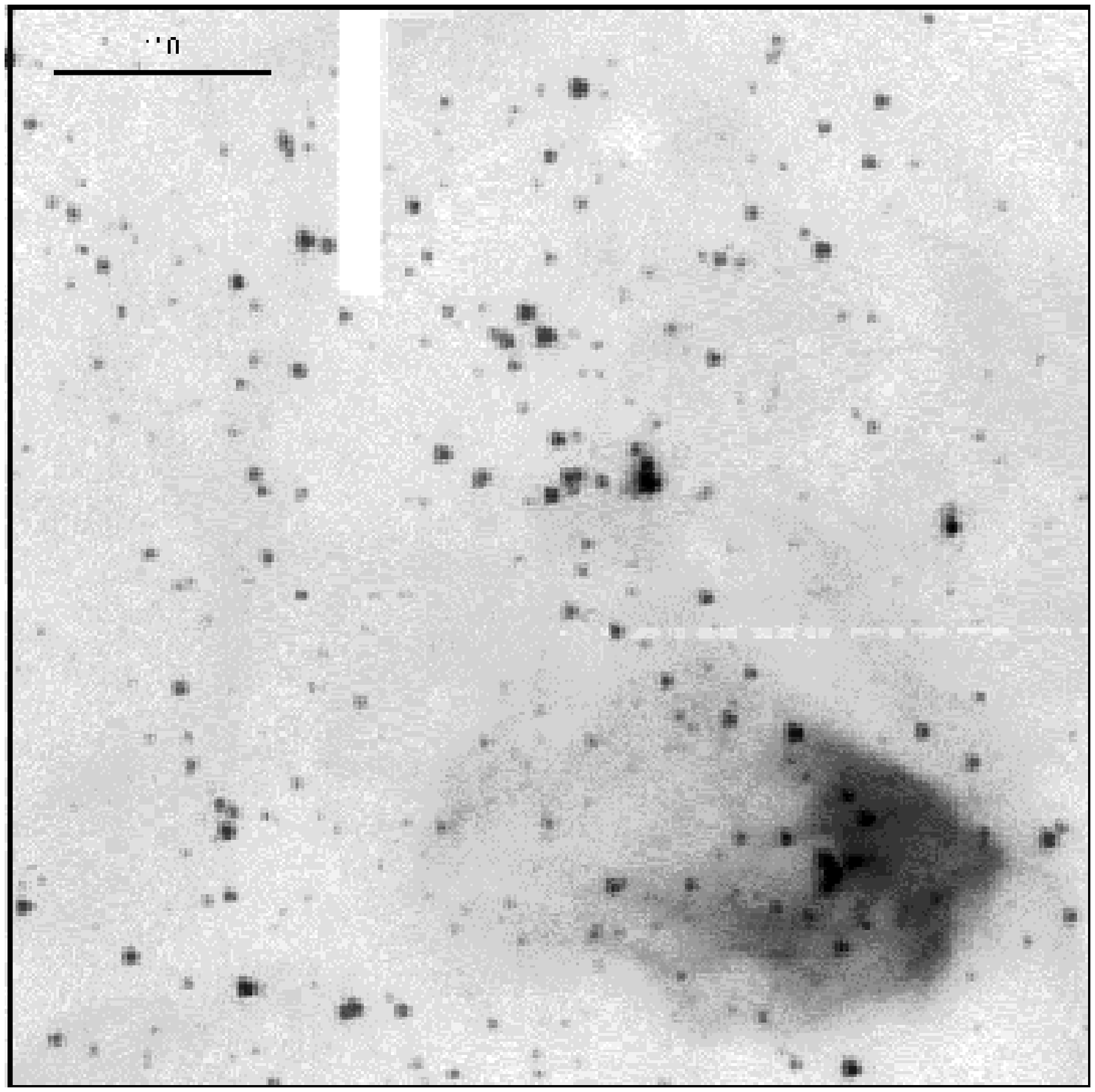}
\includegraphics[draft=False,width=0.42\textwidth]{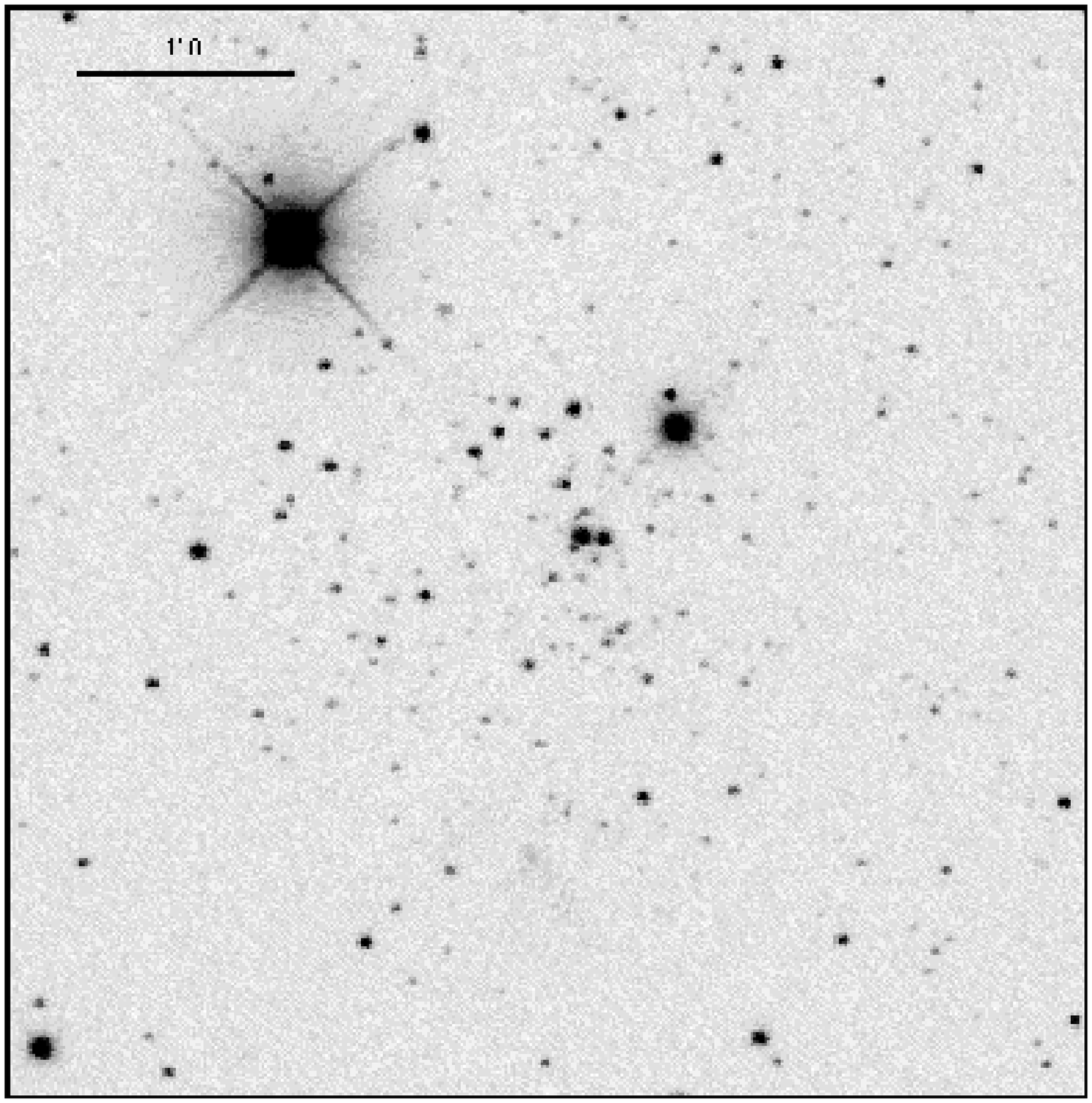}
\includegraphics[draft=False,width=0.425\textwidth]{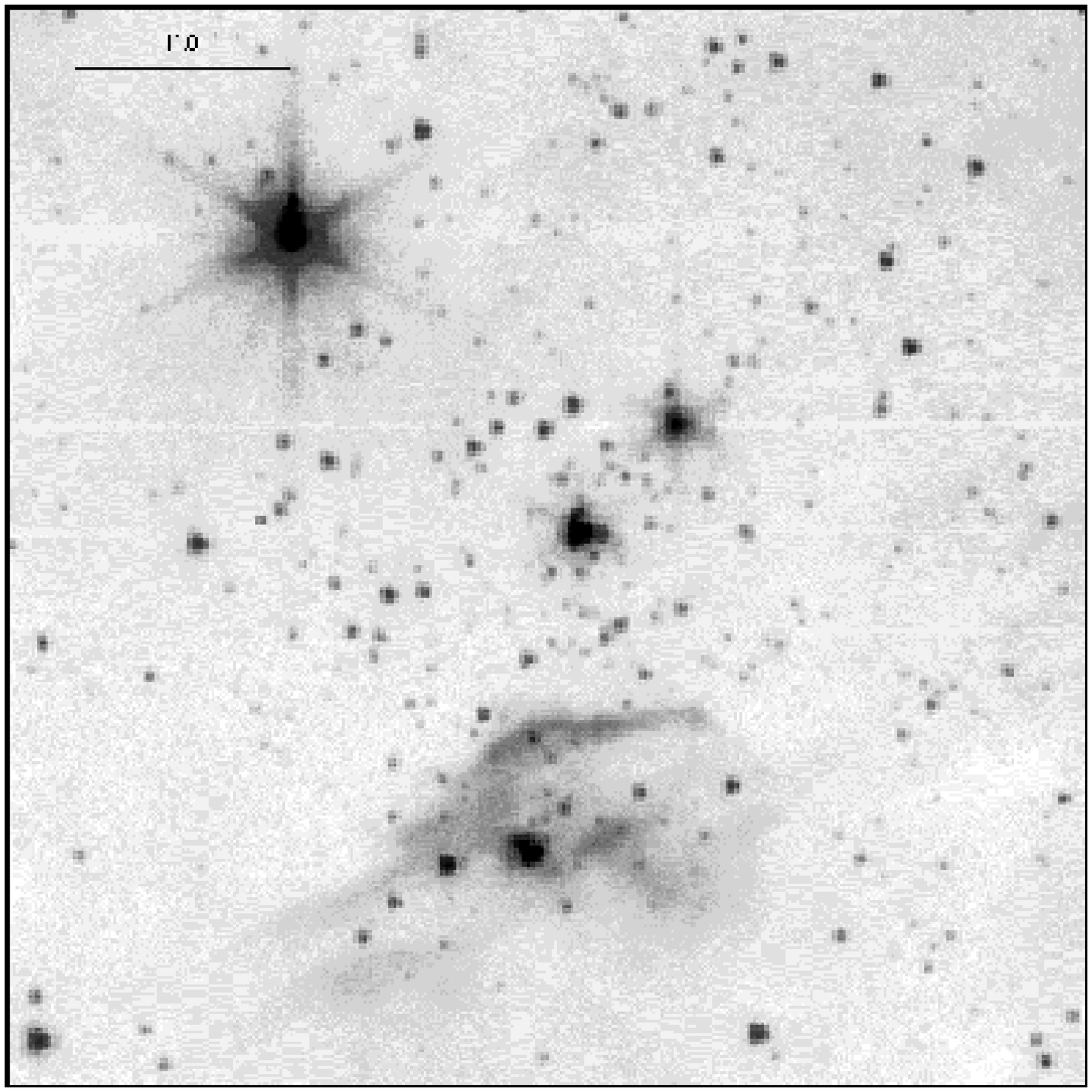}
\includegraphics[draft=False,width=0.42\textwidth]{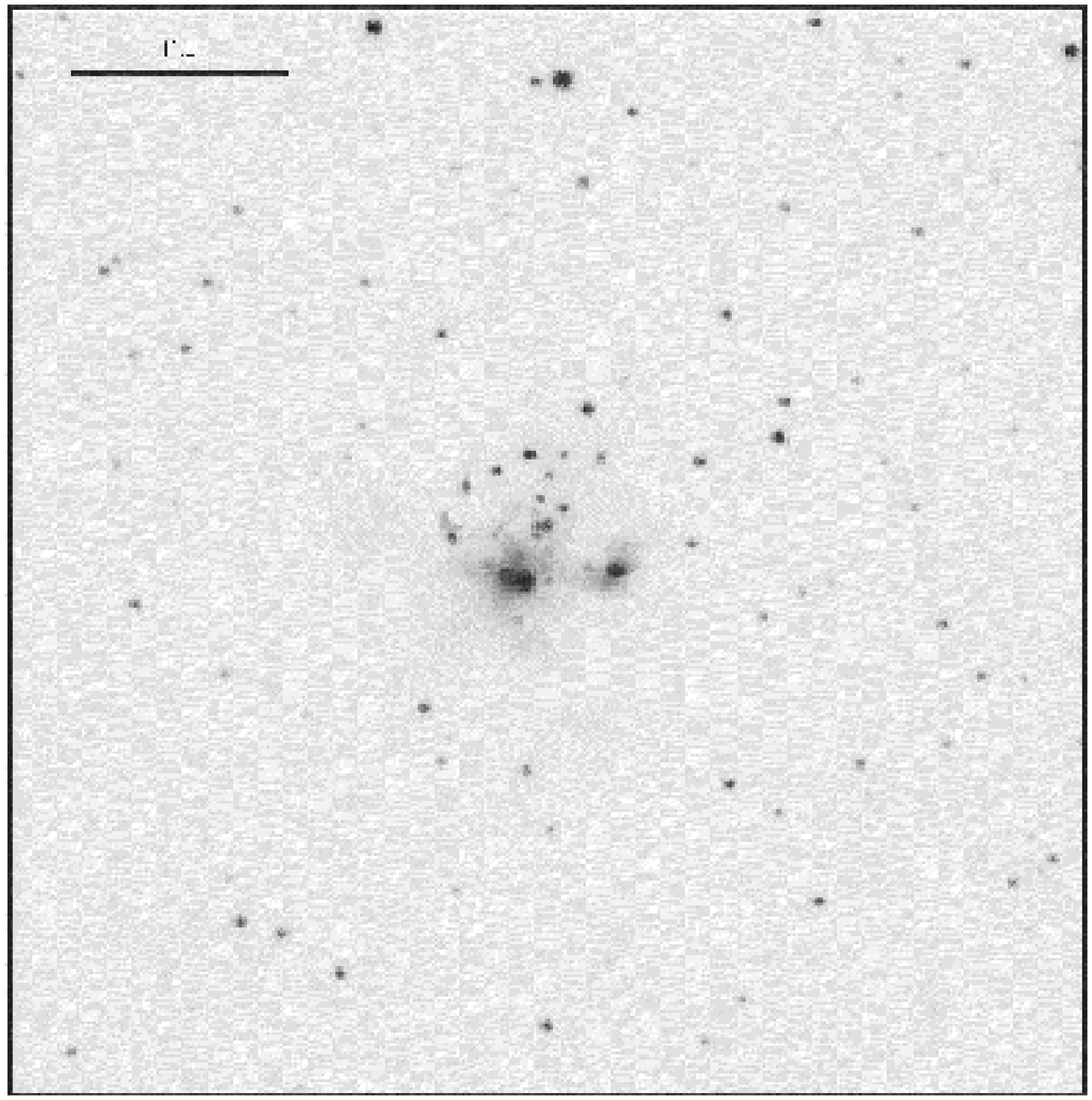}
\includegraphics[draft=False,width=0.425\textwidth]{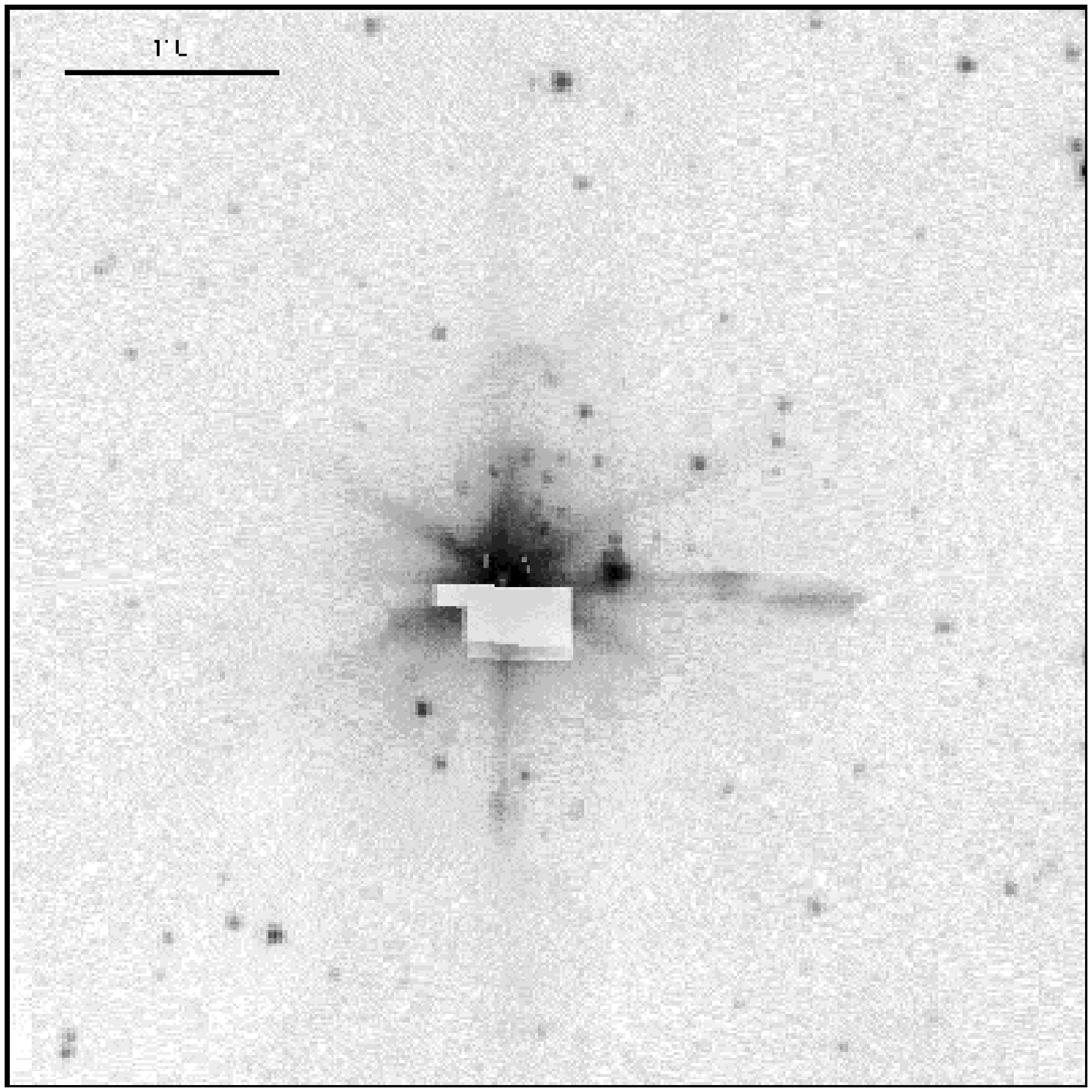}
 \caption{Same as figure \ref{f:pics1}, for embedded clusters {PL04,} {PL05} and {PL06} (top, middle and bottom, respectively).
 Note: Artifacts in cluster {PL06 IRAC} image are due to saturation of IRAC detector by the bright source {AFGL961.} \label{f:pics2} }
\end{figure}

\begin{figure}[!htbp]
\centering
\includegraphics[draft=False,width=0.42\textwidth]{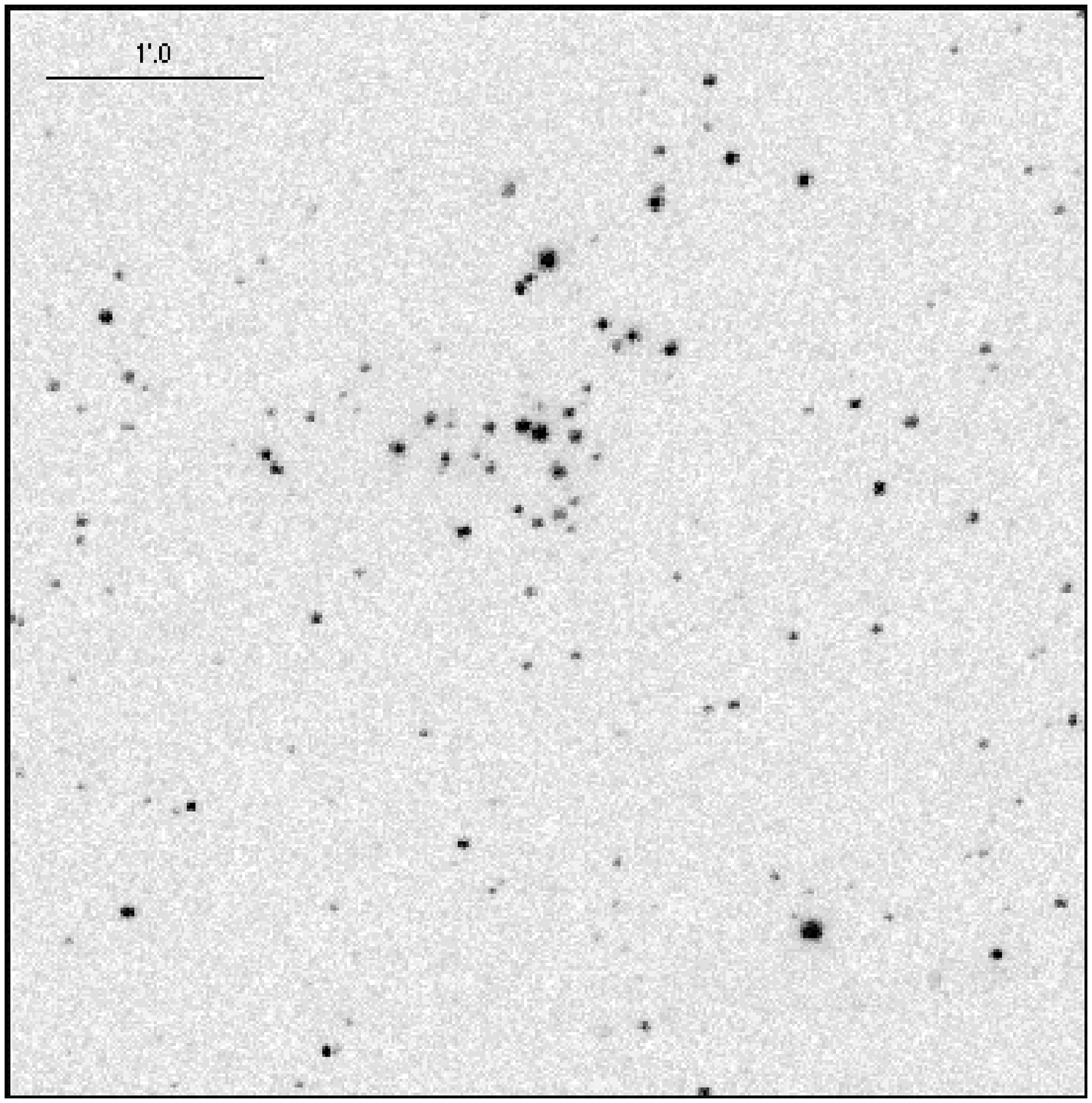}
\includegraphics[draft=False,width=0.425\textwidth]{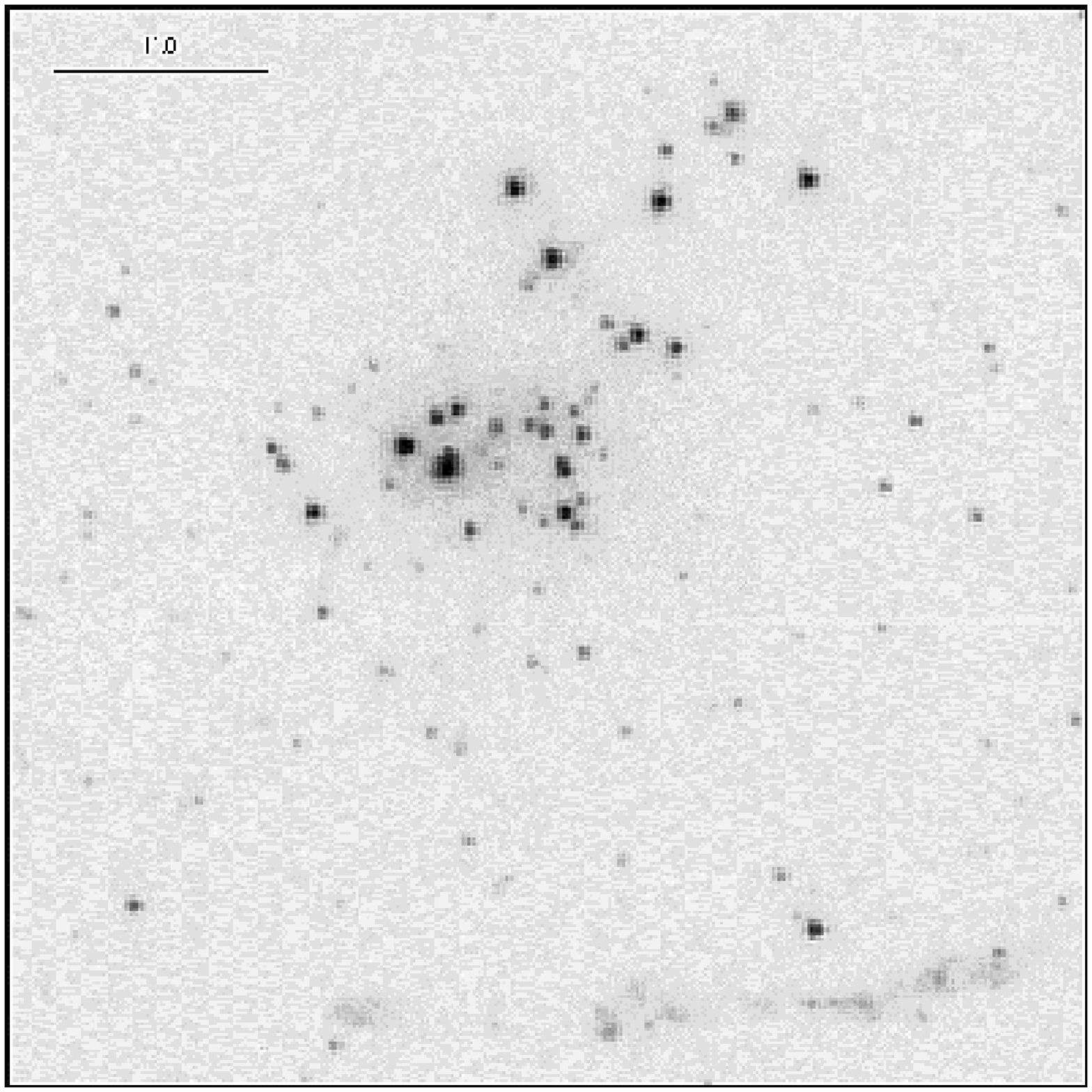}
\includegraphics[draft=False,width=0.42\textwidth]{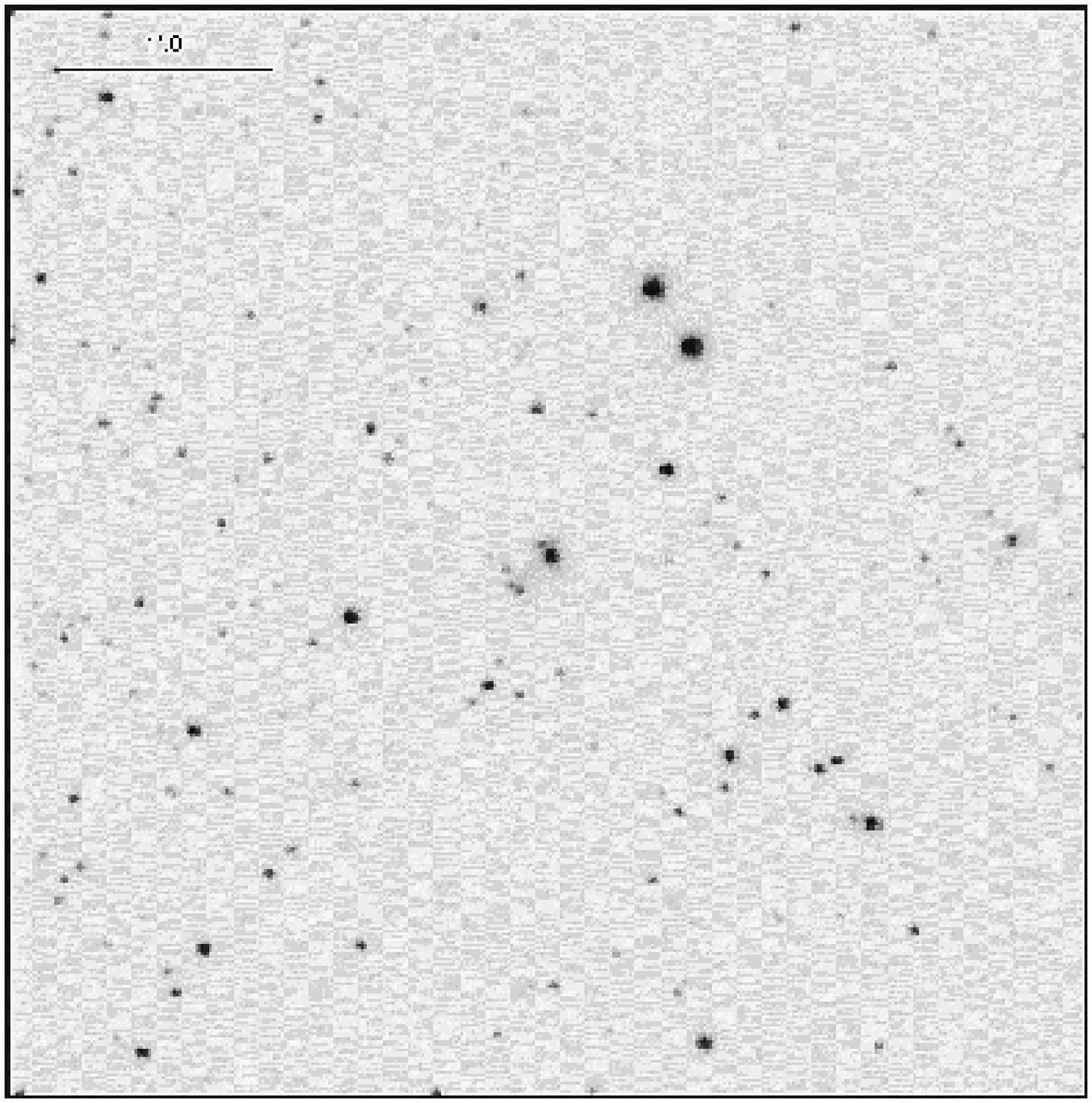}
\includegraphics[draft=False,width=0.425\textwidth]{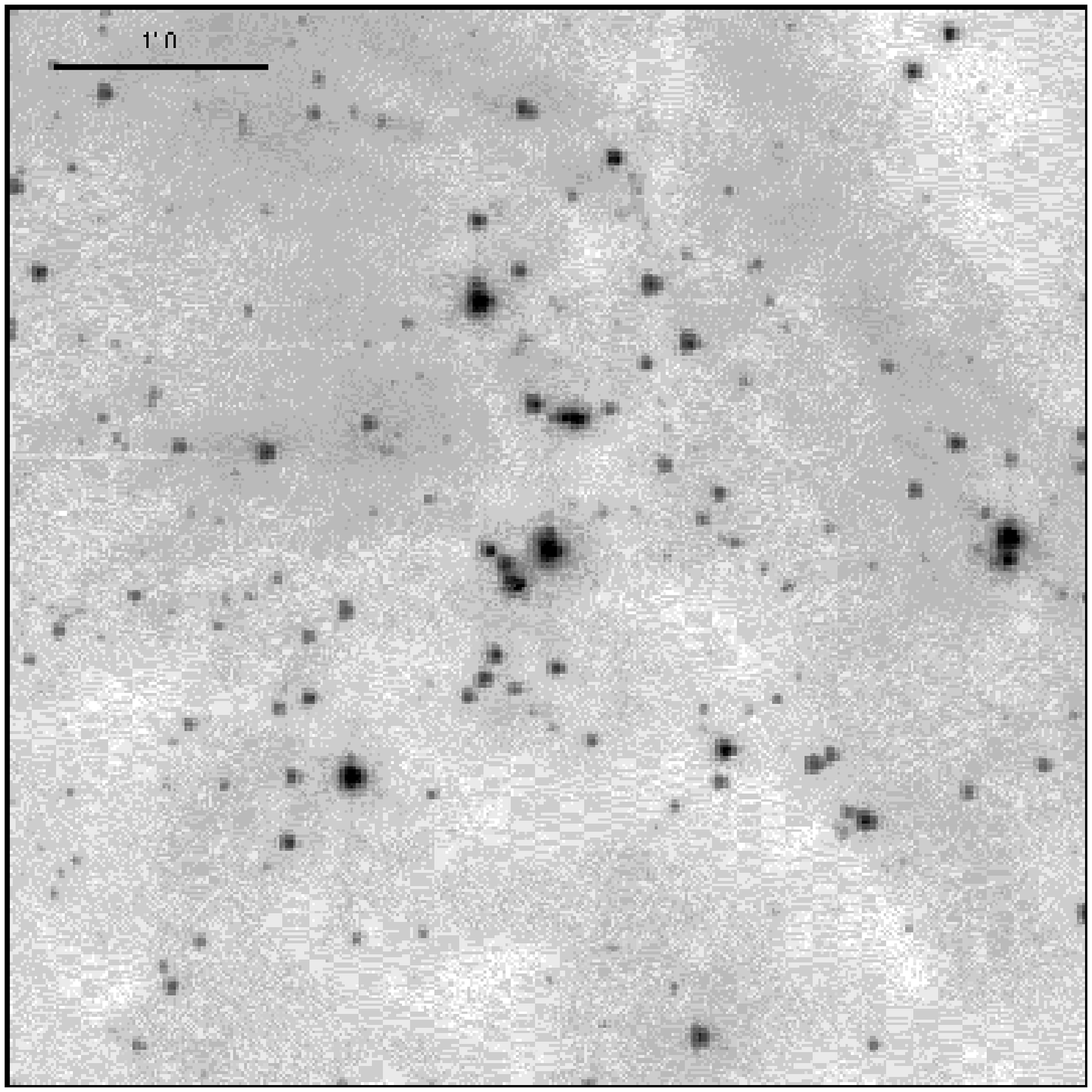}
\includegraphics[draft=False,width=0.42\textwidth]{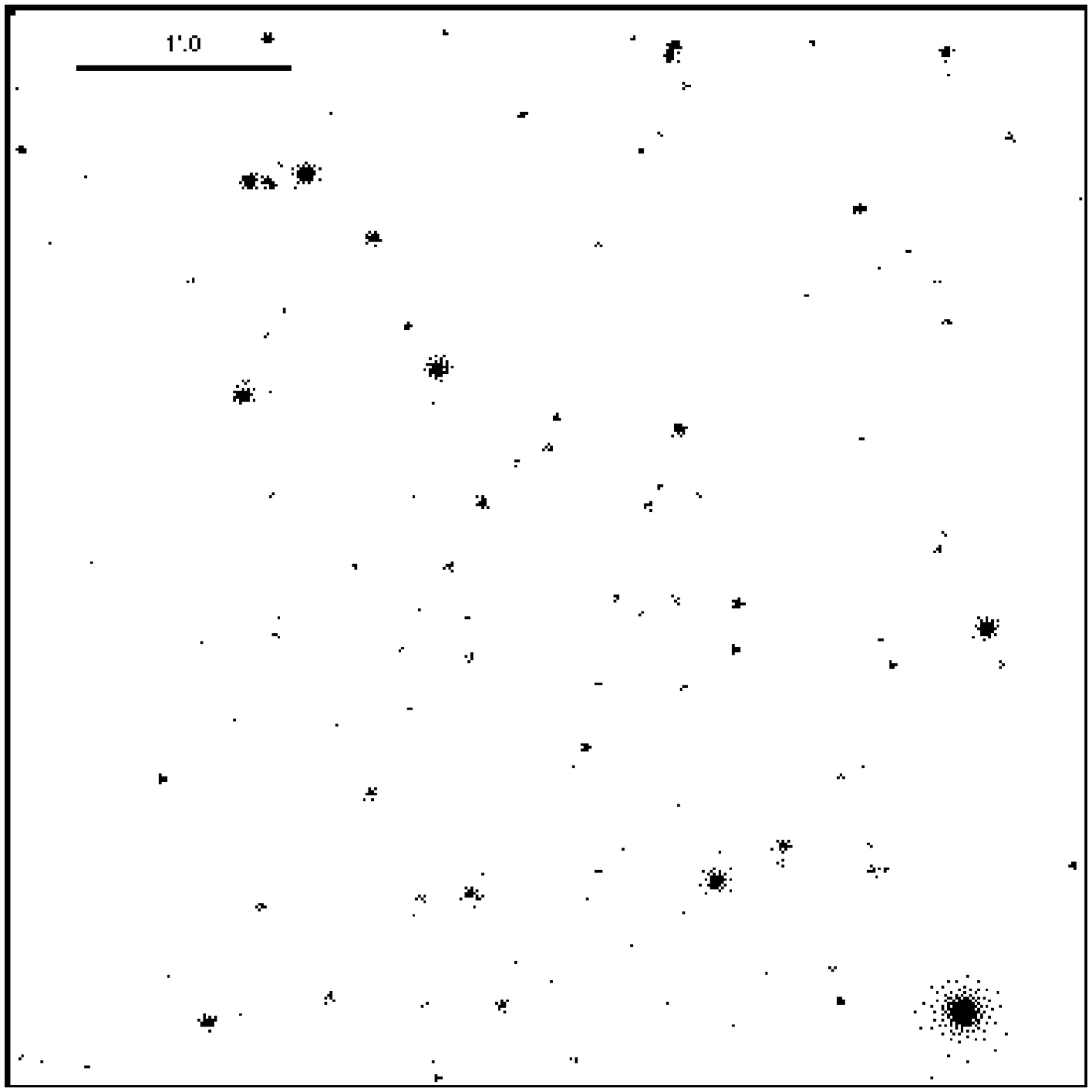}
\includegraphics[draft=False,width=0.42\textwidth]{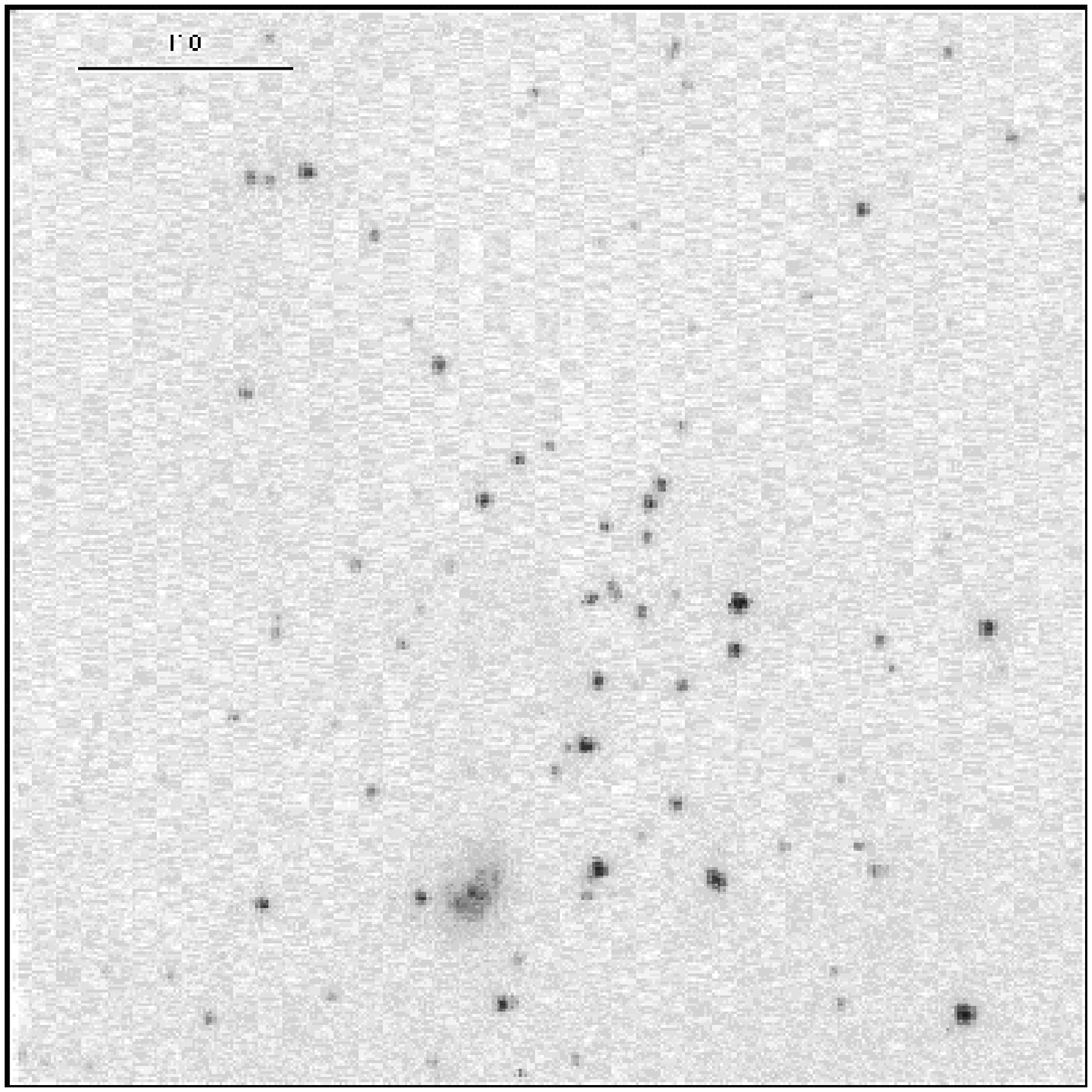}
 \caption{Same as figure \ref{f:pics1}, for embedded clusters {PL07,}
 {REFL08,} and {REFL09} (top, middle and bottom, respectively). Note: IRAC
 picture for {REFL09} was constructed with [4.5], [5.0] and [8.0] $\mu$m,
as it misses coverage in [3.6] $\mu$m. \label{f:pics3} }
\end{figure}

\begin{figure}[!htbp]
\centering
\includegraphics[draft=False,width=0.4225\textwidth]{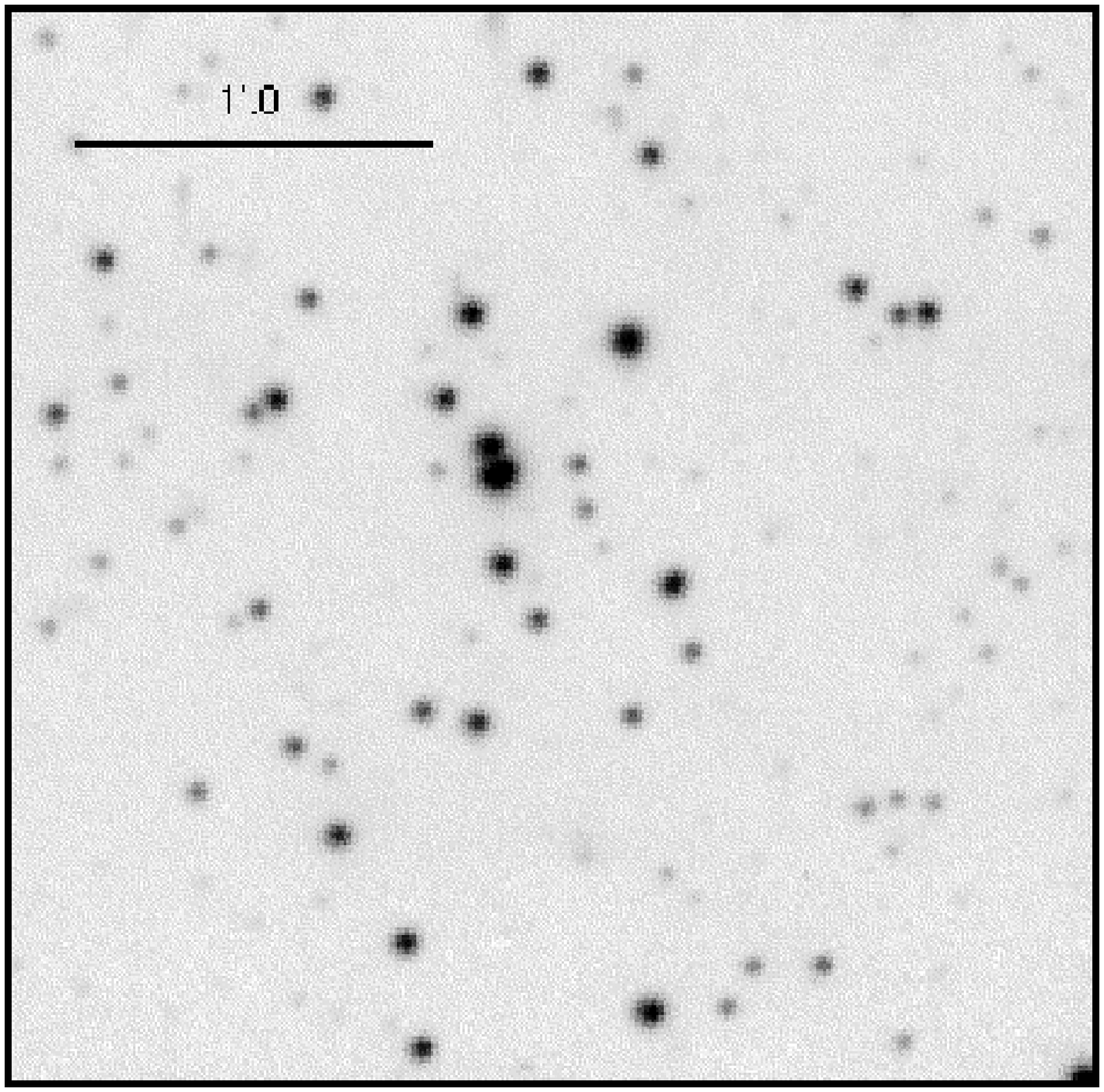}
\includegraphics[draft=False,width=0.42\textwidth]{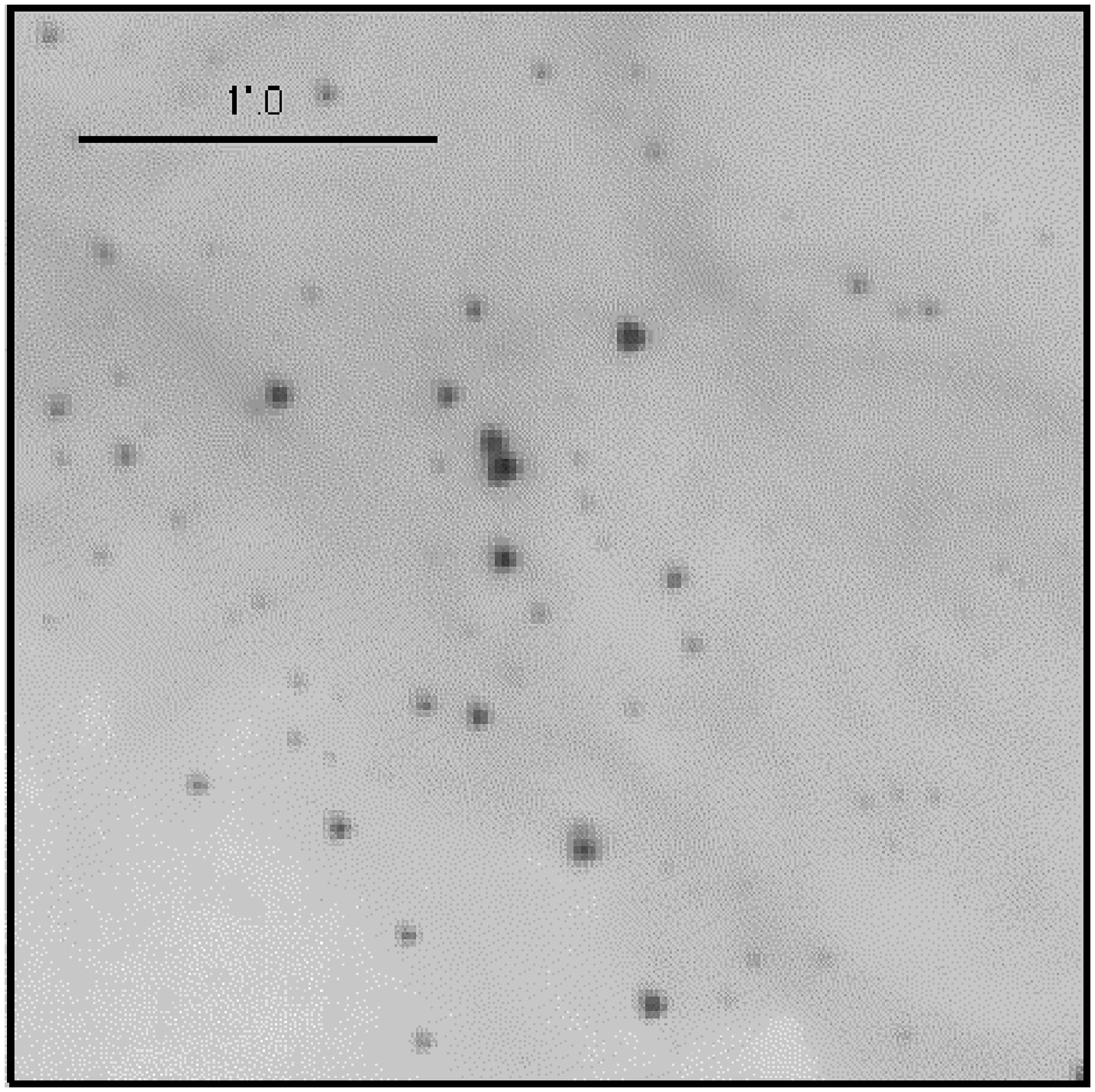}
\includegraphics[draft=False,width=0.425\textwidth]{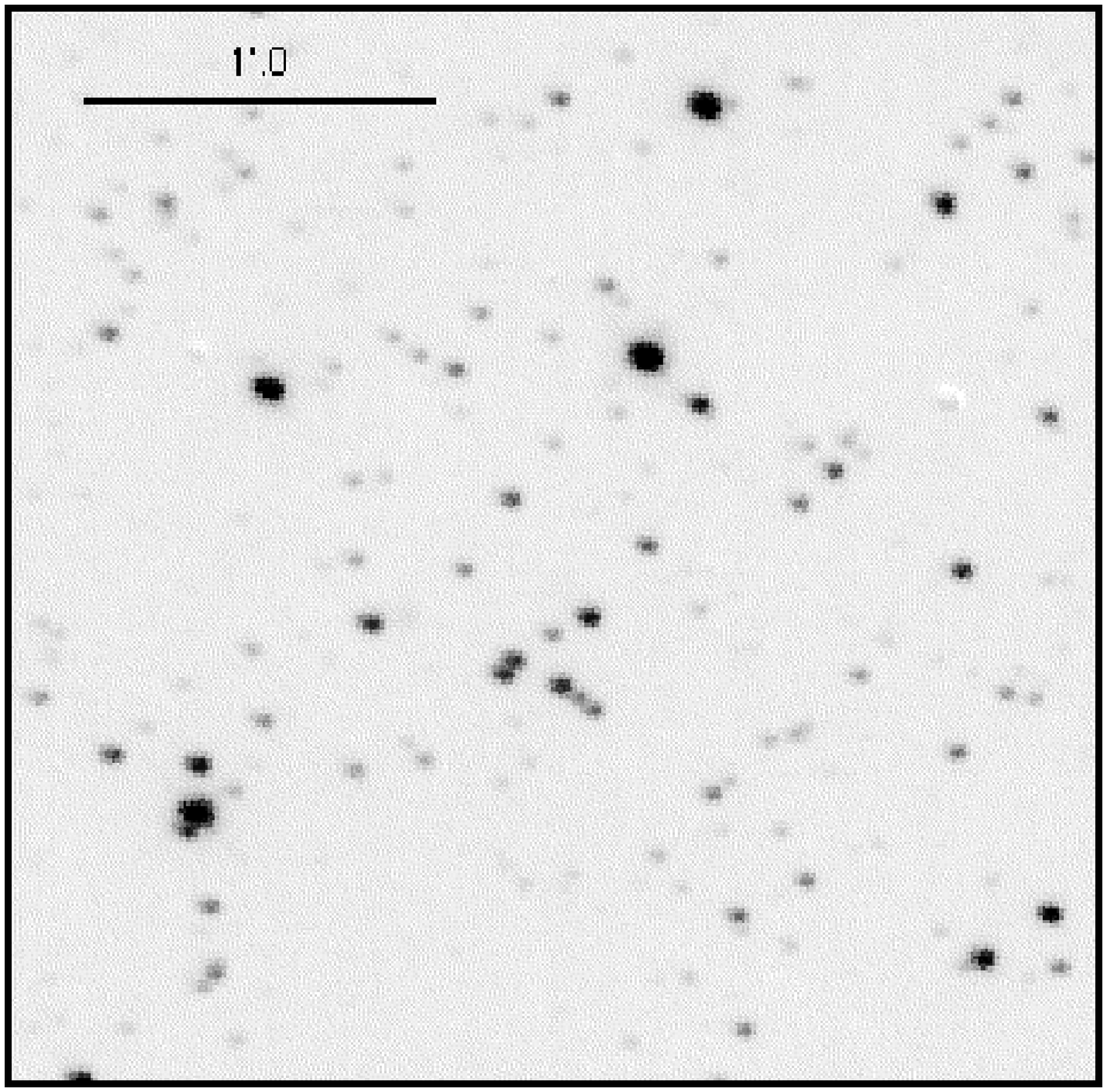}
\includegraphics[draft=False,width=0.42\textwidth]{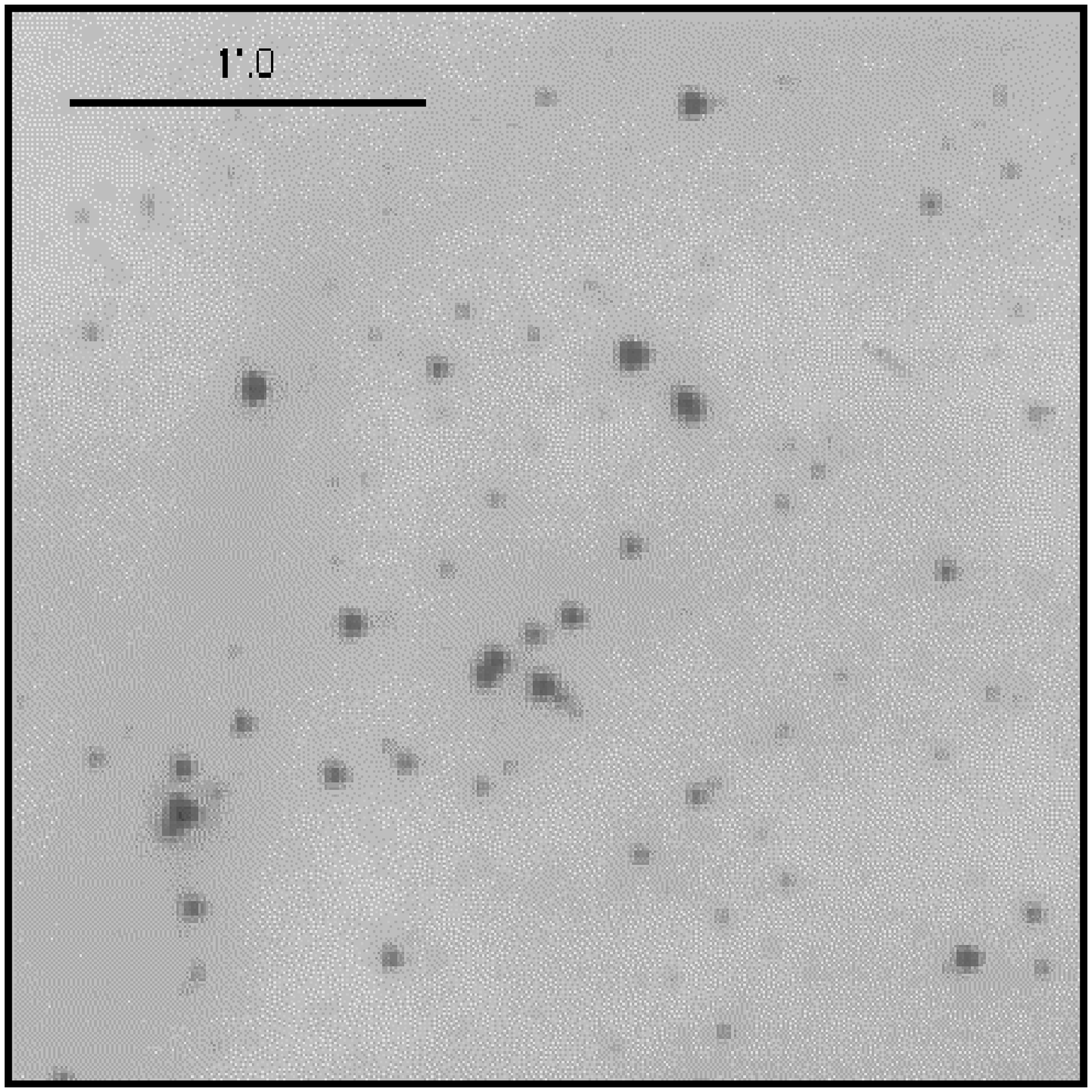}
\caption{Same as figure \ref{f:pics1}, for embedded clusters PouC and PouD (top, and bottom, respectively).
Note: The area covered in these images is 3$^\prime \times$3$^\prime$. \label{f:pics4} }
\end{figure}

In Figures \ref{f:pics1}, \ref{f:pics2}, \ref{f:pics3} and
\ref{f:pics4} we present false color images constructed with
near-infrared images from FLAMINGOS and with mid-infrared images from
IRAC for all the embedded clusters in the Rosette Molecular Cloud:
 PL01-07, {REFL08-09} and {PouC-D.} The FLAMINGOS images, with higher
resolution, define well the stellar content of the clusters, while the
IRAC images show in addition the distribution of hot dust and gas
around the embedded clusters.

Table \ref{t:clusters} contains information about the location of the
 14 known clusters embedded in the {Rosette Molecular Cloud,} along with
the corresponding clump from \citet{wbs95}. Also indicated is the
number of near-infrared excess stars detected in each cluster region
and the cluster equivalent radii, which was calculated by estimating
the extension of each cluster from local surface density contours of
near-infrared excess sources above the background level. We also
indicate the number of Class I sources reported by
\citeauthor{poulton08}

\subsubsection{Interaction of Embedded Populations with the Local Environment}

\begin{figure}[!htb]
\includegraphics[draft=False,width=1.0\textwidth]{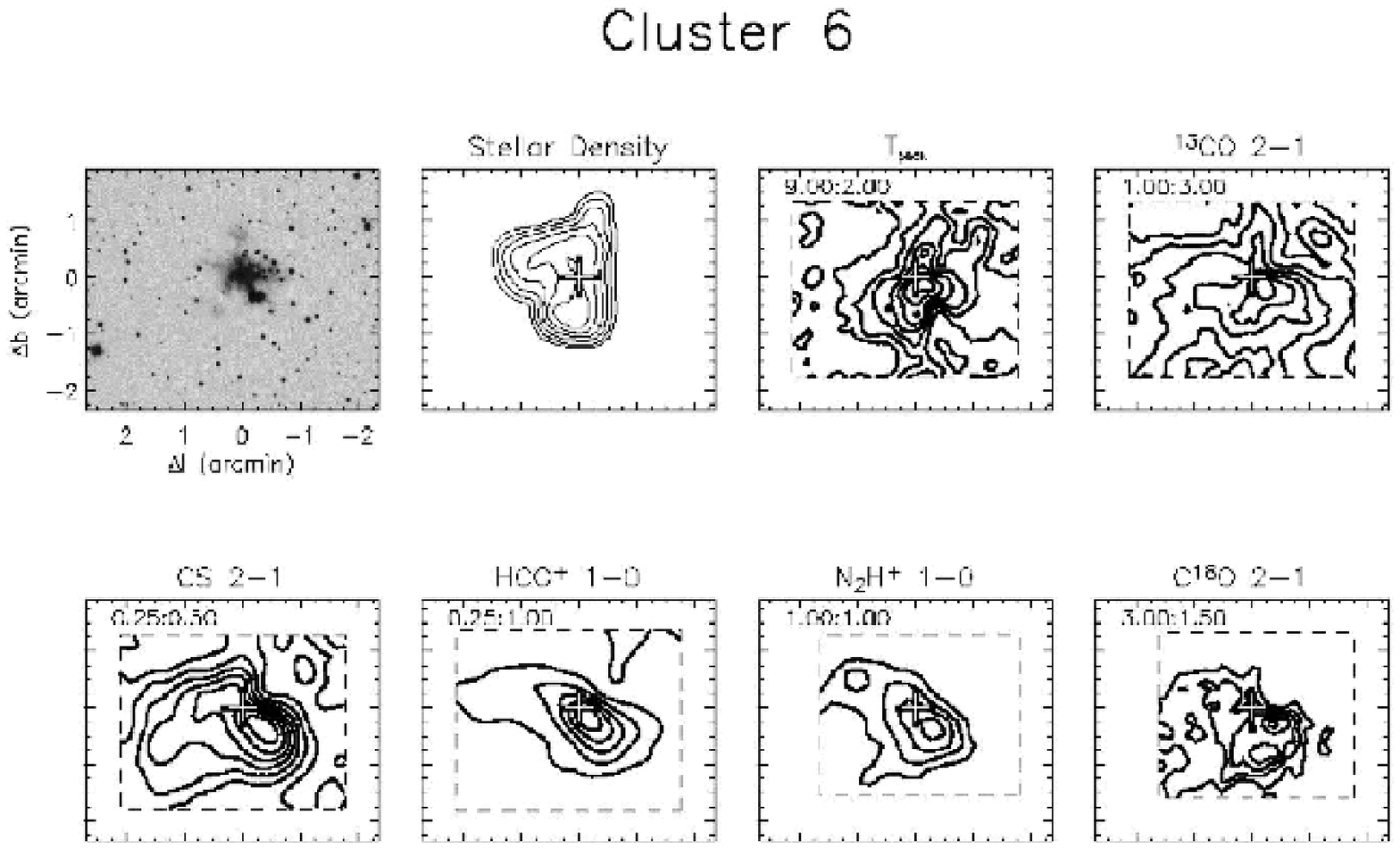}
 \caption{Maps of dense molecular gas emission towards cluster {PL06} in
 the {Rosette Molecular Cloud,} from Rom\'an-Z\'u\~niga (2006). The first
two panels show a false color near-infrared image of the cluster and a
IRX surface density map. The molecular map panels have base and step
values for the contour levels. The maps are integrated intensity
except for $^{12}$CO(2-1) which is mapped as peak
temperature. \label{f:pl06mm} }
\end{figure}

In order to give insight into the nature of the embedded populations
 of the {Rosette Molecular Cloud,} a high resolution mm-wave survey of
 clusters {PL01} to {REFL08} was performed at the IRAM 30m telescope
\citep[][Rom\'an-Z\'u\~niga, Williams \& Lada, 2008 (in
prep)]{thesis}.  The cluster regions were scanned for emission in 6
molecules: $^{12}$CO(2-1), $^{13}$CO(2-1), C$^{18}$O(2-1), CS(2-1),
HCO$^+$(1-0) and N$_2$H$^+$(1-0). All of the clusters show prominent
 clumps in most tracers, except cluster {PL02} which has weak emission in
C$^{18}$O(2-1) and N$_2$H$^+$(1-0). In Figure \ref{f:pl06mm} we show
 one example of the maps of dense gas emission maps for cluster {PL06}
 (host of {AFGL 961).}

Bulk physical properties of the clumps are calculated from these
observations. Particularly, mass, sizes and velocity dispersions
appear to be fairly similar for all of the cluster areas, which
translates to no significant variation of the properties of star
forming clumps across the cloud, as suggested by
\citet{wbs95}. However, the morphology of the gas emission is
different from clump to clump, suggesting different stages of the
clusters evolution. For example, while the gas emission around
 clusters {PL01,} {PL06,} {PL07} and {REFL08} is compact and forms a well
 defined envelope around the clusters, clusters {PL03,} {PL04} and {PL05}
appear to be partially emerging from the clumps, with the gas
apparently being swept out from some areas of the cluster and
recollected in others.

 The emergence of the massive clusters {PL04} and {PL05} might suggest a
late stage of embedded cluster evolution. The emission of dense gas
tracers for these clusters is confined to only two relatively small
areas which coincide with the most obscured and dense parts of the
respective cores. The rest of the stars in the clusters are still
embedded in the larger and more quiescent $^{13}$CO core but some of
 them are already visible in DSS optical plates. {Clusters PL04} and {PL05}
 are the largest and most extended embedded clusters in the {Rosette
 Molecular Cloud} (see Table \ref{t:clusters}) but could also be the
oldest. These clusters are located in the part of the cloud that is in
direct interaction with the shock front from the Nebula, as described
by \citet{HWB06}.

\subsubsection{The Sequential Formation Hypothesis}

\begin{figure}[!htb]
\plotone{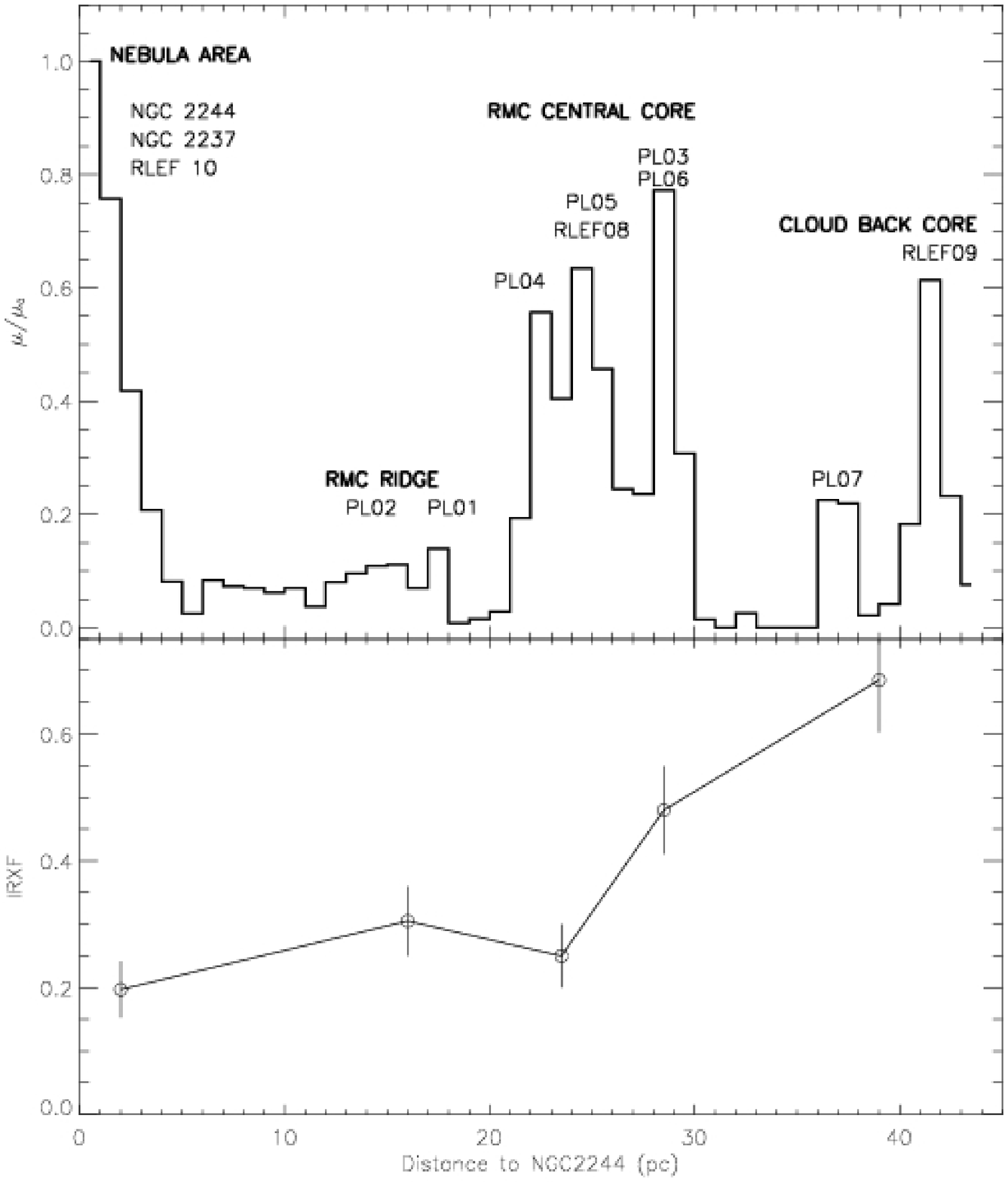}
\caption{Top panel: Distribution of near-infrared excess source
 density as a function of distance from the center of the {Rosette
 Nebula} {(NGC~2244).} The counts are made in sectors of 1.0 pc in length
and counts in each sector have been scaled and normalized to the area
 and counts in the central 1.0 pc circle in {NGC~2244.} Labels indicate
the approximate locations of embedded clusters as well as the main
'regions' of the complex, defined by Blitz \& Thaddeus (1986).  Bottom
panel: averaged near-infrared excess fractions in each of the cluster
groups defined from the top plot appear to increase with distance from
 the {Rosette Nebula.} From \citet{romanzetal1}. \label{f:nndne}}
\end{figure}

In Figure \ref{f:nndne} we show the distribution of the surface
 density of near-infrared excess sources (young stars) in the {Rosette
 Complex} as a function of the distance from the center of {NGC 2244,} as
calculated by \citet{romanzetal1}. Young clusters appear as peaks in
this distribution, and it is possible to distinguish four main groups
 of clusters. The most prominent group contains {NGC 2244,} {NGC 2237} and
 {REFL10,} which are the clusters located in the nebula. The second group
 contains clusters {PL01} and {PL02} and coincides with the cloud ridge
 (core A1-2). The third group contains clusters {PL04,} {PL05,} {REFL08} and
 {PL06,} all located at the RMC central core, and cluster {PL03} located in
 core D. The fourth and last group contains clusters {PL07} and {REFL09}
which are located at the back core of the cloud.  In the bottom panel
of Figure \ref{f:nndne} we show the average infrared excess fraction
for each of these cluster groups. The average excess fraction appears
 to increase as a function of distance from {NGC 2244,} suggesting that
the embedded clusters are progressively younger the further they are
from the Nebula.

The existence of this apparent age sequence indicates that star
 formation in the {Rosette} did indeed take place sequentially in time,
but not in the way proposed by \citet{elmelada77}, because that model
requires one episode of cluster formation to trigger the next one. The
relative age differences of clusters cannot be large enough to account
for such a scenario. Instead, it appears that the overall age sequence
of cluster formation may be primordial, possibly resulting from the
formation and evolution of the molecular cloud itself. The HII region
cannot be responsible for the sequence of cluster ages, although it
does appear to have a significant impact on the underlying sequence by
either enhancing or inhibiting the star forming process: At the cloud
ridge, the direct interaction with hot, ionized gas might contribute
 to the rapid evaporation of cluster envelopes in {PL01} and {PL02,}
possibly stopping the star formation process at an early stage;
however, recent Chandra and Spitzer telescope observations (Wang et
 al. (in prep); \citeauthor{poulton08}, 2008) suggest that {PL02} hosts
two Class 0/I sources, which is suggestive of cluster substructures
 younger than {PL04} or {PL05.} At the central core, where the main
interaction between the nebula and the molecular cloud takes place,
the result is an enhancement of the star formation efficiency,
 evidenced by the fact that the clusters in this area {(PL04,} {PL05,}
 {PL06,} {REFL08)} account for almost 50\% of the total embedded population
in the cloud. Finally, at the back core the influence of the HII
region is minimal and star formation in this part of the cloud may
have occurred spontaneously \citep{romanzetal1, poulton08}.

\section{Regions of Particular Interest}

\subsection{AFGL 961. Cluster PL06}

 Particular attention in the literature has been given to cluster {PL06,}
 associated with the bright infrared source {AFGL961.} The source was
originally identified by \citet{cohen1973}, who also obtained its
spectral energy distribution from 2 to 18~$\mu$m. This distribution
was later extended to include measurements at 53, 100 and 175~$\mu$m
by \citet{harveyetal77}. \citet{bt80} found that the CO emission
towards the source is strongly self absorbed and presents broad wings,
evidence of a violent interaction of the central source with the
molecular envelope. These characteristics were confirmed by
\citet{ladagautier82,loren81,schneiderco,HWB06} and
\citet{thesis}. The powerful CO outflow that generates the wings was
measured to have a gas kinetic energy close to $10^{47}$ ergs, and is
moving almost 19 M$_\odot$ of material \citep{ladagautier82}. The
outflow is bipolar and spectral signatures revealed that the outflow
is in fact caused by an ionized wind. The self absorption is possibly
caused by a foreground layer of colder material, and in fact, infrared
spectra at 1 $\mu$m and 1 mm obtained by \citet{cox89} revealed the
presence of ice bands towards the source, confirming the effect.

\begin{figure}[!htb]
\plotone{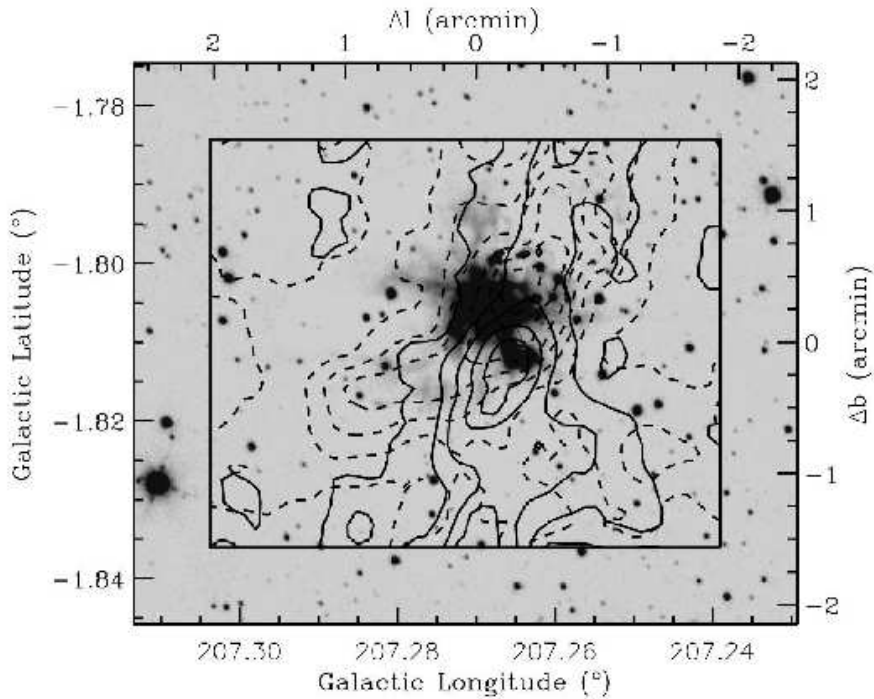}
\caption{A FLAMINGOS near infrared image overlaid with contours of
 $^{12}$CO(2-1) outflow (line wings) emission towards cluster {PL06}
 which hosts the massive PMS binary {AFGL961.} The solid contours show
integrated emission from 5 to 8 km$\cdot \mathrm{s}^{-1}$, and the
dotted contours show integrated emission from 17 to 20 km$\cdot
\mathrm{s}^{-1}$. \label{f:afgl961}}
\end{figure}

 The central source in {AFGL961} is a B type PMS binary with a separation
 of approximately 0.05 pc, hence the common nomenclature {AFGL961a} and
 {AFGL961b,} which refer to the eastern and western component,
respectively. Near infrared observations \citep{aspin98,romanzetal1}
revealed that approximately another 30 young sources are associated
with this cluster, and \citet{aspin98} was able to demonstrate that at
least 9 of these sources were associated with shock-like features
which corroborated the strong interaction with the local medium.

Figure ~\ref{f:afgl961} combines a FLAMINGOS image with $^{12}$CO(2-1)
data from the \citet{thesis} survey in the form of integrated
intensity contours. Self absorption is evident even in the averaged
spectra, and the red and blue wings appear to confirm the existence of
two different components of the gas outflow, although none of them are
highly collimated.

\subsection{Jets, Knots, Elephant Trunks and Globules}

\begin{figure}[!ht]
\plottwo{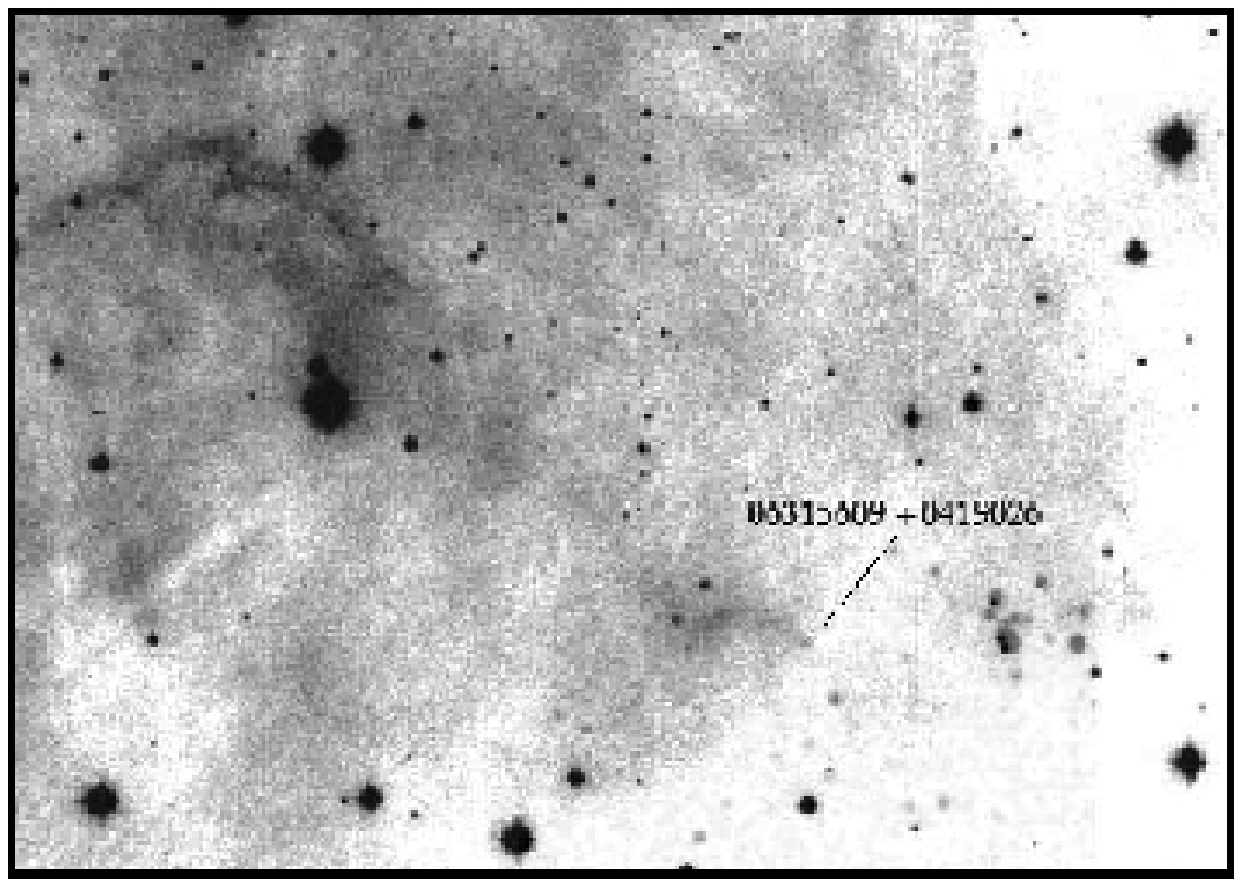}{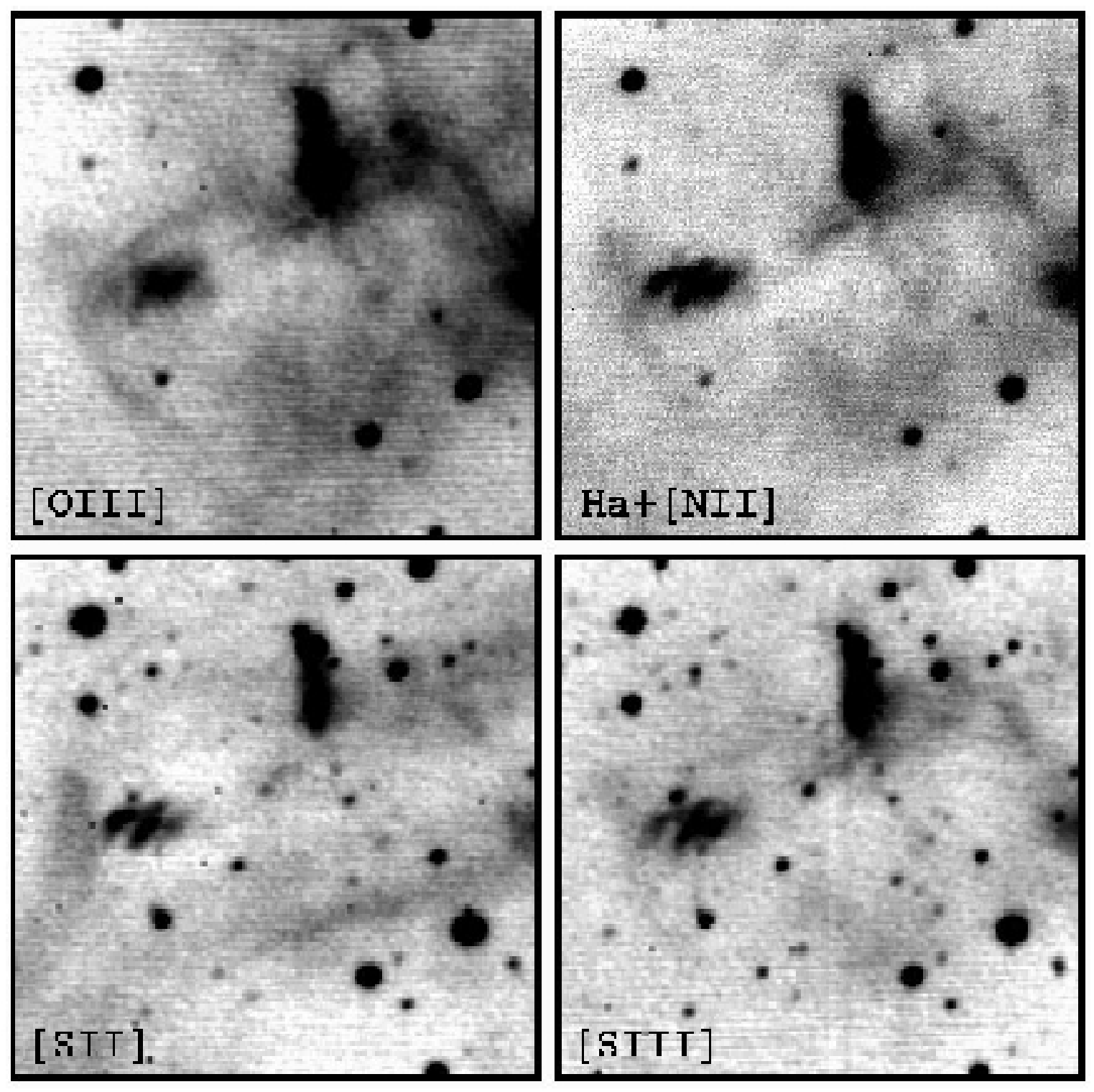}
 \caption{Large outflows and knots in the {Rosette Complex}
environment. The left side panel is a composite [SII]-2MASS image of
 object {HH~871} and its associated source Rosette Molecular Cloud-C from
\citet{phelpsybarra05}. The right panel shows a group of conspicuous
knots in the inner western edge of the HII region. From
\citet{claytonetal98}.}
\label{f:HH871}
\end{figure}

The [SII] survey of \citet{ybarraphelps04} revealed 13 regions of
excited gas, 3 associated with external excitation from winds of
 {NGC~2244} sources and 10 associated with winds from young sources in
the embedded clusters of the molecular cloud. Most of the main
clusters host one of these [SII] features, all reminiscent of HH
objects and thus suggestive of strong interaction of young sources
with their surrounding medium or at least of a high frequency of
collimated outflows from young sources. Counterpart near infrared
H$_2$ observations \citep{phelpsybarra05} revealed that one source in
the [SII] study, RMC-C (06315690+0419026) could be in fact a Class I
 object associated with a large outflow designated {HH 871.} The source
 is very close to cluster {PL01} and could be associated with it. See
Figure \ref{f:HH871}.

A series of studies by \citet{meaburnwalsh86},
\citet{claytonmeaburn95} and \citet{claytonetal98} were dedicated to
the investigation of a group of dense dense nebulous knots located at
the south eastern edge of the HII cavity in the nebula area at
approx. $(\alpha,\delta)=(6^h29^m36^s,+5^d00^\prime 00^{\prime
\prime}$, J1950). In the first paper, \citeauthor{meaburnwalsh86}
obtained H$\alpha$+[NII] filter photographs of the central cavity
areas and combined them with echelle spectroscopy data. They suggested
two possibilities for the origin of the knots: a) they are HH objects
propulsed by T Tauri sources with bipolar activity, or b) they are bow
shocks around dense globules overtaken by large wind driven shells.

In the second study, \citeauthor{claytonmeaburn95} generated a data
cube of [OIII]5007 $\mathrm{\AA}$ profiles obtained by scanning the
knot areas with a multi-slit echelle spectrometer. They noticed that
the largest knot, labeled C, is indeed very bright in this line, and
suggested again that the feature might originate from a low mass young
star. In their third study, \citeauthor{claytonetal98} studied again
the prominent knot C. The combined observations plus K band images
identified candidates for the driving sources, and proposed that the
top group of knots --- which appear to coincide with K band sources
--- are all associated with a giant Herbig-Haro flow. The large scale
of the corresponding outflow would be consistent with the idea of a
large jet associated with the source RMC-C (see Figure \ref{f:HH871}),
thus confirming the existence of parsec scale flows in the
 {Rosette.} Several FLAMINGOS sources lie in the area and the knots are
barely visible in J and H, so a very interesting follow up would be to
determine if any other faint sources are also associated with the
knots. Also, \citet{meaburn} found a possible microjet associated with
a low mass young star, and most likely observable due to the
Ly-$\alpha$ emission of the nebula.

\citet{li03} and \citet{lirector04} reported the discovery of two a
cometary jet systems with external irradiation.  These two objects
 were recently catalogued as {HH~889} and {HH~890} (see
http://casa.colorado.edu/hhcat/). These jets are immersed in the
photoionized region and thus are awash by heady UV radiation, which
apparently results in an interesting set of bow shock
 structures. Given the age of {NGC 2244,} there is an upper limit of 1-2
Myr in the timescale of the jet production. Interestingly, the objects
associated as the driving sources present no infrared excess
\citet{li05}, which suggests a very rapid disk dissipation, with
timescales of 10$^3$ to 10$^4$ yr \citep{lietal07}, leading to a rapid
transition from CTTS to WTTS stages \citep{lirector07}. These aspects
pose interesting questions about the environmental dependence of the
mechanisms of jet formation and also suggest that CTTS and WTTS
objects can be spatially mixed in regions of massive star formation.

\begin{figure}[!ht]
%\plotone{rosetteglob.ps}
\centering
\includegraphics[draft=False,width=\textwidth]{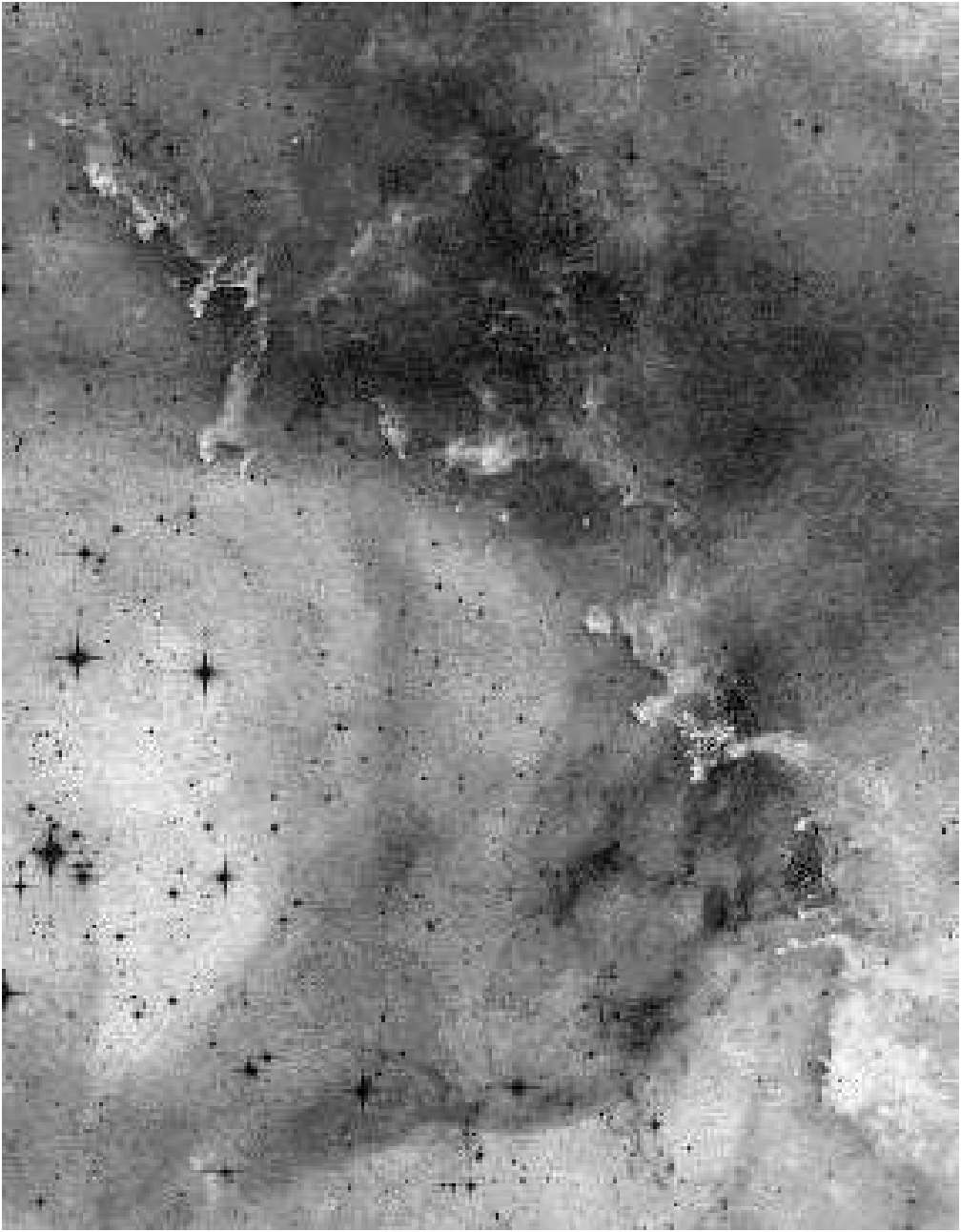}
 \caption{The northwestern edge of the cavity in the {Rosette Nebula}
showing a complex system of pillars, trunks and small
globules. Credit: Ignacio de la Cueva Torregrosa.}
\label{f:globs}
\end{figure}

 At the northwestern edge of the central cavity in the {Rosette Nebula,}
there is a large group of prominent dusty pillar structures or {\it
elephant trunks} (\citet{herbig74}, see Figure
\ref{f:globs}). \citet{schneps80} observed this region in CO and
$^{13}$CO (J=1$\rightarrow$0), and found that the morphology of the
molecular gas emission in the area follows very closely the optical
outline of the pillar features. A very interesting fact is that the
physical properties of the globules associated with the larger pillar
structures are not different from those of globules in clouds not
associated with HII regions. Individual globules in the area have
H$_2$ densities of $\sim 10^4$ cm$^{-3}$ and temperatures of
$\sim10$~K.  The trunks have velocity gradients of up to 1.1
km$\cdot$cm$^{-1}(\arcmin)^{-1}$, oriented along their direction of
elongation and these gradients might be responsible for the
morphology. The ``stretching'' of the trunks might have a time scale
of $\sim$3 to 6 $\times10^5$ yr, which is consistent with the
dynamical age of the HII region given an expansion rate of 20
km$\cdot$s$^{-1}$.

Later, the studies of \citeauthor{carlqvist98}
(\citeyear{carlqvist98,carlqvist02}) were dedicated to study the
internal structure of the elephant trunks. By constructing an
extinction map of the largest trunk from a B-band CCD image, they were
able to identify a clear helical structure, which they suggest might
originate from an alignment with a primordial helical magnetic field
(magnetic rope model). If true, this magnetic field would be present
before the HII region was formed, and it could be responsible for
maintaining the trunk-system and would explain the observed velocity
 gradient even after erosion from the winds of {NGC~2244.}

\citet{gac94} observed 4 of the densest `teardrops' located at the
tips of the elephant trunks. Using the IRAM 30m telescope they found
significant emission in $^{12}$CO(1-0 and 2-1), $^{13}$CO(1-0 and 2-1)
and CS(2-1 and 3-2). They confirmed kinetic temperatures for the
globules between 15 and 20 K and masses between 0.02 and 0.5
M$_\odot$. The $^{12}$CO line profiles had significant outflow (wing)
components, possibly indicating the pressure exerted by the ionized
gas in the HII region on the neutral gas of the teardrops.  Recently,
\citet{gahm07} made a complete survey of the entire chain of dark
trunks in the northwest edge of the central cavity. They observed 23
fields in H$\alpha$ using the 2.6m Nordic Optical Telescope, obtaining
images of filaments and globules at a resolution of 0.188 pix$^{-1}$
(about 300 AU at $d=1600$~pc). The observations revealed a total of
145 individual small globules, many of them located in isolation near
large pillars. They were able to obtain density profiles for the
globulettes by converting from $A_\alpha$ extinction, and showed that
most objects agreed well with $r^{-1.5}$ configurations typical of
protostellar conditions. The total densities of the globules varied
from $3\times10^3$ to $1\times 10^5$~cm$^{-3}$ and values of total
mass assuming uniform density were found to agree well with those
obtained for the few counterparts from the survey of
\citeauthor{gac94}. \citeauthor{gahm07} found that many objects had
radii under 10,000 AU, with a distribution peak around 25,000 AU. The
distribution of mass for the globulettes above a detection limit of 1
M$_J$ peaked at about 13 M$_J$, with only a few objects exceeding 60
M$_J$. No near-infrared counterparts have been found to be
convincingly associated with the globulettes, possibly indicating that
none of these objects is a genuine proplyd or a protostar. Virial
analysis yields the result that above 60\% of the globulettes have
conditions for gravitational contraction. Furthermore, the free-fall
times were estimated to be of a few times 10$^5$~yr, or about one
order of magnitude shorter than the photoevaporation times, estimated
to be of the order of 3 to 4 Myr. This large photoevaporation
 timescale is surprisingly large in the harsh environment of {NGC 2244,}
but may be due to a process of slow compression that follows the
erosion of the outer layers by photoevaporation. One of the most
 interesting conclusions from this study is that the small {Rosette
 Nebula} globules may be the predecessors to brown dwarfs or planetary
mass objects, reinforcing the idea that these kind of low mass objects
can form in isolation following a process of contraction similar to
stars, and not only by ejection or disruption of protoplanetary disks.

\section{Final Remarks}

 After 80 years of studies, the {Rosette Complex} is known as one of the
most important astrophysical laboratories for the study of the many
aspects of star formation.

 A solid basis for the investigation of the {Rosette Complex} has been
successfully set. We have now a very detailed picture of the physical
properties of the HII region and the molecular cloud, and we have
located the majority of the stars that have formed across the
 complex. {The Rosette} has been actively forming stars for the last 2 to
 3 Myr (based on the age of {NGC~2244)} and it will probably continue
doing so, as there is still sufficient molecular material ($>1\times
10^5$~M$_\odot$), placed in a heavily stimulated environment: The
large photodissociation region excavated by the OB association
 {NGC~2244} is in clear interaction with the adjacent molecular cloud,
which after several years of investigations has revealed an abundant
embedded population. Also, it appears that another population of
young, low mass stars could form in the surroundings of the central
 cavity {(NGC 2237,} {REFL10} and the small globules, if they are truly
predecesors to low mass objects). Could these be evidence of triggered
formation?

The formation of embedded clusters and associations in the adjacent
molecular cloud was clearly affected by the interaction with the
nebula, but we still have to understand the extent of that
interaction. The different cluster formation episodes appear to
 suggest that the current star forming properties of the {Rosette
 Molecular Cloud} could be related to the early evolution of the cloud,
and not only to the interaction with the HII region.

In the near future it will be necessary to investigate the possible
differences in the mechanisms of cluster formation, the variations in
the star formation efficiency across the cloud, the observed
properties of clusters and the definitive role of the local
environment. All of this information will be of enormous value for the
global study of star formation.

 Now that the second generation of clusters in the {Rosette} has been
identified, it is expected to be studied in even greater detail using
infrared spectroscopy. Such investigations will be of great help to
constrain the relative ages of the embedded populations and to
reconstruct the history of formation in the region. Also, the
near-future development of observational technologies and continuous
improvements in the modeling of the interstellar medium and formation
of clusters are promising, to say the least, for one of the most
notorious targets in modern astronomy.

\medskip
\par {\bf Acknowledgment}. We would like to thank Jonathan Williams
for insightful comments during the preparation of this chapter and
Joanna Levine for editing an early version of the manuscript. We
greatly acknowledge the referee Leo Blitz and the editor Bo Reipurth
for providing corrections that vastly improved the chapter text. We
also thank Harvey Liszt, Leisa Townsley, Jonathan Williams, Robert
Gendler, and Ignacio de la Cueva Torregrosa for generously sharing
images which we used in some of the figures.

Carlos Rom\'an-Z\'u\~niga wants to acknowledge CONACYT, Mexico for a
fellowship that sponsored his doctoral studies at the University of
Florida.

Elizabeth Lada acknowledges support from the NASA grant NNG05D66G
issued through the LTSA program to the University of Florida.  Data
obtained with the instrument FLAMINGOS and presented in this work were
collected under the NOAO Survey Program, "Towards a Complete
Near-Infrared Spectroscopic Survey of Giant Molecular Clouds" (PI:
E. Lada) which is supported by NSF grants AST97-3367 and AST02-02976
to the University of Florida. FLAMINGOS was designed and constructed
by the IR instrumentation group (PI: R. Elston) at the University of
Florida, Department of Astronomy, with support from NSF grant
AST97-31180 and Kitt Peak National Observatory.

This publication makes use of data products from the Two Micron All
Sky Survey, which is a joint project of the University of
Massachusetts and the Infrared Processing and Analysis
Center/California Institute of Technology, funded by the National
Aeronautics and Space Administration and the National Science
Foundation.

%%% THE BIBLIOGRAPHY
%%%
%%% CONSULT SECTION 3 OF "INSTRUCTIONS FOR AUTHORS" FOR HOW TO USE NATBIB.
%%% AUTHORS ARE ENCOURAGED TO USE EITHER THE "THEBIBLIOGRAPY" ENVIRONMENT
%%% BY UNCOMMENTING (DELETING THE "%" SYMBOL) THE COMMANDS BELOW, OR BY
%%% USING THE BIBTEX ENVIRONMENT. TO FIND OUT WHICH IS APPLICABLE TO YOUR
%%% CONTRIBUTION, CONSULT THE VOLUME EDITORhttp://www.433eros.com/spanish/headlines/y2005/22jul_perseids2005_spanish.htmS FOR YOUR PROCEEDINGS.
%%%

\end{document}